\documentclass[11pt,a4paper]{article}
\usepackage{hyperref}
\usepackage{amsmath,amsfonts,amssymb,amsthm,mathtools}
\usepackage{authblk}
\usepackage{latexsym,graphicx}
\usepackage{dsfont}
\usepackage{amscd}
\usepackage[dvipsnames]{xcolor}
\usepackage{extarrows} 
\numberwithin{equation}{section} 
\usepackage{yfonts}
\usepackage{authblk}
\usepackage{appendix}

\usepackage{comment}

\usepackage{tikz}
	 \usetikzlibrary{calc} 
	\usetikzlibrary{matrix,arrows,decorations.pathmorphing} 
\usepackage{tikz-3dplot}

\allowdisplaybreaks

\def\Xint#1{\mathchoice
{\XXint\displaystyle\textstyle{#1}}%
{\XXint\textstyle\scriptstyle{#1}}%
{\XXint\scriptstyle\scriptscriptstyle{#1}}%
{\XXint\scriptscriptstyle\scriptscriptstyle{#1}}%
\!\int}
\def\XXint#1#2#3{{\setbox0=\hbox{$#1{#2#3}{\int}$}
\vcenter{\hbox{$#2#3$}}\kern-.5\wd0}}

\def\pvint{\Xint-}
\newcommand{\pvsum}{\sideset{}{'}\sum}

\newcommand{\wedgecircold}{\mathrel{
\scalebox{0.8}{\hspace{-3.65pt}\setbox0\hbox{$\wedge$}
							\rlap{\hbox to \wd0{\hss \hspace{3.65pt}\raisebox{5.5pt}{$\circ$} \hss}}\box0} }}

\newcommand{\dotcirc}{\mathrel{ \!
\tikz[baseline={([yshift=-2.7pt]current bounding box.center)},scale=0.10]{
\draw (0,0) circle (1);
\filldraw[black] (0,0) circle (0.11);
}				
							\! }}
							 
\newcommand{\wedgecirc}{\mathrel{\!
\tikz[baseline={([yshift=-2.7pt]current bounding box.center)},scale=0.10]{
\def\a{40};
\pgfmathsetmacro{\c}{cos(\a) };
\pgfmathsetmacro{\s}{sin(\a) };
\draw (0,0) circle (1);
\draw[line width=0.4pt] (-\c,-\s ) -- (0,1);
\draw[line width=0.4pt] (\c,-\s ) -- (0,1);
}				
							\! }}

\newcommand{\re}{\mathrm{Re} }
\newcommand{\im}{\mathrm{Im} }
\newcommand{\aR}{a^\mathrm{R}}

\newcommand{\aI}{a^{\mathrm{I}}} 
\newcommand{\bsR}{\bs^\mathrm{R}}
\newcommand{\bsI}{\bs^{\mathrm{I}}} 

\newcommand{\sgn}{\mathrm{sgn}}
\newcommand{\ee}{\mathrm{e}}
\newcommand{\ii}{\mathrm{i}}
\newcommand{\dd}{\mathrm{d}}

\newcommand{\R}{{\mathbb R}}
\newcommand{\C}{{\mathbb C}}
\newcommand{\Z}{{\mathbb Z}}

\newcommand{\sutwo}{\mathfrak{su}(2)}

\newcommand{\ba}{\mathbf{a}}
\newcommand{\bb}{\mathbf{b}}

\newcommand{\be}{\mathbf{e}}
\newcommand{\bm}{\mathbf{m}}
\newcommand{\bn}{\mathbf{n}}
\newcommand{\bs}{\mathbf{s}}
\newcommand{\bt}{\mathbf{t}}
\newcommand{\bu}{\mathbf{u}}
\newcommand{\bv}{\mathbf{v}}

\newcommand{\bA}{\mathbf{A}}
\newcommand{\bB}{\mathbf{B}}
\newcommand{\bS}{\mathbf{S}}
\newcommand{\bU}{\mathbf{U}}
\newcommand{\bX}{\mathbf{X}}
\newcommand{\bsigma}{\boldsymbol{\sigma}}

\newcommand{\ua}{\underline{\ba}}
\newcommand{\ub}{\underline{\bb}}
\newcommand{\uu}{\underline{\bu}}
\newcommand{\uv}{\underline{\bv}}
\newcommand{\uA}{\underline{\bA}}
\newcommand{\uB}{\underline{\bB}}
\newcommand{\uU}{\underline{\bU}}

\newcommand{\cB}{\mathcal{B}}

\newcommand{\cH}{\mathcal{H}} 
\newcommand{\cI}{\mathcal{I}} 
\newcommand{\cL}{\mathcal{L}}
\newcommand{\cN}{\mathcal {N}}
\newcommand{\cT}{\mathcal{T}}

\newcommand{\eps}{\epsilon} 
\newcommand{\tT}{\tilde{T}}
\newcommand{\talpha}{\tilde{\alpha}}
\newcommand{\tV}{\tilde{V}}

\title{\Large{The non-chiral intermediate Heisenberg ferromagnet equation}}

\author[1]{Bjorn K. Berntson}
\author[2]{Rob Klabbers}
\author[1]{Edwin Langmann}

\affil[1]{Department of Physics, KTH Royal Institute of Technology, SE-106 91 Stockholm, Sweden}
\affil[2]{Institut f\"{u}r Physik, Humboldt-Universit\"{a}t zu Berlin,
Zum Gro{\ss}en Windkanal 2, 12489 Berlin, Germany}

\date{October 12, 2021}

\begin{document}

\begin{flushright}

	\footnotesize
	\textsc{nordita} 2021-058 \\
	HU-EP-21/41 \\
	HU-Mathematik-2021-04
\end{flushright}

\vspace{-15pt}
\bigskip

{\let\newpage\relax\maketitle}

\maketitle

\begin{abstract}
We present and solve 
 a soliton equation which we call the non-chiral intermediate Heisenberg ferromagnet (ncIHF) equation. 
This equation, which depends on a parameter $\delta>0$, describes the time evolution of two coupled spin densities propagating on the real line, and in the limit $\delta\to\infty$ it reduces to two decoupled half-wave maps (HWM) equations of opposite chirality. 
We show that the ncIHF equation is related to the $A$-type hyperbolic spin Calogero-Moser (CM) system in two distinct ways: 
(i) it is obtained as a particular continuum limit of a Inozemtsev-type spin chain related to this CM system, (ii) it has multi-soliton solutions obtained by a spin-pole ansatz and with parameters satisfying the equations of motion of a complexified version of this CM system. 
The integrability of the ncIHF equation is shown by constructing a Lax pair. 
We also propose a periodic variant of the ncIHF equation related to 
the $A$-type elliptic spin CM system.
\end{abstract} 

\section{Introduction}
\label{sec:intro}

The relationship between Calogero-Moser-Sutherland (CMS\footnote{We use the abbreviation CMS when we mean both the classical and quantum systems, and we use CM or CS when we mean specifically classical or quantum CMS systems, respectively.}) systems and quantum field theory is a fruitful and symbiotic one yielding new insights into both domains. For example, on the one hand, the conformal blocks of certain conformal field theories (CFTs) can be computed using eigenfunctions of CS-type Hamiltonians \cite{Estienne:2011qk, Isachenkov:2016gim}, whereas, on the other hand, one can describe the eigenvalues of CS Hamiltonians using the superpotential of $\mathcal{N}=2$ supersymmetric gauge theory \cite{Nekrasov:2009rc}. In the last decade, attempts to deepen this relationship have been made by studying whether certain continuum limits of CMS systems (as well as their Ruijsenaars-Schneider generalizations) can be connected to supersymmetric gauge theory \cite{bonelli2014,koroteev2016,koroteev2018,gorsky2020}.

\begin{figure}
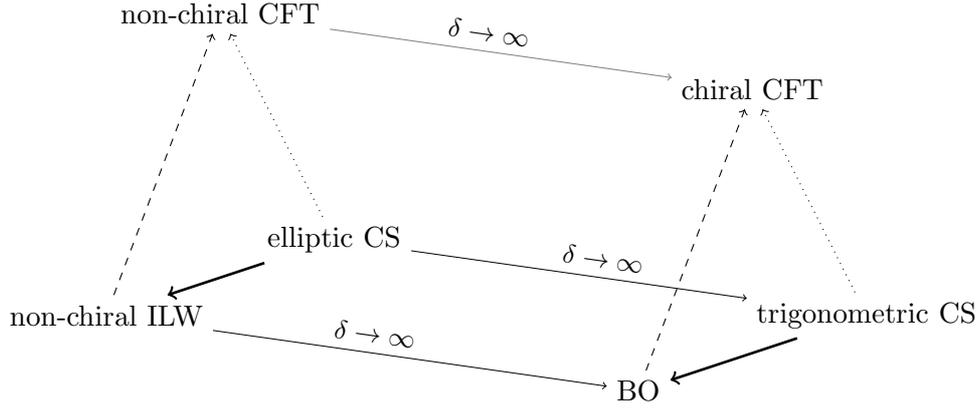

\centering
\tikz[scale=1]{
\def\x{7};
\def\y{-1};
\def\xshift{1.5};
\def\yshift{4};
\def\CMSxshift{3};
\def\CMSyshift{1};
\node (A) at (0,0) {non-chiral ILW};
\node (B) at (\x,\y) {BO};
\draw[->,line width=0.1pt] (A) -- (B);
\node[above,rotate=atan(\y/\x)] at (\x/2,\y/2) {\small $\delta \to \infty$};

\node (C) at (\x+\xshift,\y+\yshift) {chiral CFT};
\node (D) at (\x+\CMSxshift,\y+\CMSyshift) {trigonometric CS};
\draw[->,line width=1pt] (D) -- (B);
\draw[->,dashed] (B) -- (C);
\draw[->,dotted] (D) -- (C);

\node (E) at (\xshift,\yshift) {non-chiral CFT};
\node (F) at (\CMSxshift,\CMSyshift) {elliptic  CS};
\draw[->,line width=1pt] (F) -- (A);
\draw[->,dashed] (A) -- (E);
\draw[->,dotted] (F) -- (E);

\draw[->,line width=0.1pt] (F) -- (D);
\node[above,rotate=atan(\y/\x)] at (\x/2+\CMSxshift,\y/2+\CMSyshift) {\small $\delta \to \infty$};

\draw[->,line width=0.1pt, gray] (E) -- (C);
\node[above,rotate=atan(\y/\x)] at (\x/2+\xshift,\y/2+\yshift) {\small $\delta \to \infty$};

}
\caption{A diagram indicating the various relationships between the models discussed here: the thick lines indicate particular continuum limits, the dashed lines first quantizations and the dotted lines second quantizations; the lines marked by $\delta \to \infty$ indicate trigonometric limits.} 

 \label{fig:models}
\end{figure}

In particular, the trigonometric CM model\footnote{Unless stated otherwise, all CM and CS systems mentioned here and below are of $A$-type.} has a continuum limit given by the Benjamin-Ono (BO) equation \cite{Polychronakos1995,abanov2005}, which is a well-known integro-differential equation describing internal waves in deep water \cite{Benjamin1967,Ono1975}. 
Remarkably, both of these integrable systems can be related to a standard chiral CFT: this CFT accommodates a second quantization of the corresponding CS model \cite{jevicki1992,Azuma1994,Iso1995,Awata1995,carey1999} which, at the same time, is a (first) quantization of the BO equation \cite{abanov2005,abanov2009}. 
Recently, two of the authors (BKB and EL) together with Lenells introduced a non-chiral intermediate long-wave (ILW) equation providing a natural generalization of the above \cite{berntson2020}.
In particular,  the non-chiral ILW equation is a generalization of the BO equation in that it depends on a parameter $\delta>0$ and reduces to two decoupled BO equations of opposite chirality on the limit $\delta\to\infty$. 
Moreover, the quantum version of the non-chiral ILW equation is identical to a previously known second quantization of the elliptic CS system \cite{Langmann2000} (the elliptic CS system reduces to the trigonometric CS system in the limit $\delta\to\infty$); this second quantization is within a non-chiral CFT which reduces to two decoupled chiral CFTs in the limit $\delta\to \infty$.
See Fig.~\ref{fig:models} for a diagram of the relations between the different models. 

In this paper, we make steps towards generalizing this setup to the spin case, starting from the hyperbolic and elliptic spin CM models. It is known that an appropriate continuum limit in the rational case gives rise to an integrable equation describing the dynamics of a spin density on the real line known as the half-waves maps (HWM) equation \cite{zhou2015,lenzmann2018,lenzmann2020}. The integrability of the HWM equation was proved by G\'erard and Lenzmann \cite{gerard2018} by constructing its Lax pair representation. Moreover, it possesses multi-soliton solutions governed by spin CM dynamics, as recently found by us in \cite{berntsonklabbers2020}.
In the present paper, by taking a continuum limit of the hyperbolic spin CM system, we derive a non-chiral generalization of this HWM equation which, at the same time, can be interpreted as a spin analogue of the non-chiral ILW equation \cite{berntson2020}. 
For reasons elaborated in the paragraph containing \eqref{eq:IHF} below, we call this equation the \emph{non-chiral intermediate Heisenberg ferromagnet equation} (ncIHF). 
Inspired by \cite{gerard2018}, we find a Lax pair for the ncIHF equation. We also construct the multi-soliton solutions of the ncIHF equation
via a spin-pole ansatz with parameters whose dynamics is governed by the hyperbolic spin CM system,  in generalization of a result for the HWM equation obtained by us in \cite{berntsonklabbers2020}. 
We mention that the HWM equation has a periodic generalization related to the trigonometric CM system, and all results mentioned for the HWM equation are known also in the periodic case \cite{zhou2015,lenzmann2018,lenzmann2020,gerard2018,berntsonklabbers2020}; however, while we introduce a periodic generalization of the ncIHF equation related to the elliptic spin CM model below, the focus of the present paper is on the real-line case. 

The ncIHF equation describes the time evolution of two $S^2$-valued functions $\bu$ and $\bv$ on the real line, i.e., 
$\bu=\bu(x,t)$ and $\bv=\bv(x,t)$ with $x\in\mathbb{R}$ the spatial coordinates, $t\in\R$ time,  
$\bu=(u^1,u^2,u^3)\in\mathbb{R}^3$ such that  $\bu^2:= (u^1)^2+(u^2)^2+(u^3)^2=1$ 
and similarly for $\bv$.  
It is given by  the following coupled system of equations, 
\begin{equation}
\label{eq:ncIHF}
\begin{split}
\bu_t=&+ \bu\wedge T\bu_x - \bu\wedge \tT\bv_x, \\
\bv_t=&- \bv\wedge T\bv_x + \bv\wedge \tT\bu_x , 
\end{split}
\end{equation}
with the integral operators\footnote{We define $T$ and $\tT$ for on $\C$-valued functions $f$ on $\R$; the extension to $\C^3$-valued functions $\bu=(u^1,u^2,u^3)$ on $\R$ is determined by linearity: $T\bu=(Tu^1,Tu^2,Tu^3)$ and similarly for $\tT$.}
\begin{equation} 
\label{eq:TTh}
\begin{split} 
(Tf)(x)\coloneqq &\frac{1}{2\delta}\pvint_{\R}\coth\left(\frac{\pi}{2\delta}(x'-x)\right)f(x') \,\dd{x'}, \\
(\Tilde{T} f)(x)\coloneqq &\frac{1}{2\delta}\int_{\R}\tanh\left(\frac{\pi}{2\delta}(x'-x)\right)f(x') \,\dd{x'}, 
\end{split} 
\end{equation} 
where $\bu\wedge \bv:=(u^2v^3-u^3v^2,u^3v^1-u^1v^3,u^1v^2-u^2v^1)$ is the usual wedge product between three-vectors, $\bu_t\coloneqq \frac{\partial}{\partial t}\bu$ etc., and 
$\pvint$ indicates the Cauchy principal value prescription of the singular integral. 
Note that \eqref{eq:ncIHF} is consistent with the assumption that $\bu$ and $\bv$ are $S^2$-valued functions: if $\bu^2=\bv^2=1$ holds true at time $t=0$, then it is true for all times (see \eqref{eq:dotuusquare}). 
As explained in Section~\ref{subsec:continuum_limit}, it is natural to interpret $\bu$ and $\bv$ as spin densities which have opposite chiralities. Boundary conditions for the functions $\bu$ and $\bv$ are given in \eqref{eq:BC}. 

The ncIHF equation is a Hamiltonian system with the Hamiltonian 
\begin{equation}
\label{eq:ncIHFHamiltonian}
\mathcal{H}=-\frac1{2}\int_{\R} \left(\bu\cdot T\bu_{x}+\bv\cdot T\bv_x-\bu\cdot \tT\bv_x-\bv\cdot\tT\bu_x\right)\,\dd{x}
\end{equation}
and the non-trivial Poisson brackets
\begin{equation}
\label{eq:ncIHFPoissonbrackets}
\begin{split} 
\{u^a(x),u^b(x')\} &=-\eps_{abc}u^c(x)\delta(x-x'),\\
\{v^a(x),v^b(x')\} &=+\eps_{abc}v^c(x)\delta(x-x'); 
\end{split} 
\end{equation}
we use indices $a,b,c=1,2,3$ to distinguish the components of 3-vectors, $\eps_{abc}$ is the completely anti-symmetric symbol such that $\eps_{123}=+1$,  
and we use the usual summation convention so that $(\bu\wedge\bv)^{a}=\eps_{abc}u^{b}v^{c}$ and $\bu\cdot\bv=u^{a}v^{a}$. 
A proof that the Hamiltonian system \eqref{eq:ncIHFHamiltonian}--\eqref{eq:ncIHFPoissonbrackets} yields \eqref{eq:ncIHF} can be found in Appendix~\ref{app:EOM}.
 
For the physical interpretation of \eqref{eq:ncIHF}, it is useful to represent the operators in \eqref{eq:TTh} in Fourier space \cite[Eq.\ (A3)]{berntson2020}:  
\begin{equation}
\label{eq:TT}
\begin{split} 
(\widehat{Tf})(k) &=\mathrm{i}\coth(k\delta)\hat{f}(k),\\ 
(\widehat{\tT f})(k) &=\frac{\mathrm{i}}{\sinh(k\delta)}\hat{f}(k),
\end{split} 
\end{equation}
where $\hat{f}(k)=\int_{\R}f(x)\mathrm{e}^{-\ii kx}\,\dd{x}$ ($k\in\R$). This makes manifest that, in the limit $\delta\to \infty$, \eqref{eq:ncIHF} reduces to 
\begin{equation}
\label{eq:HWM}
\begin{split}
\bu_t=&+ \bu\wedge H\bu_x , \\
\bv_t=&- \bv\wedge H\bv_x , 
\end{split}
\end{equation}
with  $H$ the Hilbert transform: $(\widehat{Hf})(k)=\ii\, \mathrm{sgn}(k)\hat{f}(k)$ or, equivalently, 
\begin{equation}
\label{eq:Hilbert}
(Hf)(x)\coloneqq \frac{1}{\pi}\pvint_{\R} \frac{f(x')}{x'-x}\,\dd{x'}.
\end{equation}
The first equation in \eqref{eq:HWM} is the HWM equation \cite{zhou2015,lenzmann2018}, and the second is obtained from the first by the parity transformation $\bu\to \bv=P\bu$ where $(P\bu)(x,t):=-\bu(-x,t)$. 
Thus, \eqref{eq:HWM} is a system of decoupled HWM equations, which both are chiral in the sense that (i) they differ by a sign and (ii) one is transformed into the other by a parity transformation (we discuss the physical interpretation of this chirality in Section~\ref{sec:onesoliton}). However, the system \eqref{eq:ncIHF} is \emph{non-chiral} in the sense that it is invariant under the parity transformation $(\bu,\bv)\to (P\bv,P\bu)$ exchanging also $\bu$ and $\bv$. 

We recall that the ILW equation can be obtained from the BO equation by replacing the Hilbert transform with the integral transform $T$ in \eqref{eq:TT}: $H\to T$ \cite{berntson2020}. 
This suggests to introduce the following equation generalizing the HWM equation $\bu_t=\bu\wedge H\bu_x$: 
\begin{equation} 
\label{eq:IHF}
\bu_t = \bu\wedge T\bu_x. 
\end{equation} 
As will be discussed, this is a natural generalization of the HWM equation since it is related to Inozemtsev's elliptic generalization of the Haldane-Shastry spin chain \cite{inozemtsev1990} in a similar way as the HWM equation is related to the Haldane-Shastry spin chain \cite{zhou2015,lenzmann2018,lenzmann2020}. One can show that \eqref{eq:IHF} reduces to the Heisenberg ferromagnet equation $\bu_t = \bu\wedge \bu_{xx}$ in a certain limit $\delta\to 0^+$ (see Appendix~\ref{app:HFlimit}), and we therefore call it the {\em intermediate Heisenberg ferromagnet} (IHF) equation. 

We call \eqref{eq:ncIHF} the non-chiral IHF (ncIHF) equation since it is related to the IHF equation \eqref{eq:IHF} in the same way as the non-chiral ILW equation is related to the ILW equation \cite{berntson2020}. 
Moreover, we expect that the non-chiral IHF equation can be quantized within a non-chiral CFT which, at the same time, is a second quantization of the elliptic spin CS system, in generalization of results discussed above \cite{berntson2020} (see also Fig. \ref{fig:models}). As will be shown, the IHF equation can be formally obtained by the reduction $\bu=-\bv$ of the ncIHF equation; however, due to a mismatch of allowed boundary conditions, our Lax pair for the ncIHF equation does {\em not} reduce to a Lax pair of the IHF equation, and our soliton solutions of the ncIHF equation do {\em not} reduce to soliton solutions of the IHF equation. Thus, while we have suggestive arguments that the IHF equation is an integrable interpolation between the HWM equation and the Heisenberg ferromagnet equation, we do not have a proof.

The focus in the present paper is on the ncIHF equation on the real line.
However, we expect that our results can be generalized to the periodic case: the {\em periodic ncIHF equation} is also defined by \eqref{eq:ncIHF} but with the $S^2$-valued functions $\bu=\bu(x,t)$ and $\bv=\bv(x,t)$ restricted by the periodicity conditions $\bu(x+L,t)=\bu(x,t)$ and  $\bv(x+L,t)=\bv(x,t)$, with $L>0$ the spatial period; clearly, one can interpret this as an equation on a circle of circumference $L$. 
The integral operators $T$ and $\tT$ in the periodic case have the same Fourier space representation \eqref{eq:TT} as in the real-line case but with the Fourier variables $k$ restricted to integer multiples of $2\pi/L$ \cite{berntson2020}. 
In position space, these integral operators are given by \cite{berntsonlangmann2021}
\begin{equation} 
\label{eq:TTe} 
\begin{split} 
(Tf)(x)&\coloneqq \frac1{\pi}\pvint_{-L/2}^{L/2}\zeta_1(x'-x)f(x')\,\dd{x}' ,\\
(\tT f)(x)&\coloneqq   \frac1{\pi}\int_{-L/2}^{L/2}\zeta_1(x'-x+\ii\delta)f(x')\,\dd{x}',  
\end{split}
\end{equation}
with
\begin{equation} 
\label{eq:zeta}
\zeta_1(z) \coloneqq \lim_{M\to\infty}\sum_{m=-M}^M\frac{\pi}{L}\cot\left(\frac{\pi}{L}(z+2\ii \delta m)\right) = \zeta(z)-\frac{\eta_1}{\omega_1}z \quad (z\in\C)
\end{equation} 
where  $\zeta(z)$ is the Weierstrass function with half-periods $(L/2,\ii\delta)$ and $\eta_1/\omega_1=\zeta(L/2)/(L/2)$ \cite{DLMF}.
We introduce the periodic ncIHF equation in the present paper since the derivation of the ncIHF equation as a continuum limit of a spin CM system is easier in the periodic case $L<\infty$; the corresponding result for the ncIHF equation on the real line is then obtained by taking the limit $L\to\infty$ (note that $\zeta_1(z)\to \coth(\pi z/2\delta)$ as $L\to\infty$). 
We expect that our results below can be generalized to the periodic case. However, these generalizations are non-trivial and thus left to future work. 

\paragraph{Plan of the rest of the paper.}
Section~\ref{sec:summary} contains a 
summary of our main results, including technical details.  
In Section~\ref{sec:model}, we show how to derive the ncIHF equation as
a continuum limit of the hyperbolic spin CM model and discuss some of its basic properties. 
In Section~\ref{sec:Lax}, we derive a Lax pair for the ncIHF equation. 
In Section~\ref{sec:solitons}, we derive a family of $N$-soliton solutions. 
In Section \ref{sec:constraintsandexamples},  we present a novel way to solve the constraints appearing in our $N$-soliton solutions, discuss the geometric and physical interpretation of our soliton solutions, and give examples. 
In our final Section~\ref{sec:discussion} we give an informal summary of our results and discuss directions for future research. 
In the appendices, we derive a Hamiltonian formulation for the ncIHF equation (Appendix \ref{app:EOM}), show how \eqref{eq:ncIHF} is formally related, via \eqref{eq:IHF}, to the well-known Heisenberg ferromagnet equation (Appendix \ref{app:HFlimit}), collect some identities we use throughout the paper (Appendix \ref{app:propalpha}), 
provide some technical details related to conservation laws (Appendix \ref{app:verification}), derive a generalization of the Cotlar identity needed in the proof our Lax pair results (Appendix~\ref{app:Cotlar}), and compute the energy of one-soliton solutions  (Appendix \ref{app:eps1}). 

\section{Notation and summary}
\label{sec:summary}
We collect notation used in the rest of the paper (Section~\ref{subsec:def}), and we summarize our main results (Section~\ref{subsec:summary}). 
\subsection{Notation} 
\label{subsec:def}
The symbol $\coloneqq$ indicates definitions. We use the shorthand notations 
\begin{equation} 
\sum_{k\neq j}^N\coloneqq  \sum_{\substack{k=1\\k\neq j}}^N,\quad \pvsum_{j,k=1}^N\coloneqq \sum_{\substack{j,k=1\\j\neq k}}^N,\quad \dot X\coloneqq \frac{\dd}{\dd{t}}X,\quad \partial_x\coloneqq \frac{\partial}{\partial x}, 
\end{equation} 
and $f^\prime(z)\coloneqq \partial_z f(z)$.
We write $\re(a)$ and $\im(a)$ for the real and imaginary parts of $a\in\C$, $\ii\coloneqq\sqrt{-1}$, and $*$ denotes complex conjugation. 
For $\bs=(s^1,s^2,s^3)\in\C^3$, $\re(\bs)$, $\im(\bs)$ and $\bs^*$ are defined component-wise, i.e., $\re(\bs)\coloneqq (\re(s^1),\re(s^2),\re(s^3))$, etc. 
For $a\in\C$ and $\bs\in\C^3$, we use the notation
\begin{equation} 
\label{eq:aRaI} 
\aR\coloneqq\re(a),\quad \aI\coloneqq\im(a),\quad \bsR \coloneqq\re(\bs),\quad \bsI\coloneqq\im(\bs). 
\end{equation} 

To simplify notation, we often use the parameter 
\begin{equation} 
\label{eq:kappa} 
 \kappa\coloneqq \frac{\pi}{2\delta}
\end{equation} 
instead of $\delta$, and we introduce the special functions 
\begin{equation}
\label{eq:alpha}
 \alpha(z) \coloneqq \kappa \coth(\kappa z), \qquad  \talpha(z) \coloneqq  \alpha(z+\ii\delta) = \kappa\tanh(\kappa z) 
\end{equation}
and 
\begin{equation}
\label{eq:V}
V(z)\coloneqq \frac{\kappa^2}{\sinh^2(\kappa z)},\qquad \tV(z)\coloneqq V(z+\ii\delta) = -\frac{\kappa^2}{\cosh^2(\kappa z)}
\end{equation}
for $z\in\mathbb{C}$. Note that $V(z)$ defines the two-body interaction potential of hyperbolic CMS systems, and the functions $\alpha(z)$ and $\talpha(z)$ appear as kernels in the definition of the integral operators $T$ and $\tT$ \eqref{eq:TTh}, 
\begin{equation} 
\label{eq:TTalpha} 
\begin{split} 
(Tf)(x) & = \frac1{\pi}\pvint_{\R}\alpha(x'-x)f(x') \,\dd{x'}, \\
(\tT f)(x)& = \frac1{\pi}\int_{\R}\tilde\alpha(x'-x)f(x') \,\dd{x'}.
\end{split}
\end{equation} 
Moreover, the functions $\alpha(z)$ and $V(z)$ satisfy well-known identities playing a fundamental role in the theory of CMS systems  \cite{calogero2001}; we collect these identities in Appendix~\ref{app:propalpha}, for the convenience of the reader.

We also need the elliptic generalizations of the CMS potentials $V(z)$ and $\tV(z)$ in \eqref{eq:V}; these are given by 
\begin{equation} 
\label{eq:wp1} 
\wp_1(z)\coloneqq-\zeta^\prime_1(z) =\wp(z)+\frac{\eta_1}{\omega_1},
\end{equation} 
where $\wp(z)$ is the Weierstrass $\wp$-function with half-periods $(L/2,\ii\delta)$ \cite{DLMF}, and 
\begin{equation} 
\label{eq:twp1} 
\tilde{\wp}_1(z)\coloneqq \wp_1(z+\ii\delta), 
\end{equation} 
respectively, with $\zeta_1(z)$ defined in \eqref{eq:zeta}. 

To state and prove our results about Lax pairs,  we use the well-known bijection between real three-vectors $\bu=(u^1,u^2,u^3)$ and $\sutwo$-matrices $\uu$ defined by the Pauli matrices $\boldsymbol{\sigma}\coloneqq(\sigma_1,\sigma_2,\sigma_3)$, 
\begin{equation} 
\uu \coloneqq  \bu\cdot\bsigma = \left(\begin{array}{cc} u^3 & u^1-\ii u^2 \\ u^1+\ii u^2 & -u^3 \end{array}\right).  
\end{equation} 
This allows us to write the ncIHF equation \eqref{eq:ncIHF} in matrix form as 
\begin{equation}
\label{eq:ncIHF1}
\begin{split}
\uu_t=&\frac{1}{2\ii}\big[\uu,T\uu_x-\tT\uv_x\big], \\
\uv_t=&\frac{1}{2\ii}\big[\uv, \tT\uu_x-T\uv_x\big], 
\end{split}
\end{equation}
where $[\cdot,\cdot]$ is the commutator and $T\uu_x \coloneqq \underline{T\bu_x}$, etc.\ (this follows from the well-known identity $[\uu,\uv]=2\ii\underline{\bu\wedge\bv}$ for all $\bu,\bv\in\R^3$). 

\subsection{Summary of results}
\label{subsec:summary} 
We summarize our results, including technical details, and discuss their significance. We also mention open questions for future research. 
Items (i), (ii) and (iii) are independent from each, and each of them can be read without the others. 
Items (iv)--(vi) are supplementary results to (i)--(iii). Item (vii) summarizes what we know about the phenomenology of our multi-soliton solutions, based on analytic arguments and numerical results. 

\begin{itemize} 
\item[(i)] {\bf Continuum limit of elliptic spin chain.} 
We propose a variant of the  Inozemtsev spin chain \cite{inozemtsev1990} and argue that it can be obtained from the elliptic spin CM system using some variant of the Polychronakos freezing trick \cite{polychronakos1993}:
\begin{equation} 
\label{eq:spinchain2}
\begin{split} 
H = &\; -\frac{J}2 \pvsum_{j,k=1}^N \wp_1((j-k)L/N)\left( \bsigma_j\cdot\bsigma_k +  \tilde{\bsigma}_j\cdot\tilde{\bsigma}_k \right) \\ 
 &\; +\frac{J}2 \sum_{j,k=1}^N \tilde{\wp}_1((j-k)L/N)\left(  \bsigma_j\cdot\tilde{\bsigma}_k +  \tilde{\bsigma}_j\cdot\bsigma_k \right) , 
 \end{split} 
\end{equation} 
with $S=1/2$ quantum spin operators $\bsigma_j=(\sigma_j^1,\sigma_j^2,\sigma_j^3)$ and $\tilde{\bsigma}_j=(\tilde{\sigma}_j^1,\tilde{\sigma}_j^2,\tilde{\sigma}_j^3)$ satisfying the commutator relations
\begin{equation} 
\label{eq:spincommutators2} 
[\sigma_j^{a},\sigma_k^b]=2\ii\delta_{jk}\eps_{abc}\sigma^c_j,\quad [\tilde{\sigma}_j^{a},\tilde{\sigma}_k^b]=-2\ii\delta_{jk}\eps_{abc}\tilde{\sigma}^c_j 
\end{equation} 
for $a,b,c=1,2,3$ and $j,k=1,\ldots,N$ (the other commutators of spin operators vanish), with $J>0$ a coupling constant.
This model describes a quantum spin system on a lattice with $N$ sites distributed equidistantly over a circle with circumference $L$; on each site $j=1,\ldots,N$, there are two kinds of spins, $\bsigma_j$ and $\tilde{\bsigma}_j$; spins of the same kind interact via the long-range coupling $\wp_1(z)$, which becomes singular at short distances $z$, and spins of different kinds interact via the non-singular, long-range coupling $-\tilde{\wp}(z)$, which vanishes as $\delta\to\infty$. Note the different signs in the definition of the Poisson brackets for the different kinds of spins; obviously, one could get conventional  spin operators by changing $\tilde{\bsigma}_j\to -\tilde{\bsigma}_j$, but our conventions make manifest that the system is ferromagnet in the limit $L\to\infty$: the ground state corresponds to all spins pointing in the same direction (since $V(x)$ and $-\tV(x)$ are both are non-negative for all real $x$). 

We follow \cite{zhou2015,lenzmann2020} and derive the Hamiltonian defining the periodic ncIHF equation as a continuum limit of a classical version of the the elliptic spin chain \eqref{eq:spinchain2}--\eqref{eq:spincommutators2}; see Section~\ref{subsec:continuum_limit}. This gives, as a limit $L\to\infty$, a relation between the hyperbolic spin CM system and the ncIHF equation on the real line. 

\item[(ii)] {\bf Lax pair.}  
We show that the ncIHF equation on the real line admits a Lax pair $(\cL,\cB)$ consisting of operators acting on the Hilbert space of square-integrable functions $\Psi:\R\to \C^4$. 
The explicit forms of these operators are
\begin{equation} 
\label{eq:KLdef}
(\cL\Psi)(x)=\;\frac1\pi \int_{\R} K_{\cL}(x,x')\Psi(x')\,\dd{x'}
\end{equation} 
with an integral kernel given by the following $4\times 4$ matrix,
\begin{equation}
\label{eq:KL} 
\begin{split} 
K_{\cL}(x,x')\coloneqq &\; \alpha(x'-x)\left(\begin{array}{cc} {\uu}(x')-{\uu}(x) & 0\\ 
0 & -{\uv}(x')+{\uv}(x)\end{array}\right) 
 \\ &\; 
+\talpha(x'-x)\left(\begin{array}{cc} 0 &-{\uv}(x')+{\uu}(x)\\
{\uu}(x')-{\uv}(x) & 0\end{array}\right)
\end{split} 
\end{equation} 
and, similarly, 
\begin{equation}
\label{eq:KBdef} 
(\cB\Psi)(x)=\;\frac{1}{2\pi\ii}\pvint_{\R} \big(K_{\cB,1}(x,x')\Psi_{x'}(x')+K_{\cB,2}(x,x')\Psi(x')+K_{\cB,3}(x,x')\Psi(x)\big)\,\dd{x'} 
\end{equation}
with 
\begin{equation}
\label{eq:KB}
\begin{split}
K_{\cB,1}(x,x')\coloneqq &\; \alpha(x'-x)\left(\begin{array}{cc} {\uu}(x')+{\uu}(x) & 0\\ 
0 & -{\uv}(x')-{\uv}(x)\end{array}\right) \\
&\; +\talpha(x'-x)\left(\begin{array}{cc} 0 &-{\uv}(x')-{\uu}(x)\\
{\uu}(x')-{\uv}(x) & 0\end{array}\right), \\
K_{\cB,2}(x,x')\coloneqq &\;  \alpha(x'-x)\left(\begin{array}{cc} -{\uu}_{x'}(x')& 0 \\ 
0& {\uv}_{x'}(x')\end{array}\right) + \talpha(x'-x)\left(\begin{array}{cc} 0  & {\uv}_{x'}(x') \\ 
-{\uu}_{x'}(x')& 0 \end{array}\right),  \\
K_{\cB,3}(x,x')\coloneqq &\;  \alpha(x'-x)\left(\begin{array}{cc} {\uu}_{x'}(x')& 0 \\ 
0& -{\uv}_{x'}(x')\end{array}\right) + \talpha(x'-x)\left(\begin{array}{cc} -\uv_{x'}(x')  & 0 \\ 
0 & \uu_{x'}(x') \end{array}\right); 
\end{split}
\end{equation}
see Section~\ref{sec:Lax} for a mathematical formulation and derivation of this result.
 
This Lax pair gives an infinite number of conservation laws for the ncIHF equation, as usual: $\cI_n=\mathrm{tr}(\cL^n)$ for $n=2,3,\ldots$ are conserved;\footnote{We exclude $n=1$ since $\cI_1=0$.} in particular, $\cI_2$ is essentially the ncIHF Hamiltonian \eqref{eq:ncIHFHamiltonian}; see \eqref{eq:In}  and \eqref{eq:Itwo} for precise statements.

\item[(iii)] {\bf Multi-soliton solutions.} We construct multi-soliton solutions of the ncIHF equation with a pole ansatz depending on variables satisfying the time evolution equations of the hyperbolic spin CM system and with certain constraints on the initial conditions of these variables. 

To be specific, our multi-soliton result is as follows: {\em For an arbitrary integer $N\geq 1$, 
\begin{equation}
\label{eq:solution}
\begin{split} 
\bu(x,t) & = \bm_{0} + \ii \sum_{j=1}^N\bs_j(t) \alpha(x -a_j(t) + \ii \delta/2) -  \ii \sum_{j=1}^N \bs_j^*(t) \alpha(x -a_j^*(t) - \ii \delta/2) ,\\
\bv(x,t) & = \bm_{0} + \ii \sum_{j=1}^N\bs_j(t) \alpha(x -a_j(t) - \ii \delta/2) -  \ii \sum_{j=1}^N \bs_j^*(t) \alpha(x -a_j^*(t) + \ii \delta/2)
\end{split} 
\end{equation} 
is a solution of the ncIHF equation \eqref{eq:ncIHF} satisfying  $\bu(x,t)^2=\bv(x,t)^2=1$ provided the variables  $(a_j(t),\bs_j(t))\in \C\times\C^3$, $\delta/2<\im(a_j(t))<3\delta/2$,  satisfy the following two requirements: (a) they evolve in time according the the following equations of motion 
\begin{equation}
\label{eq:CM}
\begin{split}
\dot{\bs}_j &= -2 \sum_{k\neq j}^N \bs_j \wedge \bs_kV(a_j- a_k), \\
\ddot a_j &= -2\sum_{k\neq j}^N\bs_j\cdot\bs_k V^\prime(a_j-a_k), 
\end{split}\quad (j=1,\ldots,N), 
\end{equation}
(b) the initial conditions are given by 
\begin{equation} 
\label{eq:ajsjt=0} 
\begin{split} 
&a_j(0)=a_{j,0} ,\quad \bs_j(0)=\bs_{j,0},\\
&\dot{a}_j(0)=  -\frac{\bs_{j,0}^*\wedge \bs_{j,0}}{\bs_{j,0}^*\cdot\bs_{j,0}} \cdot \bigg( \ii \bm_{0}-\sum_{k\neq j}^{N} \bs_{k,0} \alpha(a_{j,0}-a_{k,0}) + \sum_{k =1 }^{N} \bs_{k,0}^*\talpha(a_{j,0}-a_{k,0}^*)  \bigg)  
\end{split} 
\end{equation} 
for $j=1,\ldots,N$, with $\bm_{0}\in\R^3$, $(a_{j,0},\bs_{j,0})\in\C\times\C^3$, $\delta/2<\im(a_{j,0})<3\delta/2$, satisfying the constraints
\begin{equation} 
\label{eq:constraintsbs}
 \bs_{j,0}^2=0,\quad \bs_{j,0}\cdot \bigg( \ii \bm_{0}-\sum_{k\neq j}^{N} \bs_{k,0} \alpha(a_{j,0}-a_{k,0}) + \sum_{k =1 }^{N} \bs_{k,0}^*\talpha(a_{j,0}-a_{k,0}^*)  \bigg)=0,
\end{equation} 
and 
\begin{equation} 
\label{eq:constraintsbm}
\bm_{0}^2 + 4\kappa^2\Bigg(\sum_{j=1}^N \im\, \bs_{j,0}\Bigg)^2 = 1; 
\end{equation} 
}see Section~\ref{sec:solitonderivation} for a derivation of this result. 

Note that \eqref{eq:CM} are the equations of motion of the hyperbolic spin CM system; cf.\ \eqref{eq:hCMEOM} in the limiting case $L\to\infty$ where $\wp_1(z)$ reduces to $V(z)$. 

\item[(iv)] {\bf Complex-valued multi-soliton solutions.} Our derivation of multi-soliton solutions gives, without extra work, a larger class of solutions depending on two sets of $\C\times\C^3$-valued variables $(a_j(t),\bs_j(t))$, $j=1,\ldots,N$,  and $(b_k(t),\bt_j(t))$, $k=1,\ldots,M$, for arbitrary variable numbers $N,M\geq 0$ such that $N+M\geq 1$; the special case  stated in (iii) above, which corresponds to $N=M\geq 1$, $b_j(t)=a_j^*(t)\coloneqq a_j(t)^*$, $\bt_j(t)=\bs_j^*(t)\coloneqq \bs_j(t)^*$, gives the real-valued solutions most interesting from a physics point of view. In general, one obtains $\C^3$-valued solutions $\bu(x,t)$ and $\bv(x,t)$ of \eqref{eq:ncIHF} satisfying $\bu(x,t)^2=\bv(x,t)^2=\rho^2$ for arbitrary fixed $\rho\in\C\setminus\{0\}$; see Section~\ref{sec:complexsolutions} for details. 

\item[(v)] {\bf Energy of multi-soliton solutions.} We obtain the following remarkably simple formula for the  total energy of the real $N$-soliton solution, 
$E_{N}\coloneqq \cH$ with $\cH$ in \eqref{eq:ncIHFHamiltonian} for $\bu$, $\bv$ in \eqref{eq:solution}--\eqref{eq:constraintsbm}: 
\begin{equation} 
\label{eq:EN}
E_N= -\pi   \sum_{j,k=1}^N \left(   \bs_j\cdot\bs_k^* \tV(a_j-a_k^*) +   \bs_j^*\cdot\bs_k \tV(a_j^*-a_k) \right); 
\end{equation} 
see Section~\ref{sec:energy}.\footnote{Note that the minus sign on the right-hand side in \eqref{eq:EN} cancels the one in $\tV(z)=-\kappa^2/\cosh^2(\kappa z)$.} 
Viewing \eqref{eq:EN} as a Hamiltonian, it would be interesting to study the possible dynamical systems associated with it; we leave the investigation of this to future work. It also would be interesting to extend this investigation to our complex-valued solutions in Section~\ref{sec:complexsolutions}.

\item[(vi)] {\bf Method to generate initial conditions for multi-soliton solutions.}
We derive the following result which provides a simple method to solve the constraints \eqref{eq:constraintsbs}--\eqref{eq:constraintsbm} and thus find allowed initial conditions for the $N$-soliton solutions  \eqref{eq:solution}--\eqref{eq:ajsjt=0}: {\em (a) For fixed $\bn_0\coloneqq \bm_{0}/|\bm_{0}|\in S^2$, pick $N$ initial poles and $N$ unit vectors in $\R^3$ arbitrarily: $a_{j,0}=\aR_j+\ii\aI_j$ with $\aR_j\in\R$, $\delta/2<\aI_j<3\delta/2$ and  $\bn_{j,3}\in S^2$  for $j=1,\ldots,N$. (b) For each $j=1,\ldots,N$,  choose two orthogonal unit vectors $\bn_{j,1},\bn_{j,2}$ in the plane orthogonal to $\bn_{j,3}$: $(\bn_{j,1},\bn_{j,2})\in S^2\times S^2$ such that $\bn_{j,1}\cdot\bn_{j,2}=0$ and $\bn_{j,1}\wedge\bn_{j,2}=\bn_{j,3}$, and set  $\bn_{j,12}\coloneqq \bn_{j,1}+\ii\bn_{j,2}$. (c) Construct the complex $N\times N$ matrices $A=(A_{jk})_{j,k=1}^N$, $B=(B_{jk})_{j,k=1}^N$ and the complex $N$-vector $C=(C_j)_{j=1}^N$ with entries
\begin{equation}
\label{matrixdefinitions1}
\begin{split}
A_{jk}=&-\ii\bn_{j,12}^*\cdot \bn_{k,12}\tanh(\kappa(a_{j,0}^*-a_{k,0})),  \\  
B_{jk}=& \ii (1-\delta_{jk})\bn_{j,12}^*\cdot\bn_{k,12}^*\coth(\kappa(a_{j,0}^*-a_{k,0}^*)),
\end{split}
\end{equation}
and $C_j= 2\bn_0\cdot \bn_{j,12}^*$, compute the complex $N$-vector $X=(X_j)_{j=1}^N$ determined by the linear system
\begin{equation}
\label{combinedlinearsystem1}
    \left(\begin{array}{cc}
    A& B \\
    -B^* & -A^* 
    \end{array}\right)\left(\begin{array}{c}
         X  \\ X^*
    \end{array}\right)=\left(\begin{array}{c} C \\ -C^* \end{array}\right)
\end{equation}
where $A^*=(A_{jk}^*)_{j,k=1}^N$ etc.,\footnote{Note that the $2N\times 2N$ matrix in \eqref{combinedlinearsystem1} is self-adjoint.} and compute 
\begin{equation} 
\label{eq:msolution} 
\bX_j= X_j \bn_{12,j} \quad (j=1,\ldots,N),\quad 
m = \bigg( 1 + \bigg( \sum_{j=1}^N \im(\bX_j)\bigg)^2 \bigg)^{-1/2} .
\end{equation} 
(d) Then the following provides a solution of the constraints \eqref{eq:constraintsbs}--\eqref{eq:constraintsbm}, 
\begin{equation} 
\label{eq:bsjbbb}
\bs_{j,0} = \frac{m}{2\kappa}\bX_j \quad (j=1,\ldots,N),\quad \bm_{0}=m\bn_0, 
\end{equation}  
and this solution is independent of the choice $\bn_{j,1},\bn_{j,2}$ in (b)}; see Section~\ref{sec:constraints}. 

In particular, for the one-soliton case ($N=1$), \eqref{combinedlinearsystem1} gives $X_1=C_1/A_{11}$, and the equations of motion \eqref{eq:CM} 
are solved by $a_1(t)=a_{1,0}+\dot{a}_{1}(0)t$, $\bs_1(t)=\bs_{1,0}$, leading to a fully explicit result; see Section~\ref{sec:onesoliton}. We also give numerical examples and plots for two- and three-soliton solution; see Section~\ref{sec:examples}. 

We mention that, in \cite{berntsonklabbers2020}, a more complicated numerical method was used to solve the corresponding constraints for the HWM equation, and our result is therefore new even in that case. 
Moreover, for fixed $\bn_0\in S^2$, a $N$-soliton solution of the ncIHF equation has $4N$ degrees of freedom: $a_{j,0}\in \C$ and $\bn_{j,3}\in S^2$ for $j=1,\ldots,N$; this is the same as for the HWM equation \cite{berntsonklabbers2020}. 

It would be interesting to extend these results to the complex-valued multi-soliton solutions in Section~\ref{sec:complexsolutions}, but this is left to future work. 

\item[(vii)] {\bf Physical properties of solitons.} We present a detailed discussion of one-soliton solutions ($N=1$), including their physical interpration; see Section~\ref{sec:onesoliton}. We also present analytical and numerical results suggesting that, in a $N$-soliton solution for $N>1$, one has $N$ well-defined one-solitons characterized by their velocities, $v_j\in(-1,1)$, energies, $E_j>0$, and rotation radii $R_j\in(0,1]$  ($j=1,\ldots,N$): at times where all solitons are well-separated, soliton $j$ moves with constant velocity $v_j$,  has constant energy $E_j$, and $\bu$ and $\bv$ rotate on a well-defined circle $\subset S^2$ with radius $R_j$; after time intervals when soliton collisions occur, the individual solitons involved in a collision re-emerge with unchanged velocities, energies, and rotation radii; see Sections~\ref{sec:examples}--\ref{sec:Nsolitons}. However, different from the HWM equation \cite{berntsonklabbers2020}, the orientations of the rotation circles are changed in a non-trivial way by soliton collisions.
Moreover, while the local vacua far away from all solitons are always the same in the HWM equation: $\bu\approx \bn_0$ for a fixed $\bn_0\in S^2$ \cite{berntsonklabbers2020}, the local vacua $\bu\approx \bv\approx \bm_j^+\in S^2$ for the ncIHF equation are different in different regions $j=0,\ldots ,N$ between two adjacent solitons (with $j=0$ and $j=N$ the regions $x\to -\infty$ and $x\to +\infty$, respectively), and these local vacua $\bm_j^+$ can be changed by soliton collisions.  

Some of the observations described above have the status of conjectures; it would be interesting to prove them by analytical means. 
It also would be interesting to extend this to the complex-valued solutions in Section~\ref{sec:complexsolutions}. 
\end{itemize}

\section{Derivation and properties of the model}
\label{sec:model}

We derive the ncIHF equation as a continuum limit of a classical spin chain related to the elliptic spin CMS model (Section~\ref{subsec:continuum_limit}) and establish some of its basic properties (Section~\ref{subsec:properties}). We also discuss possible reductions of the ncIHF equation (Section~\ref{subsec:reductions}). 

\subsection{Continuum limit of classical Inozemtsev-type spin chains}
\label{subsec:continuum_limit}

\paragraph{Elliptic spin Calogero-Moser-type systems.} 
For fixed $N\in\mathbb{Z}_{\geq 1}$, the $A_{N-1}$ elliptic spin CM system is defined by the Hamiltonian 
\begin{equation}
\label{eq:hCMHamiltonian}
H_{\mathrm{CM}}=\frac12 \sum_{j=1}^N p_j^2+\frac{g^2}{2}\pvsum_{j,k=1}^N  \bS_j\cdot\bS_k \wp_1(x_j-x_k),
\end{equation}
with $g>0$ a coupling parameter and $\wp_1(z)$ the elliptic CMS potential in \eqref{eq:wp1}. 
The dynamical variables of these models are the positions $x_j$ of particles on a circle of circumference $L>0$, i.e., $x_j\in\R$ with $x_j$ and $x_j+L$ identified, 
the canonical momenta $p_j\in\R$ of these particles, and spins $\bS_j$ of radius $S_0>0$ associated with these particles, i.e., $\bS_j\coloneqq (S_j^1,S_j^2,S_j^3)\in \R^3$ such that $\bS_j^2=S_0^2$. 
These variables obey the following non-trivial Poisson brackets, 
\begin{equation}
\label{eq:hCMPoissonbrackets}
\{x_j,p_k\}=\delta_{jk},\qquad \{S_j^a,S_k^b\}=\delta_{jk}\eps_{abc}S_j^c
\end{equation}
with the Kronecker delta $\delta_{jk}$. 
For reference we note the following equations of motion implied by these definitions, 
\begin{equation} 
\label{eq:hCMEOM} 
\begin{split} 
\dot \bS_j &=  -\frac{g^2}{2}\sum_{k\neq j}^N  \bS_j\wedge\bS_k \wp_1(x_j-x_k), \\
\ddot x_j &= -\frac{g^2}{2}\sum_{k\neq j}^N \bS_j\cdot\bS_k \wp_1'(x_j-x_k), 
\end{split} \quad (j=1,\ldots,N)
\end{equation} 
with $\dot x_j=p_j$. 

We use a trick, due to Calogero \cite{calogero1976}, to obtain from (\ref{eq:hCMHamiltonian}--\ref{eq:hCMPoissonbrackets}) a model of two different types of particles and spins.
For that, we double the degrees of freedom by replacing $N$ by $2N$, and for the second half of variables we rename, shift particle positions into the complex plane, and change the sign of spins as follows, 
\begin{equation} 
\label{eq:doubling}
\tilde{x}_j\coloneqq x_{N+j}+\ii\delta,\quad \tilde{p}_j\coloneqq p_{N+j},\quad \tilde{\bS}_j\coloneqq -\bS_{N+j}\quad (j=1,\ldots,N).
\end{equation}
By making these changes in \eqref{eq:hCMHamiltonian} and recalling the definition of $\tilde{\wp}_1$ in \eqref{eq:twp1}, we obtain the following variant of the elliptic spin CM system, 
\begin{equation} 
\label{eq:hCMHamiltonian1}
\begin{split} 
H_{\mathrm{CM}}=&\;\frac12 \sum_{j=1}^{N} p_{j}^2 +\frac{g^2}{4}\pvsum_{j,k=1}^N \bS_{j}\cdot\bS_{k} \wp_1(x_{j}-x_{k}) \\ 
&\;+\frac{1}{2} \sum_{j=1}^{N} \tilde{p}_{j}^2 +\frac{g^2}{4}\pvsum_{j,k=1}^N \tilde{\bS}_{j}\cdot\tilde{\bS}_{k} \wp_1(\tilde{x}_{j}-\tilde{x}_{k}) \\
&\; -\frac{g^2}{4}\sum_{j,k=1}^N \left[ \bS_{j}\cdot\tilde{\bS}_{k}\tilde{\wp}_1(x_{j}-\tilde{x}_{k}) + \tilde{\bS}_{j}\cdot\bS_{k} \tilde{\wp}_1(\tilde{x}_{j}-x_{k}) \right], 
\end{split} 
\end{equation} 
and the Poisson brackets \eqref{eq:hCMPoissonbrackets} imply  $\{x_{j},p_{k}\}=\{\tilde{x}_{j},\tilde{p}_{k}\}=\delta_{jk}$ and 
\begin{equation}
\label{eq:hCMPoissonbrackets1}
\{S_{j}^a,S_{k}^b\}=\delta_{jk}\eps_{abc}S_{j}^c,\quad \{\tilde{S}_{j}^a,\tilde{S}_{k}^b\}=-\delta_{jk}\eps_{abc}\tilde{S}_{j}^c
\end{equation}
for $j,k=1,\ldots,N$. 
We thus obtain a model of two different kinds of particles and spins, with variables of the second kind marked by tilde; variables of the same kind interact via the singular, repulsive  interaction potential $\wp_1(z)$, and variables of different kinds interact via the non-singular, weakly attractive interaction potential $\tilde{\wp}_1(z)$; note that, in the limit $L\to\infty$, $\wp_1(z)$ and $\tilde{\wp}_1(z)$ converge to $V(z)=\kappa^2/\sinh^2(\kappa z)$ and $\tV(z)=-\kappa^2/\cosh^2(\kappa z)$, respectively. 
It is important to note that the integrability of the standard spin CM model \eqref{eq:hCMHamiltonian}--\eqref{eq:hCMPoissonbrackets} implies that the modified spin CM model \eqref{eq:hCMHamiltonian1}--\eqref{eq:hCMPoissonbrackets1}  is integrable as well \cite{calogero1976}. 

\paragraph{Continuum limit of classical Inozemtsev spin chains.} 
The Inozemtsev spin chain \cite{inozemtsev1990}  is defined by the Hamiltonian\footnote{Our definition of the Inozemtsev Hamiltonian is convenient for taking a continuum limit but differs from the original conventions \cite{inozemtsev1990}; more specifically: (i) we use $\wp_1$ instead of $\wp$, which contributes an inconsequential additive term to the Hamiltonian, (ii) 
the argument of $\wp_1$ is scaled by the factor $L/N$, and our first period of $\wp_1$ is $L$ instead of $N$ in \cite{inozemtsev1990} and, using that $\wp(c z|c\omega_1,c\omega_2)=c^{-2}\wp(z|\omega_1,\omega_2)$ for $c>0$ \cite[Eq.~23.10.17]{DLMF}, we see that (ii) amounts to a redefinition of $J$ and $\delta$.}
\begin{equation} 
\label{eq:spinchain}
\begin{split} 
H_{\mathrm{Inozemtsev}} = & -\frac{J}2 \pvsum_{j,k=1}^N \wp_1((j-k)L/N)\bsigma_j\cdot\bsigma_k 
 \end{split} 
\end{equation} 
with $S=1/2$ quantum spin operators $\bsigma_j=(\sigma_j^1,\sigma_j^2,\sigma_j^3)$ satisfying the commutator relations
\begin{equation}
\label{eq:spincommutators}
[\sigma_j^{a},\sigma_k^b]=2\ii\delta_{jk}\eps_{abc}\sigma^c_j
\end{equation} 
for $a,b,c=1,2,3$ and $j,k=1,\ldots,N$, with $J$ a real coupling constant; we are interested in the ferromagnetic case where $J>0$. 
This model can be obtained from a quantum version of the elliptic spin CM model using Polychronakos's freezing trick \cite{finkel2014}, and it reduces to the Haldane-Shastry spin chain \cite{haldane1988,shastry1988} in the limit $\delta\to\infty$.
Zhou and Stone \cite{zhou2015} proposed a classical version of the Haldane-Shastry spin chain,  and they derived the HWM equation from this classical spin chain model in a  continuum limit; see also \cite{lenzmann2018}. 
This suggests the IHF equation \eqref{eq:IHF} can be obtained in a similar way from the Inozemtsev spin chain \eqref{eq:spinchain}--\eqref{eq:spincommutators}; we expect that this can be substantiated by a variant of our arguments given below.

In the same way as the Inozemtsev spin model \eqref{eq:spinchain}--\eqref{eq:spincommutators} can be obtained from the elliptic spin CM model \eqref{eq:hCMHamiltonian}--\eqref{eq:hCMPoissonbrackets} by freezing \cite{finkel2014}, we believe that the modified Inozemtsev model \eqref{eq:spinchain2}--\eqref{eq:spincommutators2} can be obtained from the modified spin CM model \eqref{eq:hCMHamiltonian1}--\eqref{eq:hCMPoissonbrackets1} (it would be interesting to make this precise, but this goes beyond goes beyond the scope of the current paper). 
We now show that the ncIHF equation is related to this modified Inozemtsev spin chain in the same way as the HWM equation is related to the Haldane-Shastry spin chain \cite{zhou2015,lenzmann2020}.

The classical version of the modified Inozemtsev spin chain \eqref{eq:spinchain2}--\eqref{eq:spincommutators2} is given by the Hamiltonian
\begin{equation} 
\label{eq:spinchain3}
\begin{split} 
H_{\mathrm{spin}} = &\; \frac{J}2 \pvsum_{j,k=1}^N \wp_1((j-k)L/N)\left( 2S_0^2-\bS_j\cdot\bS_k -  \tilde{\bS}_j\cdot\tilde{\bS}_k \right) \\ 
 &\; -\frac{J}2 \sum_{j,k=1}^N \tilde{\wp}_1((j-k)L/N)\left(  2S_0^2-\bS_j\cdot\tilde{\bS}_k -  \tilde{\bS}_j\cdot\bS_k \right) 
 \end{split} 
\end{equation} 
with $\bS_j=(S_j^{1},S_j^{2},S_j^{3})$ and $\tilde{\bS}_j=(\tilde{S}_j^{1},\tilde{S}_j^{2},\tilde{S}_j^{3})$ variables in $\R^3$ satisfying $\bS_j^2=\tilde{\bS}_j^2=S_0^2$ for some fixed constant $S_0>0$, and the Poisson brackets \eqref{eq:hCMPoissonbrackets1}. Note that we find it convenient to add an irrelevant constant to the naive classical analogue of the Hamiltonian \eqref{eq:spinchain2}; this allows us to write
\begin{equation} 
\label{eq:spinchain4}
\begin{split} 
H_{\mathrm{spin}} 
  = &\; \frac{J}4 \pvsum_{j,k=1}^N \wp_1((j-k)L/N)\left[ (\tilde{\bS}_k-\tilde{\bS}_j)^2 +(\bS_k-\bS_j)^2 \right] \\ 
 &\;-\frac{J}4 \sum_{j,k=1}^N \tilde{\wp}_1((j-k)L/N)\left[ (\tilde{\bS}_k-\bS_j)^2 + (\bS_k-\tilde{\bS}_j)^2\right]. 
 \end{split} 
\end{equation} 

We define\footnote{The minus signs in the definitions of $\bu$ and $\bv$ are needed since our conventions for the ncIHF equation requires a different sign in the Poisson brackets of $\bu$ and $\bv$ than the one following from the Poisson brackets commonly used for spin systems.} 
\begin{equation} 
\label{eq:SSttouv}
\bu(x_j) \coloneqq -\frac{\tilde{\bS}_j}{\Delta x},\quad \bv(x_j)  \coloneqq -\frac{\bS_j}{\Delta x},\quad x_j  \coloneqq j\Delta x-L/2\quad (j=1,\ldots,N),\quad \Delta x \coloneqq\frac{L}{N} 
\end{equation} 
and set $S_0=\Delta x$ so that, in the formal continuum limit $N\to\infty$, we obtain $L$-periodic spin density variables $\bu(x)$ and $\bv(x)$ of $x\in\R$  satisfying the Poisson brackets \eqref{eq:ncIHFPoissonbrackets}, together with the constraints  $\bu(x)^2=\bv(x)^2=1$. Using this notation, the Hamiltonian \eqref{eq:spinchain4} can be written as
\begin{equation} 
\label{eq:spinchain5}
\begin{split} 
H_{\mathrm{spin}} 
  = & \frac{J}{4} \pvsum_{j,k=1}^N \wp_1(x_j-x_k)\left[ \left(\bu(x_k)-\bu(x_j)\right)^2 +\left(\bv(x_k)-\bv(x_j)\right)^2 \right](\Delta x)^2 \\ 
 &-\frac{J}{4} \sum_{j,k=1}^N \tilde{\wp}_1(x_j-x_k)\left[\left(\bu(x_k)-\bv(x_j)\right)^2 +  \left(\bv(x_k)-\bu(x_j)\right)^2\right](\Delta x)^2,  
\end{split} 
\end{equation} 
with the Riemann sums converging to integrals in the limit $N\to\infty$: $\lim_{N\to\infty}H_{\mathrm{spin}} \eqqcolon \cH$ with\footnote{The principal value prescription in the second integral below can be omitted for differentiable functions $\bu$ and $\bv$, but we keep it to simplify some of our arguments.} 
\begin{equation} 
\begin{split} 
\cH = & \frac1{4\pi} \int_{-L/2}^{L/2} \pvint_{-L/2}^{L/2}  \wp_1(x'-x)\left[  \left(\bu(x)-\bu(x')\right)^2+\left(\bv(x)-\bv(x')\right)^2  \right]\,\dd{x'}\,\dd{x}  \\ & 
-\frac1{4\pi}\int_{-L/2}^{L/2} \int_{-L/2}^{L/2} \tilde{\wp}_1(x'-x)\left[  \left(\bu(x)-\bv(x')\right)^2+\left(\bv(x)-\bu(x')\right)^2\right]\,\dd{x'}\,\dd{x}; 
\end{split} 
\end{equation} 
note that we fixed the time scale by setting $J=1/\pi$, without loss of generality. 

We insert $\wp_1(x'-x)=-\partial_{x'}\zeta_1(x'-x)$ to compute, by integration by parts,
\begin{equation}
\begin{split} 
 \frac1{4\pi} \pvint_{-L/2}^{L/2} \wp_1(x'-x)\left( \bu(x)- \bu(x')\right)^2\, \dd x'  =&\;  \frac1{4\pi} \pvint_{-L/2}^{L/2} \zeta_1(x'-x)\partial_{x'}\left(\bu(x)-\bu(x')\right)^2\, \dd x' \\
=&\; -\frac1{2\pi} \pvint_{-L/2}^{L/2} \zeta_1(x'-x)\bu(x)\cdot\bu_{x'}(x')\, \dd x' \\
 =&\; -\frac12\bu(x)\cdot (T\bu_{x})(x),
\end{split}
\end{equation}
with $T$ as in \eqref{eq:TTe} (note that  the boundary terms vanish for differentiable $L$-periodic functions $\bu(x')$, and $\bu(x')\cdot\bu_{x'}(x')=0$) and, similarly, 
\begin{equation} 
\frac1{4\pi} \int_{-L/2}^{L/2} \tilde{\wp}_1(x'-x)\left( \bu(x)- \bv(x')\right)^2\, \dd x'  = -\frac12\bu(x)\cdot (\tT\bv_{x})(x),
\end{equation} 
with $\tT$ as in \eqref{eq:TTe}. Thus, 
\begin{equation} 
\mathcal{H} = -\frac1{2} \int_{-L/2}^{L/2}\left(  \bu\cdot T\bu_x +   \bv\cdot T\bv_x -   \bu\cdot \tT\bv_x -   \bv \cdot \tT\bu_x  \right) \, \dd x . 
\end{equation}  
In the limit $L\to\infty$, we obtain the Hamiltonian  \eqref{eq:ncIHFHamiltonian}.
The equations of motion $\bu_t=\{\cH,\bu\}$ and $\bv_t=\{\cH,\bv\}$ following from this Hamiltonian and the Poisson brackets \eqref{eq:ncIHFPoissonbrackets} 
are precisely the ncIHF equation \eqref{eq:ncIHF}  (this is shown in Appendix~\ref{subapp:HE}). 

To avoid misunderstanding, we mention that our derivation of the Hamiltonian \eqref{eq:ncIHFHamiltonian} from the classical version of the modified Inozemtsev model above is not rigorous. 
However, we expect that, by generalizing arguments developed in \cite{lenzmann2020} for the Haldane-Shastry case, it should be possible to make our derivation mathematically precise. 
  
\subsection{Properties of the ncIHF equation}
\label{subsec:properties}
We establish several basic properties of the ncIHF equation \eqref{eq:ncIHF}.
\paragraph{Norm conservation.} 
The ncIHF equation \eqref{eq:ncIHF} preserves the quantities 
$\bu(x,t)^2$ and $\bv(x,t)^2$  
in time. This can be shown by direct computation:
\begin{equation}
\label{eq:dotuusquare}
(\bu^2)_t=2\bu\cdot\bu_t=2\bu\cdot (\bu\wedge T\bu_x-\bu\wedge\tT\bv_x)=0, 
\end{equation}
and similarly for $\bv$. Note that this holds true in both the real-line and periodic cases. 

\paragraph{Boundary conditions.} 
To make the definition of the ncIHF equation \eqref{eq:ncIHF} complete,  we propose the following conditions at spatial infinity (where the limits are understood in the sense of Schwartz --- 
 see \eqref{eq:bubvdecomposition} for a more precise statement), 
\begin{equation} 
\label{eq:BC} 
\lim_{x\to\pm\infty} \bu(x,t) = \lim_{x\to\pm\infty} \bv(x,t) \eqqcolon \bm_{\pm\infty} ,  
\end{equation} 
i.e., the functions $\bu(x,t)$ and $\bv(x,t)$ converge to the same asymptotic values $\bm_{+\infty}$ and $\bm_{-\infty}$ as $x\to +\infty$ and $x\to-\infty$, respectively, and these asymptotic values are constant in time. 

To give a physical motivation for these conditions, we note that the ncIHF Hamiltonian \eqref{eq:ncIHFHamiltonian} can be written as
\begin{equation}
\label{eq:ncIHFHamiltonian1}
\begin{split} 
\mathcal{H}=\frac1{4\pi} \int_{\R}\pvint_{\R} \Bigl( & V(x'-x)\left[\left( \bu(x')-\bu(x)\right)^2+\left(\bv(x')-\bv(x)\right)^2\right]  \\ 
 - & \tV(x'-x)\left[\left( \bu(x')-\bv(x)\right)^2+\left(\bv(x')-\bu(x)\right)^2\right] \Bigr)  \,\dd{x'} \,\dd{x}, 
\end{split} 
\end{equation}
as shown in Appendix~\ref{subapp:alternativecH}. Since $V(x'-x)\geq 0$ and $-\tV(x'-x)\geq 0$, the Hamiltonian \eqref{eq:ncIHFHamiltonian1} is manifestly non-negative, i.e., energy is bounded from below by zero. Moreover, \eqref{eq:ncIHFHamiltonian1} makes manifest that the ncIHF equation describes a ferromagnetic system with  ground states given by $\bu(x,t)=\bv(x,t)=\bm_0$, for arbitrary constant $\bm_0\in S^2$. Indeed,  it is easy to check that these are the solutions of \eqref{eq:ncIHF} that minimize the energy. Thus, it is natural to require that the allowed non-equilibrium states, where $\bu$ and $\bv$ vary in space and time, are such that they converge to vacuum states $\bm_{\pm\infty}$ at spatial infinity $x\to\pm\infty$. 

\paragraph{Conservation of total spin.} It is well-known that the spin CM model defined in \eqref{eq:hCMHamiltonian}--\eqref{eq:hCMPoissonbrackets} conserves total spin $\sum_{j=1}^N \bS_j$. Recalling \eqref{eq:doubling} and \eqref{eq:SSttouv}, the 
corresponding continuum quantity is
\begin{equation}
\label{eq:totalspin}
\bS\coloneqq \int_{\R} (\bu-\bv)\,\dd{x}
\end{equation}
and, since conservation laws are often preserved under continuum limits, one expects that this is conserved in time. 
This is indeed true: one can show by a direct computation that $\bS$ is conserved provided $\bu$ and $\bv$ evolve in time according to \eqref{eq:ncIHF} and behave at spatial infinity as in \eqref{eq:BC} (see  Appendix~\ref{subapp:verificationbS} for details). 
 
\paragraph{Further conservation laws.}
As shown in Section~\ref{sec:Lax}, the ncIHF equation \eqref{eq:ncIHF} has the Lax pair $(\cL,\cB)$ given in \eqref{eq:KLdef}--\eqref{eq:KBdef}. As a consequence, \eqref{eq:ncIHF} admits an infinite number of 
conservation laws, 
\begin{equation}
\cI_n\coloneqq \mathrm{tr} (\cL^n)\quad (n=2,3,\ldots), 
\end{equation}
with $\mathrm{tr}(\cdot)$ the appropriate Hilbert space trace. More explicitly, one obtains from \eqref{eq:KLdef}--\eqref{eq:KL}: 
\begin{equation}
\label{eq:In} 
\begin{split}  
\cI_2=&\; \frac1{\pi^2}\int_{\R^2} \mathrm{tr}_{\mathbb{C}^4}\big(K_{\cL}(x,x')K_{\cL}(x',x)\big)\,\dd{x'}\,\dd{x},  \\
\cI_3=&\; \frac1{\pi^3}\int_{\R^3} \mathrm{tr}_{\mathbb{C}^4}\big(K_{\cL}(x,x')K_{\cL}(x',x'')K_{\cL}(x'',x)\big)\,\dd{x'}'\,\dd{x'}\,\dd{x},  \\
 & \vdots \\
\cI_n=&\; \frac1{\pi^n}\int_{\R^n} \mathrm{tr}_{\mathbb{C}^4}\big(K_{\cL}(x_1,x_2)K_{\cL}(x_2,x_3)\cdots K_{\cL}(x_n,x_1)\big)\,\dd{x_n}\,\dd{x_{n-1}}\cdots \,\dd{x_1} ,
\end{split} 
\end{equation} 
for $n=4,5,\ldots$, with $K_{\cL}(x,x')$ \eqref{eq:KL} and $\mathrm{tr}_{\mathbb{C}^4}(\cdot)$ the usual trace of $4\times 4$ matrices. 

One can show that $\cI_2$ is a linear combination of Hamiltonian $\cH$ \eqref{eq:ncIHFHamiltonian} and the square of the the total spin $\bS$ \eqref{eq:totalspin}; the precise statement is 
\begin{equation} 
\label{eq:Itwo} 
\cI_2 =  \frac{8}{\pi}\cH  - \frac{4}{\pi^2}\kappa^2 \bS^2, 
\end{equation} 
which can be shown by a straightforward generalization of an argument for the HWM equation in \cite{gerard2018} (the interested reader can find details in Appendix~\ref{subapp:verificationcH}). 

\subsection{Reductions and limits}
\label{subsec:reductions}
It is generally assumed in the theory of integrable systems that integrability is preserved under reductions and limits \cite{calogero1991}. It is therefore interesting to consider such degenerate equations arising from the ncIHF equation \eqref{eq:ncIHF}.

Setting $\bv=-\bu$ in \eqref{eq:ncIHF}, we obtain
\begin{equation}
\label{eq:ncIHFchiralreduction}
\bu_t=\bu\wedge(T\bu_x+\tT\bu_x).
\end{equation}
\label{eq:hyperbolicidentity}
The standard hyperbolic identity 
\begin{equation}\tanh(z)+\coth(z)=2\coth(2z)
\end{equation}
together with \eqref{eq:TTh} shows that $T+\tT=T_{\delta\to \frac{\delta}{2}}$; hence, \eqref{eq:ncIHFchiralreduction} is 
the IHF equation \eqref{eq:IHF} (with $\delta$ changed to $\delta/2$);  this  is an  intermediate variant of the Heisenberg ferromagnet (HF) equation 
\begin{equation}
\label{eq:HF}
\bu_t=\bu\wedge\bu_{xx}
\end{equation}
in the sense of \cite{tamizhmani1999}; see also \cite[Chapter~4]{ablowitz1991}:  the IHF equation with the boundary conditions 
\begin{equation}
\label{eq:BC1} 
\lim_{x\to\pm\infty} \bu(x)=\pm \bm_{\infty}
\end{equation} 
 reduces to \eqref{eq:HF} in the limit $\delta\to 0^+$ (a derivation of this fact can be found in Appendix~\ref{subapp:IHF}). 

One might hope that the ncIHF equation has an interesting limit $\delta\to 0^+$ as well. 
However, as substantiated in Appendix~\ref{subapp:ncIHF}, this is not the case.

As already discussed in the introduction, the ncIHF equation reduces to two decoupled HWM equations of opposite chirality in the limit $\delta\to\infty$; see \eqref{eq:HWM}. 

Since the reduction $\bv=-\bu$ of the ncIHF equation leads to the IHF equation, one might expect that the results about the former reduce to results of the latter. 
This is not the case for the ncIHF on the real line: this equation is defined with the boundary conditions that $\bu$ and $\bv$ converge to the same asymptotic values as $x\to\pm\infty$ (see \eqref{eq:BC}), and this is clearly incompatible with the reduction $\bv=-\bu$.

It is interesting to note that these limits of the ncIHF equation are very similar to what is known to be true for the intermediate long-wave (ILW) equation: while the $\delta\to 0^+$ of the ILW equation leads to the Korteweg-de Vries equation \cite{kodama1981}, the non-chiral ILW (ncILW) equation does not have a non-trivial such limit \cite[Appendix B]{berntsonlangmann2020}. Moreover, in the limit $\delta\to\infty$, the non-chiral ILW equation reduces to two decoupled Benjamin-Ono equations of opposite chirality \cite{berntson2020}. 
One seeming difference is that the ncILW equation cannot be reduced to the ILW equation \cite{berntson2020}, whereas the ncIHF equation has the reduction $\bv=-\bu$ to the IHF equation; however, as discussed, the latter reduction is in conflict with the imposed boundary conditions and thus, as far as we can see, of no use: perhaps, one should regard this reduction as spurious.

\section{Lax pair} 
\label{sec:Lax} 
Following work of G\'erard and Lenzmann on the HWM equation \cite{gerard2018}, we obtain a Lax pair for the ncIHF equation  \eqref{eq:ncIHF}. 

\subsection{Result} 
\label{sec:Laxresult} 
\paragraph{Notation.} To give a precise formulation of our Lax pair result, we work in the Hilbert space $L^2(\R,\C^4)$ of square-integrable functions 
$\Psi:\R\to \C^4$, $x\mapsto \Psi(x)$  and write such functions $\Psi$ as two-vectors with components in $L^2(\R,\C^2)$: 
\begin{equation} 
\label{eq:Psi}
 \Psi = \left(\begin{array}{c}\psi_1 \\ \psi_2 \end{array}\right)\quad (\psi_1,\psi_2\in L^2(\R,\C^2)). 
\end{equation} 
Using this notation, we associate a two-vector $\uA$ with a pair $\ua_1,\ua_2$ of $\sutwo$-valued functions on $\R$ and define a corresponding linear operator $\mu_{\uA}$ on $L^2(\R,\C^4)$ as follows, 
\begin{equation} 
\label{eq:muA}
\uA\coloneqq  \left(\begin{array}{c}\ua_1\\ \ua_2\end{array}\right),\quad 
(\mu_{\uA}\Psi)(x) \coloneqq  \left(\begin{array}{cc}\ua_1(x) & 0 \\ 0 & \ua_2(x) \end{array}\right)\left(\begin{array}{c} \psi_1(x) \\ \psi_2(x) \end{array}\right) = 
\left(\begin{array}{c}\ua_1(x) \psi_1(x) \\ \ua_2(x) \psi_2(x) \end{array}\right).
\end{equation} 
We also introduce the matrix operator 
\begin{equation} 
\label{eq:cT}
 \cT\coloneqq  \left(\begin{array}{cc} T & -\tT\\ \tT & -T  \end{array}\right)
\end{equation} 
which acts naturally on two-vectors as in \eqref{eq:muA}: 
\begin{equation} 
\label{eq:cTA}
\cT\uA \coloneqq \left(\begin{array}{cc} T & -\tT\\ \tT & -T  \end{array}\right)\left(\begin{array}{c}\ua_1\\ \ua_2\end{array}\right) = \left(\begin{array}{c} T\ua_1-\tT\ua_2\\ \tT\ua_1-T\ua_2\end{array}\right) . 
\end{equation} 
In particular, we represent the pair of functions $\uu,\uv$ in the ncIHF equation \eqref{eq:ncIHF1} as 
\begin{equation} 
\label{eq:uU} 
\uU \coloneqq  \left(\begin{array}{c}\uu\\ \uv\end{array}\right)
\end{equation} 
and, using the notation above, we can write  \eqref{eq:ncIHF1}  as time evolution equation for the operator $\mu_{\uU}$, 
\begin{equation} 
\label{eq:ncIHF3}
\frac{\dd}{\dd{t}}\mu_{\uU} = \frac{1}{2\ii}[\mu_{\uU},\mu_{\cT\uU_x}]
\end{equation} 
where $\frac{\dd}{\dd{t}}\mu_{\uU}=\mu_{\uU_t}$, with $[\cdot,\cdot]$ the usual commutator of Hilbert space operators.

We note that $\partial_x$ and $\cT$ \eqref{eq:cT} naturally define linear operators on $L^2(\R,\C^4)$ (the definition of $\cT\Psi$ is as in \eqref{eq:cTA} but with $\uA$ in \eqref{eq:muA} replaced by $\Psi$ in \eqref{eq:Psi}, of course); to not clutter our notation, we use the same symbol $\cT$ for this Hilbert space operator as for the operator defined in \eqref{eq:cTA}. 

We are now ready to give a mathematically precise formulation of our Lax pair result.

\paragraph{Lax pair.} {\em The ncIHF equation \eqref{eq:ncIHF3} implies that the linear operators 
\begin{equation} 
\label{eq:LB1}
\cL\coloneqq [\cT,\mu_{\uU}],\quad \cB\coloneqq \frac{1}{2\ii}\left(\mu_{\uU}\cT\partial_x + \cT\partial_x \mu_{\uU}-\mu_{\cT\uU_x}  \right)
\end{equation} 
on $L^2(\R,\C^4)$ satisfy the Lax equation 
\begin{equation} 
\label{eq:Laxeq} 
\frac{\dd}{\dd{t}}\cL = [\cB,\cL]. 
\end{equation} 
}

Inserting the definition of $(\mu_{\uU}\Psi)(x)$ in \eqref{eq:muA} and 
\begin{equation} 
(\cT \Psi)(x) = \frac1\pi\pvint\left(\begin{array}{cc}\alpha(x'-x) &  -\tilde\alpha(x'-x) \\  \tilde\alpha(x'-x) & -\alpha(x'-x) \end{array} \right)\left(\begin{array}{c}\psi_1(x') \\ \psi_2(x') \end{array}\right)  \,\dd{x'}, 
\end{equation} 
one obtains the representations of $(\cL,\cB)$ in \eqref{eq:KLdef}--\eqref{eq:KB} by straightforward computations. 
 
\paragraph{Self-adjointness.} It is interesting to note that, different from the HWM equation case \cite{gerard2018},  the operators $\cL$ and $\cB$ are not (formally) self-adjoint and anti-self-adjoint, respectively, but instead they obey the conditions 
\begin{equation} 
\cL^\dag =\Lambda\cL\Lambda,\quad \cB^\dag = -\Lambda\cB\Lambda 
\end{equation} 
where $\dag$ indicates the Hilbert space adjoint and
\begin{equation} 
\label{eq:Lambda}
\Lambda \coloneqq \left(\begin{array}{rr} I & 0 \\ 0 & -I \end{array}\right) 
\end{equation} 
is a non-trivial grading operator, i.e., $\Lambda^\dag=\Lambda$ and $\Lambda^2=I$ (this follows from the fact that, while the integral operators $T$ and $\tT$ are anti-self-adjoint, the matrix operators  $\cT$ is not, but rather obeys $\cT^\dag=-\Lambda\cT\Lambda$). 
Thus, the conservation law $\mathrm{tr}(\cL^2)$ is different from $\mathrm{tr}(\cL^\dag\cL)$; in fact, we checked that the latter is divergent, while the former is finite and, essentially, the Hamiltonian; see \eqref{eq:Itwo}. 
We believe that the Lax operator $\cL$ not being (formally) self-adjoint is an important feature of the ncIHF equation which deserves to be better understood. 

\subsection{Derivation of result}
\label{sec:derivationlax}
We present a derivation of \eqref{eq:LB1}--\eqref{eq:Laxeq}, following closely the proof of the limiting case $\delta\to\infty$ of this result by G\'erard and Lenzmann \cite{gerard2018}. 
Remarkably, by introducing a suitable notation, the generalizations of all steps but one in this proof are straightforward; in one step, we need a generalization of a product rule for Hilbert transforms known as the Cotlar identity \cite{gerard2018} which we prove in Appendix~\ref{app:Cotlar}. 

\paragraph{Vector notation.} We define the following component-wise product of two-vectors (which is a variant of the so-called Hadamard product), 
\begin{equation} 
\label{eq:Hadamard}
F\coloneqq \left(\begin{array}{c} f_1\\ f_2\end{array}\right),\quad G\coloneqq \left(\begin{array}{c} g_1\\ g_2\end{array}\right),\quad F\circ G \coloneqq
\left(\begin{array}{c} f_1 g_1\\  f_2g_2\end{array}\right)
\end{equation} 
for $\C$-valued functions $f_j$ and $g_j$ on $\R$ ($j=1,2$). This definition has the following natural extension to two-vectors with components that are $\sutwo$- or $\C^2$-valued  functions on $\R$, 
\begin{equation} 
 \uA\circ\uB \coloneqq \left(\begin{array}{c}\ua_1\, \ub_1\\ \ua_2\, \ub_2\end{array}\right),\quad \uA\circ \Psi \coloneqq  \left(\begin{array}{c}\ua_1\, \psi_1 \\ \ua_2\, \psi_2 \end{array}\right)
\end{equation} 
with $\uA$ as in \eqref{eq:muA} and $\Psi$ as in \eqref{eq:Psi}, for all $\sutwo$-valued functions $\ua_j,\ub_j$ on $\R$ ($\uB$ is defined as $\uA$ in  \eqref{eq:muA} but with $\ua_j$ replaced by $\ub_j$). With this notation, the operators $\mu_{\uA}$ defined in \eqref{eq:muA} have the properties 
\begin{equation}
\mu_{\uA}\Psi = \uA\circ\Psi,\quad \mu_{\uA{\phantom '}}\mu_{\uB} = \mu_{\uA\circ\uB}
\end{equation} 
which will be helpful in the following derivation. 

\paragraph{Identities.} We start with three known identities which are important in the derivation. 
First, 
\begin{equation} 
\cT^2=-I
\end{equation} 
with $I$ the identity operator  \cite{berntson2020}; this is true since the operators $T$ and $\tT$ in \eqref{eq:TT} satisfy the identities  $T^2-\tT^2=-I$ and $T\tT=\tT T$. 
Second, 
\begin{equation} 
\mu_{\bU}^2=I 
\end{equation}  
since $\uu^2=\uv^2=I$ is implied by $\bu^2=\bv^2=1$ and well-known properties of the Pauli matrices \cite{gerard2018}. 
Third, $\partial_x\cT=\cT\partial_x$   \cite{berntson2020} since, by \eqref{eq:TT}, $\partial_x$ commutes with $T$ and $\tT$. 

\paragraph{Derivation.} We compute, using the ncIHF equation \eqref{eq:ncIHF3} and the Jacobi identity $[A,[B,C]]=[[A,B],C]+[B,[A,C]]$ for Hilbert space operators $A,B,C$, 
\begin{equation} 
\frac{\dd}{\dd{t}}\cL = [\cT,\frac{\dd}{\dd{t}}(\mu_{\uU})] = \frac{1}{2\ii}[\cT,[\mu_{\uU},\mu_{\cT\uU_x}]] = \frac{1}{2\ii}[[\cT,\mu_{\uU}],\mu_{\cT\uU_x}] + \frac{1}{2\ii}[\mu_{\uU},[\cT,\mu_{\cT\uU_x}]]  , 
\end{equation} 
which, by the anti-symmetry of the commutator and the definition of $\cL$, can be written as 
\begin{equation} 
\label{eq:res0} 
\frac{\dd}{\dd{t}}\cL  = -\frac{1}{2\ii}[[\cT,\mu_{\cT\uU_x}],\mu_{\uU}] -\frac{1}{2\ii}[\mu_{\cT\uU_x},\cL].
\end{equation} 
To proceed, we need the following identity,  
\begin{equation} 
\label{eq:Cotlargen} 
\cT(F\circ G) = (\cT F)\circ G + F\circ (\cT G)+\cT((\cT F)\circ (\cT G))
\end{equation} 
for all $\C$-valued functions $f_j,g_j$ on $\R$ with well-defined Fourier transform (in the classical, not distributional, sense),
using the notation in \eqref{eq:Hadamard}. 
This is the generalization of the Cotlar identity already mentioned; we prove \eqref{eq:Cotlargen} in Appendix~\ref{app:Cotlar}.
Using definitions, one can check that this identity extends to two-vectors $F$ and $G$ with components that are $\sutwo$- and $\C^2$-valued functions, respectively. 
Thus, we can use it to compute, for $\Psi\in L^2(\R,\C^4)$, 
\begin{equation} 
\label{eq:difficultstep}
\begin{split} 
[\cT,\mu_{\cT\uU_x}]\Psi =\cT((\cT\uU_x)\circ \Psi)-(\cT\uU_x)\circ (\cT\Psi) = (\cT(\cT\uU_x))\circ \Psi + (\cT\uU_x)\circ (\cT\Psi) \\ + \cT((\cT(\cT\uU_x))\circ (\cT\Psi)) -(\cT\uU_x)\circ (\cT\Psi) 
= -\uU_x\circ \Psi  - \cT(\uU_x\circ (\cT\Psi)) \\ =  \left( -\mu_{\uU_x}-\cT\mu_{\uU_x}\cT \right)\Psi 
\end{split} 
\end{equation} 
since $\cT(\cT\bU_x)=\cT^2\bU_x=-\bU_x$. Thus, we conclude that $[\cT,\mu_{\cT\uU_x}]=-\mu_{\uU_x}-\cT\mu_{\uU_x}\cT$, 
 and inserting this into \eqref{eq:res0} we obtain
\begin{equation} 
\label{eq:res1} 
\frac{\dd}{\dd{t}}\cL  = \frac{1}{2\ii}[\mu_{\uU_x}+\cT\mu_{\uU_x}\cT,\mu_{\uU}] -\frac{1}{2\ii}[\mu_{\cT\uU_x},\cL].
\end{equation} 
We proceed by computing, using $\cT^2=-I$, $\mu_{\bU}^2=I$, $\partial_x\cT=\cT\partial_x$, and the Leibniz rule, 
\begin{equation}
\begin{split}  
[\mu_{\uU}\cT\partial_x,\cL]\Psi= [\mu_{\uU}\cT\partial_x,\cT\mu_{\uU}-\mu_{\uU}\cT]\Psi= \mu_{\uU}\cT\partial_x\cT\mu_{\uU}\Psi - \cT\mu_{\uU}^2\cT\partial_x\Psi \\
- \mu_{\uU}\cT\partial_x\mu_{\uU}\cT\Psi +\mu_{\uU} \cT\mu_{\uU}\cT\partial_x\Psi 
= -\mu_{\uU}\mu_{\uU_x}\Psi - \mu_{\uU}\cT\mu_{\uU_x}\cT\Psi
\end{split} 
\end{equation} 
for all $\Psi\in L^2(\R,\C^4)$ and, similarly, 
\begin{equation}
\begin{split}  
[\cT\partial_x \mu_{\uU},\cL]\Psi = [\cT\partial_x \mu_{\uU},\cT\mu_{\uU}-\mu_{\uU}\cT]\Psi = \cT\partial_x \mu_{\uU}\cT\mu_{\uU}\Psi-\cT\mu_{\uU}\cT\partial_x \mu_{\uU}\Psi\\
- \cT\partial_x \mu_{\uU}^2\cT\Psi + \mu_{\uU}\cT^2\partial_x \mu_{\uU}\Psi =  \cT\mu_{\uU_x}\cT\mu_{\uU}\Psi - \mu_{\uU}\mu_{\uU_x}\Psi .
\end{split} 
\end{equation} 
Combining these results, and using that $\partial_x\mu_{\uU}^2=\mu_{\uU}\mu_{\uU_x}+\mu_{\uU_x}\mu_{\uU}=0$ since $\mu_{\uU}^2=I$, we obtain 
\begin{equation} 
[\mu_{\uU}\cT\partial_x+\cT\partial_x \mu_{\uU},\cL] = [\mu_{\uU_x},\mu_{\uU}] +  [\cT\mu_{\uU_x}\cT,\mu_{\uU}] = [\mu_{\uU_x}+\cT\mu_{\uU_x}\cT,\mu_{\uU}].
\end{equation} 
Inserting this into \eqref{eq:res1} gives
\begin{equation} 
\frac{\dd}{\dd{t}}\cL  = \frac{1}{2\ii}\left[\mu_{\uU}\cT\partial_x+\cT\partial_x \mu_{\uU},\cL\right]  -\frac{1}{2\ii}\left[\mu_{\cT\uU_x},\cL\right] ,
\end{equation} 
which is \eqref{eq:Laxeq} with $\cB$ in \eqref{eq:LB1}. 

\section{Soliton solutions}
\label{sec:solitons}
We derive the multi-soliton solutions of the ncIHF equation presented already in \eqref{eq:solution}--\eqref{eq:constraintsbm} (Section~\ref{sec:solitonderivation}), and we generalize this result to complex-valued multi-soliton solutions  (Section~\ref{sec:complexsolutions}). We also derive explicit formulas for the energy density and total energy of the $N$-soliton solutions (Section~\ref{sec:energy}). 

\subsection{Derivation of multi-soliton solutions} 
\label{sec:solitonderivation} 
Remarkably, after introducing a vector notation inspired by previous work on the non-chiral ILW equation \cite{berntson2020}, we can derive the multi-soliton solutions of the ncIHF equation following closely our work on the HWM equation \cite{berntsonklabbers2020}. 
  
\paragraph{Vector notation.} 
We represent the function $\bu$ and $\bv$ in the ncIHF equation \eqref{eq:ncIHF} by a two-vector, 
\begin{equation} 
\label{eq:bU} 
\bU\coloneqq \left(\begin{array}{c}\bu\\\bv\end{array}\right) 
\end{equation} 
where $\bu$ and $\bv$ as $\R^3$-valued functions of the real variables $x$ and $t$ satisfying the constraints $\bu^2=\bv^2=\rho^2$ with $\rho=1$; we introduce the parameter $\rho$ here for the sake of generality: as discussed in Section~\ref{sec:complexsolutions} below, there are also solution of \eqref{eq:ncIHF1} where $\rho\in\C\setminus\{0\}$ is arbitrary. 

In Section~\ref{sec:Lax}, we already introduced the component-wise product \eqref{eq:Hadamard}. Here, we also make use of the following component-wise products of two-vectors with three-vector components,  
\begin{equation} 
\label{eq:matrixnotation2} 
\bA\coloneqq \left(\begin{array}{c}\ba_1\\\ba_2\end{array}\right),\quad \bB\coloneqq \left(\begin{array}{c}\bb_1\\ \bb_2\end{array}\right),\quad 
\bA\wedgecirc\bB \coloneqq \left(\begin{array}{c}\ba_1\wedge\bb_1\\\ba_2\wedge\bb_2\end{array}\right),\quad 
\bA\dotcirc\bB \coloneqq \left(\begin{array}{c}\ba_1\cdot\bb_1\\\ba_2\cdot\bb_2\end{array}\right)
\end{equation} 
for $\C^3$-valued functions $\ba_j,\bb_j$ on $\R$ ($j=1,2$). We also recall the definition of the matrix operators $\cT$ \eqref{eq:cT}, which naturally acts on two-vectors $\bA$ as in \eqref{eq:matrixnotation2} ($\cT\bA$ is defined as in \eqref{eq:cTA} but with  $\uA$ in \eqref{eq:muA} replaced by $\bA$ in \eqref{eq:matrixnotation2}, of course).  

This notation allows to write $\bu^2=\bv^2=\rho^2$ and the ncIHF equation \eqref{eq:ncIHF}  as 
\begin{equation} 
\label{eq:bUbU}
\bU\dotcirc\bU = \rho^2\left(\begin{array}{c} 1\\ 1\end{array}\right)
\end{equation} 
and 
\begin{equation} 
\label{eq:ncIHF1a} 
\bU_t = \bU\wedgecirc\cT\bU_x, 
\end{equation} 
respectively. 
We also note the following computation rules following from these definitions, 
\begin{equation} 
\label{eq:rules} 
\begin{split} 
\ba F\wedgecirc  \bb G =\,  & \ba\wedge\bb F\circ G, \\ 
\ba F\dotcirc \bb G =\,  &  \ba\cdot\bb F\circ G, \\
\left(\begin{array}{c} 1\\ 1\end{array}\right) \circ F = &F\circ \left(\begin{array}{c} 1\\ 1\end{array}\right) =F
\end{split} 
\end{equation} 
for all $\ba,\bb\in \C^3$  and $F,G$ as in \eqref{eq:Hadamard} with $\C$-valued functions $f_j,g_j$  on $\R$ where, of course, 
\begin{equation} 
\ba \left(\begin{array}{c} f_1\\f_2\end{array} \right)\coloneqq \left(\begin{array}{c} \ba\, f_1\\\ba\, f_2\end{array} \right).
\end{equation} 

We introduce the shorthand notation  
\begin{equation}
\label{eq:Ar}
A_{\pm}(z) \coloneqq \left(\begin{array}{c}
\alpha(z \pm \ii \delta/2)\\ \alpha(z \mp \ii \delta/2)
\end{array}\right) \quad (z\in\C) 
\end{equation}
and 
\begin{equation}
\label{eq:strip}
\begin{split} 
R^+_\delta \coloneqq &\left\{  z \in \mathbb{C} | \delta/2 < \text{Im}(z) < 3 \delta/2 \right\}, \\
R^-_\delta \coloneqq &\left\{  z \in \mathbb{C} | -3\delta/2 < \text{Im}(z) < -\delta/2 \right\}
\end{split}
\end{equation} 
(see \eqref{eq:kappa}--\eqref{eq:alpha} for definitions of $\kappa$ and $\alpha(z)$). 
The functions in \eqref{eq:Ar} are the building blocks in our spin-pole ansatz; the regions $R^\pm_\delta$ will be used to constrain the poles. 

\paragraph{Identities.} The special functions $A_\pm(x-a)$, $x$ real and $a$ complex (suitably restricted), satisfy three identities which play a key role in our derivation of multi-soliton solutions: 
First, 
\begin{equation} 
\label{eq:AId1} 
\cT A_r(x-a)= -r\ii A_r(x-a)
\end{equation} 
for all $r\in\{+,-\}$, $x\in\R$, and  $a\in R^r_{\delta}$; this was proved in  \cite[Appendix A.1]{berntson2020} (to see this, note that the definitions of $\cT$ here and in \cite{berntson2020} differ by a similarity transformation $\cT\to \Lambda\cT\Lambda$ with $\Lambda=\Lambda^{-1}$ given in \eqref{eq:Lambda}). Second, \begin{equation} 
\label{eq:AId2} 
A_r(x-a)\circ A_{s}(x-b) = \alpha(a-b+(r-s)\ii\delta/2) \left( A_r(x-a)-A_{s}(x-b)\right) + \kappa^2  \left(\begin{array}{c} 1\\ 1\end{array}\right)
\end{equation} 
for all $r,s\in\{+,-\}$, $x\in\R$, and  $a,b\in\C$ such that $a\neq b$; this is implied by well-known identities satisfied by the special function $\alpha(z)$ (given in \eqref{eq:alphaidentities}, first and second lines). Third, 
\begin{equation} 
\label{eq:AId3} 
A_r(x-a)\circ A_r(x-a) = -\partial_xA_r(x-a) + \kappa^2 \left(\begin{array}{c} 1\\ 1\end{array}\right) 
\end{equation} 
where
\begin{equation} 
\label{eq:dAr}
\partial_x A_r(x-a) = -\partial_{a} A_r(x-a)=-\left(\begin{array}{c}
V(x-a \mp \ii \delta/2)\\ V(x-a \pm \ii \delta/2)\end{array}\right)
\end{equation} 
for all $r\in\{+,-\}$, $x\in\R$ and $a\in R^r_\delta$; this follows from well-known identities for $\alpha(z)$ (given in \eqref{eq:alphaidentities}, third line). 

As we now show, with these definitions and identities in place, we can follow our derivation for the HWM equation in \cite[Eqs. (2.1)--(2.11)]{berntsonklabbers2020} nearly line-by-line to obtain multi-soliton solutions of the ncIHF equation. 

\paragraph{Spin-pole ansatz.} Inspired by \cite[Theorem 3.1]{berntsonklabbers2020} and its proof, we make the following ansatz to solve \eqref{eq:bUbU}--\eqref{eq:ncIHF1}, 
\begin{equation}
\label{eq:hansatz}
\bU(x,t) = 
\bm_{0}\left(\begin{array}{c}1 \\ 1
\end{array}\right) + \ii \sum_{j=1}^N \bs_j(t) A_+(x-a_j(t))
- \ii \sum_{j=1}^N \bs_j(t)^* A_-(x-a_j^*(t))
\end{equation}
with $\bm_{0}\in\R^3$ constant and $(a_j(t),\bs_j(t))\in R^+_\delta\times\C^3$.  Note that, for $N=0$ and $\bm_{0}^2=1$, this is the vacuum solution. 
Moreover, the functions $\bu$ and $\bv$ defined by \eqref{eq:bU} and \eqref{eq:hansatz} are real-valued and equal to the ones given in \eqref{eq:solution}. 

To simplify computations, we find it convenient to introduce the notation
\begin{equation} 
\label{eq:shorthandreal}
\cN= 2N,\quad (a_j,\bs_j,r_j)=\begin{cases} (a_j,\bs_j,+) & (j=1,\ldots,N) \\ (a_{j-N}^*,\bs_{j-N}^*,-) & (j=N+1,\ldots,\cN) \end{cases} , 
\end{equation} 
allowing us to write \eqref{eq:hansatz} shorter as 
\begin{equation}
\label{eq:hansatz1}
\bU(x,t) = 
\bm_{0}\left(\begin{array}{c}1 \\ 1\end{array}\right) + \ii \sum_{j=1}^{\cN} r_j\bs_j(t) A_{r_j}(x-a_j(t)).
\end{equation}
Remarkably, with this notation, the computation is simpler and, at the same time, more general. 

\paragraph{Constraints.} We insert \eqref{eq:hansatz1} into \eqref{eq:bUbU} and compute, using the rules \eqref{eq:rules}, 
\begin{equation} 
\begin{split} 
\bU\dotcirc\bU = &  \bm_{0}^2 \left(\begin{array}{c}1 \\ 1\end{array}\right) +2\ii  \bm_{0}\cdot \sum_{j=1}^{\cN} r_j\bs_j A_{r_j}(x-a_j) \\ 
& - \sum_{j,k=1}^{\cN}r_jr_k\bs_j\cdot\bs_kA_{r_j}(x-a_j)\circ A_{r_k}(x-a_k), 
\end{split} 
\end{equation} 
and by inserting the identities \eqref{eq:AId2}--\eqref{eq:AId3} we find (computational details are the same as in \cite[Appendix A.1]{berntsonklabbers2020} and thus omitted), 
\begin{equation} 
\begin{split} 
\bU\dotcirc\bU = & \Bigg( \bm_{0}^2 -\kappa^2 \Bigg(\sum_{j=1}^{\cN}r_j\bs_j\Bigg)^2 
\Bigg) \left(\begin{array}{c}1 \\ 1\end{array}\right) +  \sum_{j=1}^{\cN} \bs_j^2\partial_x A_{r_j}(x-a_j) \\ & +
2\sum_{j=1}^{\cN}r_j\bs_j\cdot\Bigg(\ii\bm_{0} - \sum_{k\neq j}^{\cN}r_k\bs_k\alpha(a_j-a_k+(r_j-r_k)\ii\delta/2)\Bigg) A_{r_j}(x-a_j).
\end{split} 
\end{equation} 
Since the functions $\partial_x A_{r_j}(x-a_j)$, $A_{r_j}(x-a_j)$, and $\left(\begin{array}{c}1 \\ 1\end{array}\right)$ are linearly independent,
we conclude: {\em the ansatz \eqref{eq:hansatz1} satisfies the constraint \eqref{eq:bUbU} if and only if the following conditions hold true, 
\begin{equation} 
\label{eq:bsbs} 
\bs_j^2=0,\quad  \bs_j\cdot\Bigg( \ii\bm_{0} - \sum_{k\neq j}^{\cN}r_k\bs_k\alpha(a_j-a_k+(r_j-r_k)\ii\delta/2)\Bigg)=0 \quad (j=1,\ldots,\cN), 
\end{equation} 
and 
\begin{equation} 
\label{eq:bmbm}
 \bm_{0}^2- \kappa^2\Bigg( \sum_{j=1}^{\cN}r_j\bs_j\Bigg)^2=\rho^2. 
\end{equation} }
Since the norms $\bu^2$ and $\bv^2$ do not change with time if $\bu$ and $\bv$ satisfy the ncIHF equation, the conditions \eqref{eq:bsbs} and \eqref{eq:bmbm} are constraints on initial conditions: {\em If \eqref{eq:bsbs} and \eqref{eq:bmbm} hold true at time $t=0$, they hold true at all times $t\in\R$.} This fact can also be checked directly using the time evolution equations \eqref{eq:bsdot}--\eqref{eq:adot} derived below.   

\paragraph{Spin-pole dynamics.} To show that the ansatz \eqref{eq:hansatz1} provides solutions of the ncIHF equation \eqref{eq:ncIHF1}, we compute
\begin{equation} 
\label{eq:bUdot} 
\bU_t = \ii\sum_{j=1}^{\cN}r_j \left( \dot \bs_jA_{r_j}(x-a_j) - \bs_j\dot a_j\partial_x A_{r_j}(x-a_j) \right)
\end{equation} 
and, using \eqref{eq:AId1} and the fact that $\cT$ and $\partial_x$ commute, 
\begin{equation} 
\label{eq:cTbUx}
\cT\bU_x =  \ii \sum_{j=1}^{\cN} r_j \bs_j \cT \partial_x A_{r_j}(x-a_j) = \ii \sum_{j=1}^{\cN} r_j \bs_j \partial_x\cT A_{r_j}(x-a_j) =  \sum_{j=1}^{\cN}  \bs_j \partial_xA_{r_j}(x-a_j) . 
\end{equation} 
The latter implies, using \eqref{eq:hansatz1} again and rules \eqref{eq:rules},  
\begin{equation} 
\label{eq:bUTbU0}
\bU\wedgecirc \cT\bU_x =  \sum_{j=1}^{\cN}  \bm_{0}\wedge \bs_j \partial_xA_{r_j}(x-a_j) +\ii \pvsum_{j,k=1}^{\cN} r_j \bs_j\wedge\bs_k A_{r_j}(x-a_j)\circ \partial_xA_{r_k}(x-a_k)  . 
\end{equation} 
To proceed, we use the following identity obtained by differentiating \eqref{eq:AId2} with respect to $b$ and using $\partial_{z}\alpha(z)=-V(z)$, 
\begin{equation} 
\label{eq:Aid22}
\begin{split} 
A_r(x-a)\circ \partial_x A_{s}(x-b) = &  -\alpha(a-b+(r-s)\ii\delta/2) \partial_x A_{s}(x-b)   \\ & -V(a-b+(r-s)\ii\delta/2) \left( A_r(x-a)-A_{s}(x-b)\right) . 
\end{split} 
\end{equation}  
Inserting this in \eqref{eq:bUTbU0}, we obtain after some computations (computational details are the same as in  \cite[Appendix A.2]{berntsonklabbers2020} and thus omitted)
\begin{equation} 
\label{eq:bUTbUx}
\begin{split} 
\bU\wedgecirc \cT\bU_x = &   \ii \sum_{j=1}^{\cN} \bs_j\wedge \Bigg( \ii\bm_{0} - \sum_{k\neq j}^{\cN}r_k\bs_k\alpha(a_j-a_k+(r_j-r_k)\ii\delta/2)\Bigg) \partial_xA_{r_j}(x-a_j) \\ 
& +  \pvsum_{j,k=1}^{\cN}\ii (r_j +r_k) \bs_j\wedge\bs_k V(a_j-a_k+(r_j-r_k)\ii\delta/2)A_{r_j}(x-a_j)  . 
\end{split} 
\end{equation} 
Equating  \eqref{eq:bUdot}  and \eqref{eq:bUTbUx} we conclude, similarly as above: {\em  the ansatz \eqref{eq:hansatz1} satisfies the ncIHF equation in  \eqref{eq:ncIHF1} if and only if the following time evolution equations hold true, 
\begin{equation} 
\label{eq:bsdot}
 \dot \bs_j = -\sum_{k\neq j}^{\cN}(1+r_jr_k) \bs_j\wedge\bs_k V(a_j-a_k+(r_j-r_k)\ii\delta/2) 
\end{equation} 
and 
\begin{equation} 
\label{eq:adot}
 \bs_j\dot a_j =  -r_j\bs_j\wedge \Bigg( \ii\bm_{0} - \sum_{k\neq j}^{\cN}r_k\bs_k\alpha(a_j-a_k+(r_j-r_k)\ii\delta/2)\Bigg)
\end{equation} 
for $j=1,\ldots,\cN$.} 

\paragraph{Consistency.} One can show that  \eqref{eq:adot} is consistent with the other equations in \eqref{eq:bsbs} and \eqref{eq:bsdot}; moreover, one can replace \eqref{eq:adot} by the following equations, without loss of information, 
\begin{equation} 
\label{eq:adot1general} 
\dot a_j = -\frac{\bs_j^*\wedge\bs_j}{\bs_j^*\cdot\bs_j}\cdot  \Bigg( \ii\bm_{0} - \sum_{k\neq j}^{\cN}r_k\bs_k\alpha(a_j-a_k+(r_j-r_k)\ii\delta/2)\Bigg)\quad (j=1,\ldots,\cN) 
\end{equation} 
(both these results follow from \cite[Lemma A.3]{berntsonklabbers2020}). For later reference, we specialize this to \eqref{eq:shorthandreal}: 
\begin{equation} 
\label{eq:adot1real} 
\dot a_j = -\frac{\bs_j^*\wedge\bs_j}{\bs_j^*\cdot\bs_j}\cdot  \Bigg( \ii\bm_{0} - \sum_{k\neq j}^{N}\bs_k\alpha(a_j-a_k)+ \sum_{k=1}^{N}\bs_k^*\talpha(a_j-a_k^*)\Bigg)\quad (j=1,\ldots,N). 
\end{equation} 

\paragraph{Relation to Calogero-Moser system.} Clearly, by the shifts 
\begin{equation} 
\label{eq:shift}
a_j\to a_j-r_j \ii\delta/2\quad (j=1,\ldots,\cN) , 
\end{equation} 
the arguments in all equations \eqref{eq:bsbs}, \eqref{eq:bsdot}, and \eqref{eq:adot} simplify as follows, 
\begin{equation} 
a_j-a_k+(r_j-r_k)\ii\delta/2\to a_j-a_k\quad (j,k=1,\ldots,\cN) . 
\end{equation}  
Thus, these four equations can be written as \cite[Eqs. (A.10)--(A.21)]{berntsonklabbers2020}, and \cite[Proposition A.4]{berntsonklabbers2020} applies as it stands (this is true since the proof of this result only uses certain identities satisfied by the functions $\alpha(z)$ and $V(z)$, and these identities are true in the present case as well; see Appendix~\ref{app:propalpha}). 
We can thus conclude: {\em If the constraints \eqref{eq:bsbs} hold true, then the first order equations  \eqref{eq:bsdot}--\eqref{eq:adot} imply
\begin{equation} 
\label{eq:addot} 
\ddot a_j = -\sum_{k\neq j}^{\cN}(1+r_jr_k)V^\prime(a_j-a_k)
\end{equation} 
for $j=1,\ldots,\cN$}. Note that, if \eqref{eq:addot} holds true for the variables shifted as in \eqref{eq:shift}, it holds true for the unshifted variables as well (this is true since, for cases $r_j=-r_k$ where the shifts are non-trivial,  there are no interactions in \eqref{eq:addot}). 

By specializing the results above to \eqref{eq:shorthandreal}, we obtain the multi-soliton solutions summarized in \eqref{eq:solution}--\eqref{eq:constraintsbm}: Eqs.\ \eqref{eq:bsdot} and \eqref{eq:addot} reduce to the ones in \eqref{eq:CM}; Eq.\ \eqref{eq:adot1real}  at time $t=0$ becomes the third equation in \eqref{eq:ajsjt=0} (the first two introduce notation);  the constraints \eqref{eq:constraintsbs}--\eqref{eq:constraintsbm} follow from Eqs.\ \eqref{eq:bsbs} and \eqref{eq:bmbm} at time $t=0$. 

This completes our derivation of the multi-soliton solutions \eqref{eq:solution}--\eqref{eq:constraintsbm}. 

\paragraph{Traveling wave solutions.} It is interesting to note that, if one chooses $\bs_{j,0}=\bs$ (independent of $j=1,\ldots,N$), then the equations of motion of the hyperbolic CM model \eqref{eq:CM} are solved by  $\bs_j(t)=\bs$ (independent of $t$) and $a_j(t)=a_{j,0}+vt$ for $j=1,\ldots,N$ with $v=-\ii(\bs^*\wedge\bs)\cdot\bm_{0}/(\bs^*\cdot\bs)$, and \eqref{eq:solution}--\eqref{eq:constraintsbm} reduces to traveling wave solutions. This generalizes well-known solutions of the HWM equation \cite{zhou2015,lenzmann2018}; see also \cite{berntsonklabbers2020}.

\subsection{Complex multi-soliton solutions}
\label{sec:complexsolutions}  
(This section can be skipped without loss of continuity.) 

While the real-valued multi-soliton solutions of the ncIHF equation in  \eqref{eq:solution}--\eqref{eq:constraintsbm} are most interesting from a physics point of view, our results in Section~\ref{sec:solitonderivation} provide more general, complex-valued,  solutions which are interesting from a mathematics point of view: 
instead of \eqref{eq:shorthandreal}, one can use the  following more general specialization of the ansatz \eqref{eq:hansatz1}, 
\begin{equation} 
\label{eq:shorthandcomplex}
\cN=N+M,\quad (a_j,\bs_j,r_j)=\begin{cases} (a_j,\bs_j,+) & (j=1,\ldots,N) \\ (b_{j-N},\bt_{j-N},-) & (j=N+1,\ldots,\cN) \end{cases} 
\end{equation} 
with $(a_j,\bs_j)\in R^+_\delta\times\C^3$ for $j=1,\ldots,N$ and $(b_j,\bt_j)\in R^-_\delta\times\C^3$ for $j=1,\ldots,M$, for arbitrary variable numbers $N$ and $M$. 
Moreover, one can allow $\rho$ in \eqref{eq:bUbU} to be an arbitrary non-zero complex parameter. 
With that, our results imply the following multi-soliton solutions generalizing the ones in  \eqref{eq:solution}--\eqref{eq:constraintsbm}: 
{\em Let $\rho\in\C\setminus\{0\}$ be fixed. Then, for arbitrary variable numbers $N$ and $M$, 
\begin{equation}
\label{eq:solution2}
\begin{split} 
\bu(x,t) & = \bm_{0} + \ii \sum_{j=1}^N \bs_j(t) \alpha(x -a_j(t) + \ii \delta/2) -\ii\sum_{j=1}^M \bt_j(t) \alpha(x -b_j(t) - \ii \delta/2),\\
\bv(x,t) & = \bm_{0} + \ii \sum_{j=1}^N \bs_j(t) \alpha(x -a_j(t) - \ii \delta/2) - \ii\sum_{j=1}^M \bt_j(t) \alpha(x -b_j(t) + \ii \delta/2),
\end{split} 
\end{equation} 
is a solution of the ncIHF equation \eqref{eq:ncIHF} satisfying $\bu(x,t)^2=\bv(x,t)^2=\rho^2$ provided the variables $(a_j(t),\bs_j(t))\in R^+_\delta\times\C^3$ and $(b_j(t),\bt_j(t))\in R^-_\delta\times\C^3$ satisfy the following conditions, (i) they evolve in time according to the following equations of motion 
\begin{equation}
\label{eq:CM2a}
\begin{split} 
\begin{split}
\dot{\bs}_j &= -2 \sum_{k\neq j}^N \bs_j \wedge \bs_kV(a_j - a_k), \\
\ddot a_j &= -2\sum_{k\neq j}^N\bs_j\cdot\bs_k V^\prime(a_j-a_k), 
\end{split}\quad (j=1,\ldots,N) 
\\
\begin{split}
\dot{\bt}_j &= -2 \sum_{k\neq j}^M \bt_j \wedge \bt_kV(b_j - b_k), \\
\ddot b_j &= -2\sum_{k\neq j}^M\bt_j\cdot\bt_k V^\prime(b_j-b_k), 
\end{split}\quad (j=1,\ldots,M) , 
\end{split} 
\end{equation}
(ii) the initial conditions are given by 
\begin{equation} 
\label{eq:ajsjt=02} 
\begin{split} 
\begin{split} 
a_j(0)&=a_{j,0} ,\quad \bs_j(0)=\bs_{j,0},\\
\dot{a}_j(0)=  -\frac{\bs_{j,0}^*\wedge \bs_{j,0}}{\bs_{j,0}^*\cdot\bs_{j,0}} \cdot &\bigg( \ii \bm_{0}-\sum_{k\neq j}^{N} \bs_k \alpha(a_{j,0}-a_{k,0}) + \sum_{k =1 }^{M} \bs_k^*\talpha(a_{j,0}-b_{k,0})  \bigg)
\end{split} 
\\
\begin{split} 
b_j(0)&=b_{j,0} ,\quad \bt_j(0)=\bt_{j,0},\\
\dot{b}_j(0)= +\frac{\bt_{j,0}^*\wedge \bt_{j,0}}{\bt_{j,0}^*\cdot\bt_{j,0}} \cdot &\bigg( \ii \bm_{0}+\sum_{k\neq j}^{M} \bt_k \alpha(b_{j,0}-b_{k,0}) - \sum_{k =1 }^{N} \bs_k\talpha(b_{j,0}-a_{k,0})  \bigg) 
\end{split} 
\end{split} 
\end{equation} 
with $\bm_{0}\in\C^3$, $(a_{j,0},\bs_{j,0})\in R^+_\delta\times\C^3$ ($j=1,\ldots,N$) and $(b_{j,0},\bt_{j,0})\in R^-_\delta\times\C^3$  ($j=1,\ldots,M$) satisfying the following constraints, 
\begin{equation} 
\begin{split} 
 \bs_{j,0}^2=0,\quad \bs_{j,0}\cdot \bigg( \ii \bm_{0}-\sum_{k\neq j}^{N} \bs_k \alpha(a_{j,0}-a_{k,0}) + \sum_{k =1 }^{M} \bt_k\talpha(a_{j,0}-b_{k,0})  \bigg)=0,  \\
  \bt_{j,0}^2=0,\quad \bt_{j,0}\cdot \bigg( \ii \bm_{0}+\sum_{k\neq j}^{M} \bt_k \alpha(b_{j,0}-b_{k,0}) - \sum_{k =1 }^{N} \bs_k\talpha(b_{j,0}-a_{k,0})  \bigg)=0, 
\end{split} 
\end{equation} 
and 
\label{eq:constraints2}
\begin{equation} 
 \bm_{0}^2 -\kappa^2\Bigg(\sum_{j=1}^N\bs_{j,0}-\sum_{j=1}^M\bt_{j,0} \Bigg)^2 = \rho^2.
\end{equation} 
} 

It would be interesting to study these complex solutions in more details. However, this is beyond the scope of the present paper: 
In the following, we only discuss the real-valued solutions \eqref{eq:solution}--\eqref{eq:constraintsbm}.

\subsection{Energy of solitons} 
\label{sec:energy}
We define, using the notation in \eqref{eq:cT}, \eqref{eq:bU} and \eqref{eq:matrixnotation2}, 
\begin{equation} 
\label{eq:bE} 
\left(\begin{array}{c} \eps_{\bu} \\ -\eps_{\bv} \end{array}\right) \coloneqq -\frac12 \bU \dotcirc \cT\bU_x = -\frac12  \left(\begin{array}{c} \bu\cdot(T\bu_x-\tT\bv_x) \\ \bv\cdot(\tT\bu_x-T\bv_x)  \end{array}\right). 
\end{equation} 
By comparing with the Hamiltonian $\cH$ \eqref{eq:ncIHFHamiltonian} we observe that the functions $\eps_{\bu}$ and $\eps_{\bv}$ (of $x$ and $t$) defined in this way have a natural physical interpretation as energy density contributions  of $\bu$ and $\bv$, respectively: 
the total energy density $\eps$ such that  $\cH=\int_{\R}\eps \,\dd{x}$ is $\eps= \eps_{\bu}  +  \eps_{\bv}$; moreover, $\eps_{\bu}$ is the contribution of $\bu$, together with the interaction part between $\bu$ and $\bv$ that vanishes in the limit $\delta\to \infty$, and similarly for $\eps_{\bv}$. In the following, we compute the energy densities, $\eps_{\bu}$ and $\eps_{\bv}$, together with the total energy, $E_N=\cH$, for the real $N$-soliton solution $\bu$, $\bv$ in \eqref{eq:solution}--\eqref{eq:constraintsbm}. 

We use \eqref{eq:hansatz1}, \eqref{eq:cTbUx} and $\bs_j^2=0$ to obtain, similarly as \eqref{eq:bUTbU0}, 
\begin{equation} 
\label{eq:bUTbU2}
\bU\dotcirc \cT\bU_x =  \sum_{j=1}^{\cN}  \bm_{0}\cdot \bs_j \partial_xA_{r_j}(x-a_j) +\ii \pvsum_{j,k=1}^{\cN} r_j \bs_j\cdot\bs_k A_{r_j}(x-a_j)\circ \partial_xA_{r_k}(x-a_k)  . 
\end{equation} 
Inserting the identity in \eqref{eq:Aid22} we obtain, similarly as \eqref{eq:bUTbUx},  
\begin{equation} 
\label{eq:bUTbUx1}
\begin{split} 
\bU\dotcirc \cT\bU_x = &  - \ii \sum_{j=1}^{\cN} \bs_j\cdot\Bigg( \ii\bm_{0} - \sum_{k\neq j}^{\cN}r_k\bs_k\alpha(a_j-a_k+(r_j-r_k)\ii\delta/2)\Bigg) \partial_xA_{r_j}(x-a_j) \\ 
& -  \pvsum_{j,k=1}^{\cN}\ii (r_j -r_k) \bs_j\cdot\bs_k V(a_j-a_k+(r_j-r_k)\ii\delta/2)A_{r_j}(x-a_j)  ; 
\end{split} 
\end{equation} 
note the sign differences in \eqref{eq:bUTbUx} and \eqref{eq:bUTbUx1} which are due to that the wedge produce is antisymmetric while the dot product is symmetric. 
The first sum in \eqref{eq:bUTbUx1} vanishes due to the second set of constraints in \eqref{eq:bsbs}; in the second sum, only the terms where $r_j=-r_k$ are non-zero, which allows us to replace the shifts $(r_j-r_k)\ii\delta/2$ in the arguments of $V$ by $\ii\delta$. 
Thus, recalling $V(z+\ii\delta)=\tV(z)$, we obtain 
\begin{equation} 
\bU\dotcirc \cT\bU_x =  -\pvsum_{j,k=1}^{\cN}\ii (r_j -r_k) \bs_j\cdot\bs_k \tV(a_j-a_k)A_{r_j}(x-a_j) . 
\end{equation} 
To get the energy densities of the real $N$-soliton solution, we specialize to \eqref{eq:shorthandreal}. This gives
\begin{equation} 
\bU\dotcirc \cT\bU_x =   
-2\ii \sum_{j,k=1}^{N}\left( \bs_j\cdot\bs_k^* \tV(a_j-a^*_k)A_{+}(x-a_j) - \bs_j^*\cdot\bs_k \tV(a_j^*-a_k)A_{-}(x-a_j^*)\right) \\
\end{equation} 
or, by recalling \eqref{eq:Ar} and \eqref{eq:bE}, 
\begin{equation} 
\label{eq:epsuepsv}
\begin{split} 
\eps_{\bu}(x) = &   - 2\sum_{j,k=1}^N \im\left(   \bs_j\cdot\bs_k^* \tV(a_j-a_k^*)\alpha(x-a_j + \ii\delta/2)  \right) , \\
\eps_{\bv}(x) = &   + 2\sum_{j,k=1}^N \im\left(   \bs_j\cdot\bs_k^* \tV(a_j-a_k^*)\alpha(x-a_j - \ii\delta/2)  \right) .
\end{split} 
\end{equation}  
With that, we can compute $E_{N}=\int_{\R}\left(\eps_{\bu}+\eps_{\bv} \right)\,\dd{x}$ using the exact integral
\begin{equation} 
\int_{\R} \left( \alpha(x-a + \ii\delta/2)  - \alpha(x-a - \ii\delta/2) \right)\, \dd{x} = \ii\pi\quad (\re(a)\in\R,\quad \delta/2<\im(a)<3\delta/2), 
\end{equation} 
and we obtain the result $E_N= -\pi   \sum_{j,k=1}^N \left(   \bs_j\cdot\bs_k^* \tV(a_j-a_k^*) +   \bs_j^*\cdot\bs_k \tV(a_j^*-a_k) \right)$, which was given already in \eqref{eq:EN}.

In particular, in the one-soliton case ($N=1$), setting $a_1=a$ and $\bs_1=\bs$, we obtain 
\begin{equation} 
\label{eq:eps1} 
\eps_{1}(x) = 
E_1 f(x-\aR;\aI),\quad  f(x;\aI) = -\frac{4\kappa}{\pi}\frac{\cosh(2\kappa x)\cos(2\kappa\aI)}{\cosh(4\kappa x) + \cos(4\kappa\aI)}
\end{equation}
where 
\begin{equation} 
\label{eq:E1} 
E_1=\frac{2\pi\kappa^2\bs\cdot\bs^* }{\cos^2(2\kappa\aI)}
\end{equation} 
is the total energy of the soliton and $f(x;\aI)$ describes the spatial distribution of its energy density:  this function $f(x;\aI)$ is well-localized in $x$, and it satisfies $f(x;\aI)\geq 0$ and $\int_{\R} f(x;\aI)\, \dd{x}=1$ for $\delta/2<\aI<3\delta/2$ (see Appendix~\ref{app:eps1} for computational details on these one-soliton results). 

\section{Explicit solutions: examples and properties} 
\label{sec:constraintsandexamples}

We present a detailed discussion of one-soliton solutions (Section~\ref{sec:onesoliton}), an exact method to solve the constraints  \eqref{eq:constraintsbs}--\eqref{eq:constraintsbm} for arbitrary soliton number (Section~\ref{sec:constraints}), and examples of two- and 
three-soliton solutions (Section~\ref{sec:examples}). We also give a physical interpretation of $N$-soliton solutions, based on a combination of analytical arguments and numerical results (Section~\ref{sec:Nsolitons}). 

\subsection{Geometry of one-solitons}
\label{sec:onesoliton}
\begin{figure}[!t]
\centering
\begin{tikzpicture}
\node at (-4,0) {
\begin{tikzpicture}
\def\a{1.6};
\def\b{210};
\def\c{0.6};
	\node at (0,0) {\includegraphics[scale=0.82]{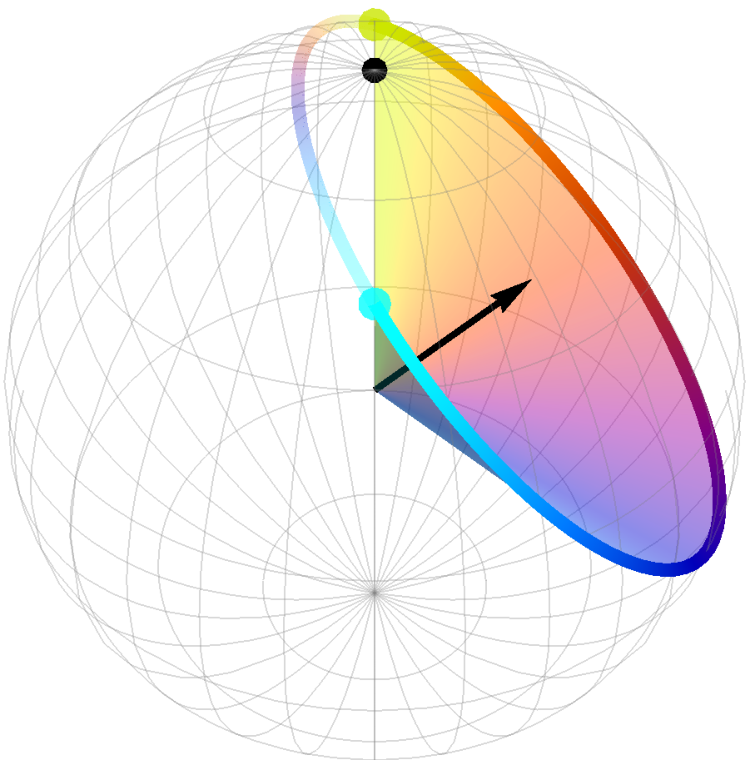}};
\node at (1.5,1) {$\mathbf{n}$};
  \draw[thick, black,->,rotate=25] (0.8,-0.3) arc
  (\b:330:\c cm and \a cm);
  \node at (3,-2.3) {$\bu$};
  \node at (-0.4,3.5) {\color{gray} $\bv$};
\end{tikzpicture}
};

\node at (4,0) {
\begin{tikzpicture}
\def\a{1.6};
\def\b{210};
\def\c{0.6};
\node at (0,0) {\includegraphics[scale=0.82	]{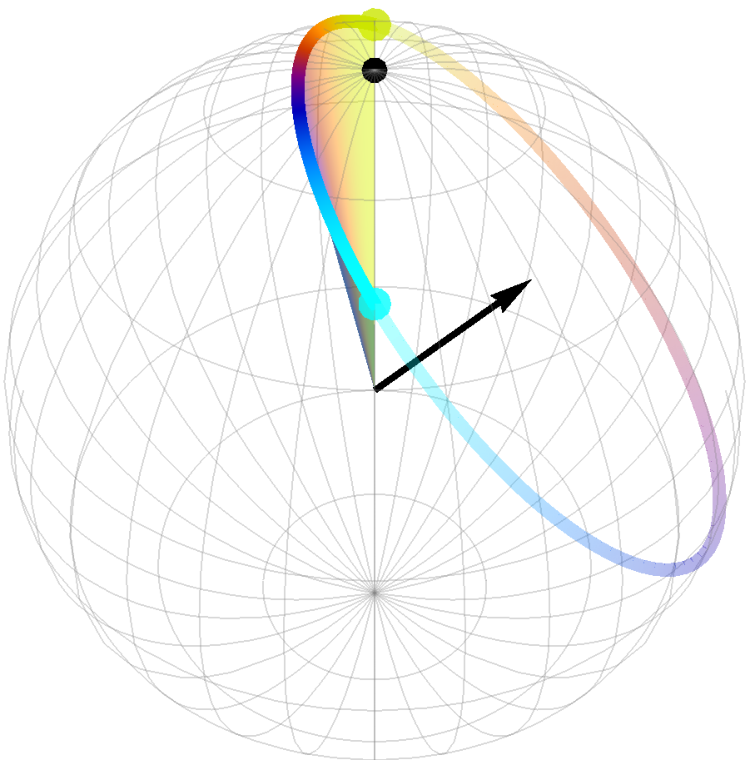}};
\node at (1.5,1) {$\mathbf{n}$};
\draw[thick, black,<-,rotate=205] (-1.6,-0.8) arc
  (\b:330:0.5 cm and \a cm);
  \node at (-0.4,3.5) {$\bv$};
    \node at (3,-2.3) {\color{gray} $\bu$};
\end{tikzpicture}
};

\node at (-0.26,-4.8) {\includegraphics[scale=1.4]{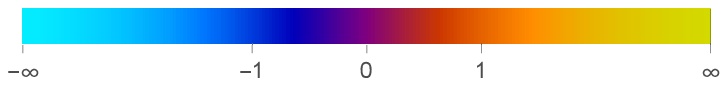}};
 \node[above] at (0,-4.3) {$x/\delta$};
\end{tikzpicture}
\caption{Spin configuration for the one-soliton solution with the initial data \eqref{eq:example1} at fixed time $t=0$. 
Shown are the reference vector $\bn_{0}$ (black dot),  the local vacua  $\bm_{\pm\infty}$ at $x\to\pm\infty$ (blue and yellow dots), the vector $\bn$ to the center of the rotation circle (black arrow), and the spin densities $\bu(x,t)$ (left plot) and $\bv(x,t)$ (right plot) as a function of $x\in\R$, sweeping out a colored cone: at fixed $x$, $\bu(x,t)$ and $\bv(x,t)$ are drawn as a colored vector from the origin with their color indicating the value of $x$ according to the legend underneath. To show how the images of $\bu$ and $\bv$ together form a circle on $S^2$, a shadow of the image of $\bv$ is plotted in the plot for $\bu$ and vice versa. The curved arrow indicates the rotation direction of $\bu$ (left plot) and $\bv$ (right plot) when going from $x=-\infty$ to $x=\infty$.}
\label{fig:one_soliton}
\end{figure}

To offer some geometric intuition, and as a warm-up to Sections~\ref{sec:constraints} and \ref{sec:Nsolitons}, we discuss the (real-valued) one-soliton solutions of the ncIHF equation, \eqref{eq:solution}--\eqref{eq:constraintsbm} for $N=1$; see Fig.~\ref{fig:one_soliton} for an illustration. 

\paragraph{One-soliton solution}
For $N=1$, the equations of motion of the hyperbolic spin CM system \eqref{eq:CM} are trivial: $\dot{\bs}_1=\mathbf{0}$ and $\ddot a_1=0$, which implies $\bs_1(t)=\bs$ and $a_1(t)=a + v t$, using the short-hand notation $\bs\coloneqq \bs_{1,0}$, $a\coloneqq a_{1,0}$ and $v\coloneqq \dot a_1(0)$. 
Thus, \eqref{eq:solution}--\eqref{eq:constraintsbm} specializes to the one-soliton solution 
\begin{equation}
\label{eq:onesoliton}
\begin{split} 
\bu(x,t) & = \bm_{0} + \ii \bs \alpha(x -a-vt + \ii \delta/2) - \ii \bs^* \alpha(x -a^*-v^*t - \ii \delta/2) ,\\
\bv(x,t) & = \bm_{0} + \ii\bs \alpha(x -a -vt- \ii \delta/2) - \ii \bs^* \alpha(x -a^*-v^*t + \ii \delta/2), 
\end{split} 
\end{equation} 
with 
\begin{equation} 
\label{eq:adot1} 
v =  -\ii \frac{\bs^*\wedge \bs}{\bs^*\cdot\bs} \cdot \bm_{0} 
\end{equation} 
(since $(\bs^*\wedge \bs)\cdot\bs^*=0$) and parameters $\bm_0\in\R^3$ and $(a,\bs)\in\C\times \C^3$ constrained by the conditions $\delta/2<\im(a)<3\delta/2$ and 
\begin{equation} 
\label{eq:con1} 
\bs^2=0,\quad \bs\cdot \left( \bm_{0}  -\ii \bs^*\talpha(a-a^*) \right) =0,\quad \bm_{0}^2+4\kappa^2 (\bsI)^2=1. 
\end{equation} 
We use the notation in \eqref{eq:aRaI} to write \eqref{eq:adot1} as 
\begin{equation}
\label{eq:adotgen} 
v = 2\frac{(\bsR\wedge\bsI)\cdot\bm}{\bs^*\cdot\bs}, 
\end{equation} 
which makes manifest that $v$ is real: $v=v^*$. 

To solve these constraints, we make use of the following fact \cite[Lemma B.1]{berntsonklabbers2020}: {\em solutions of $\bs^2=0$ in $\C^3$ are given by  
\begin{equation} 
\label{eq:ssquare} 
\bs=s(\bn_1+\ii\bn_2) 
\end{equation} 
with complex $s$ and unit vectors $\bn_1$ and $\bn_2$ in $\R^3$ that are orthogonal, and this representation of solutions is unique up to the following $\mathrm{U}(1)$-transformation, 
\begin{equation} 
\label{eq:Uone}
(s,\bn_1,\bn_2)\to (s\ee^{\ii\varphi},\bn_1\cos\varphi+\bn_2\sin\varphi, \bn_2\cos\varphi-\bn_1\sin\varphi)\quad (\varphi\in[0,2\pi)) .
\end{equation} } 

Inserting this and the parametrization 
\begin{equation} 
\label{eq:bbb} 
\bm_{0}=m\bn_0, \quad m\coloneqq |\bm_{0}|>0
\end{equation} 
into the second constraint in \eqref{eq:con1} we obtain $sm(\bn_1+\ii\bn_2)\cdot\bn_0-2\ii|s|^2\talpha(a-a^*) =0$, which gives $s^*$ and, by complex conjugation, 
\begin{equation} 
s = m \frac{\bn_0\cdot (\bn_1-\ii\bn_2)}{2\ii\talpha(a-a^*)} = -\frac{m}{2\kappa} \frac{\bn_0\cdot (\bn_1-\ii\bn_2)}{\tan(2\kappa\aI)} ; 
\end{equation}  
we inserted $\ii\talpha(a-a^*)=-\kappa\tan(2\kappa\aI)$. Thus, we find 
\begin{equation} 
\label{eq:bs} 
\bs = - \frac{m}{2\kappa}\cot(2\kappa\aI)\bn_0\cdot (\bn_1-\ii\bn_2)(\bn_1+\ii\bn_2), 
\end{equation} 
and the third constraint in \eqref{eq:con1} becomes $m^2+(2\kappa \bsI)^2=1$, allowing to compute $m$: 
\begin{equation} 
\label{eq:m} 
m = \left(1+(2\kappa\bsI/m)^2\right)^{-1/2},\quad \frac{2\kappa}{m}\bsI  =\cot(2\kappa\aI)\bn_0\wedge(\bn_1\wedge\bn_2). 
\end{equation} 
Moreover, inserting \eqref{eq:ssquare} and \eqref{eq:bbb} into \eqref{eq:adot1} yields 
\begin{equation} 
\label{eq:adot2} 
v = m \bn_0\cdot (\bn_1\wedge\bn_2) . 
\end{equation} 

To summarize: 
{\em for fixed $\bn_0\in S^2$, choose $a=\aR+\ii\aI$ with $\aR\in\R$ and $\delta/2<\aI<3\delta/2$, and $(\bn_1,\bn_2)\in S^2\times S^2$ such that $\bn_1\cdot\bn_2=0$; then \eqref{eq:onesoliton} with $\bm_{0}$, $\bs$, $v$ in \eqref{eq:bbb}, \eqref{eq:bs} and \eqref{eq:m} give an explicit one-soliton solution of the ncIHF equation \eqref{eq:ncIHF}.} 

For example, if we choose 
\begin{equation} 
\label{eq:example1} 
\bn_0=(0,0,1),\quad  a=\frac{3}{4}\ii \delta,\quad \bn_1=(1,0,0) , \quad \bn_2=\frac1{\sqrt{2}}(0,1,1), 
\end{equation} 
we obtain 
\begin{equation} 
\label{eq:initial_data_one_soliton}
\bm_{0}=\sqrt{\frac{2}{3}}(0,0,1), \quad \bs= \frac1{4\kappa}\sqrt{\frac{2}{3}}(-\sqrt{2}\ii,1,1) , \quad v= \sqrt{\frac{1}{3}}.
\end{equation} 

\paragraph{Number of degrees of freedom.}
For fixed vacuum direction $\bn_0$, a soliton seems to have 5 real degrees of freedom: the real part of the pole $a$, the imaginary part of the pole $a$, two angles fixing the real unit vector $\bn_1$, and a third angle fixing the real unit vector $\bn_2$ orthogonal to $\bn_1$. However, at closer inspection, one can see that one degree of freedom disappears due to the $\mathrm{U}(1)$-invariance in \eqref{eq:ssquare}--\eqref{eq:Uone}:  The results in \eqref{eq:bs} can be written as 
\begin{equation} 
\label{eq:bs1} 
\bs =  \frac{m}{2\kappa}\cot(2\kappa\aI)\left( (\bn_0\cdot\bn_3)\bn_3-\bn_0+\ii\bn_0\wedge\bn_3 \right) ,
\end{equation} 
with $\bn_3\coloneqq \bn_1\wedge\bn_2$ the unit vector orthogonal to the plane spanned by $\bn_1$ and $\bn_2$. This and \eqref{eq:m}--\eqref{eq:adot2} make manifest that, actually, only {\em one}  real unit vector $\bn_3$ is relevant for the one-soliton solution: a soliton has only 4 degrees of freedom. 

\paragraph{Geometric interpretation.}
We write \eqref{eq:onesoliton} as 
\begin{equation}
\label{eq:onesoliton1}
\left. \begin{array}{c} \bu(x,t)\\ \bv(x,t)\end{array} \right\} = \bm_0 -2\, \im\left( \alpha\left(z_\pm\right)\bs\right) ,\quad z_\pm = x-\aR-vt -\ii(\aI\mp\delta/2) 
\end{equation} 
where $\bs=\bsR+\ii\bsI$ with $|\bsR|=|\bsI|$ (the latter follows from \eqref{eq:ssquare}).  Clearly, this describes spatial configurations of $\bu$ and $\bv$ moving with the same velocity, $v$, to the left ($v<0$) or right ($v>0$), with the limiting case $v=0$ corresponding a static soliton. 
Since $\alpha(z)\to\pm \kappa$ as $\re(z)\to\pm\infty$, the common asymptotic vacuum for $\bu$ and $\bv$ at spatial infinity $x\to\pm\infty$ (defined in \eqref{eq:BC})  is 
\begin{equation}
\label{eq:onesoliton2}
\bm_{\pm\infty}= \bm_0 \mp 2\kappa\bsI . 
\end{equation} 
Moreover, it is natural to regard the position $x=\aR$ at $t=0$ as the center of the soliton since, there, the deviations of $\bu$ and $\bv$  from their asymptotic values $\bm_{\pm\infty}$ are maximal
\begin{equation} 
\label{eq:uCMvCM}
 \left. \begin{array}{c} \bu_{\mathrm{c}} \coloneqq \bu(\aR,0)\\ \bv_{\mathrm{c}} \coloneqq  \bv(\aR,0)\end{array} \right\} = 
 \bm_{0}+ 2\kappa\bsR\cot(\kappa\aI\mp\pi/4); 
\end{equation} 
note that $\cot(\kappa\aI\mp\pi/4)\lessgtr 0$ for $\delta/2<\aI<3\delta/2$. 
For generic values of $x$ and $t$, we define $\varphi_\pm\in[0,2\pi)$ so that  $\alpha(z_\pm)=|\alpha(z_\pm)|\ee^{\ii\varphi_\pm}$ and use $\im(\ee^{\ii\varphi_\pm}\bs)=\bsI\cos\varphi_\pm +\bsR\sin\varphi_\pm$ to write \eqref{eq:onesoliton1} as 
 \begin{equation}
\label{eq:onesoliton3}
\left. \begin{array}{c} \bu(x,t)\\ \bv(x,t)\end{array} \right\} = \bm_{0} -2\left|\alpha\left(z_\pm\right)\right|\left( \bsI\cos\varphi_\pm +\bsR\sin\varphi_\pm \right).   
\end{equation} 
This makes manifest that, as one varies position $x$ and/or time $t$, the spin densities $\bu$ and $\bv$ are always on the plane containing the point $\bm_{0}$ and spanned by the vectors $\bsR$ and $\bsI$. This, together with results discussed above, implies that $\bu$ and $\bv$ rotate on the circle obtained by intersecting this plane with $S^2$, and the center and radius of this circle are given by 
\begin{equation} 
\label{eq:bn} 
\bn =  \frac12\left( \bu_{\mathrm{c}} + \bv_{\mathrm{c}} \right) = \bm_0 -2\kappa\bsR\tan(2\kappa \aI) 
\end{equation}  
and 
\begin{equation} 
\label{eq:R} 
R = \frac12  \left|  \bu_{\mathrm{c}} - \bv_{\mathrm{c}} \right| = -\frac{2\kappa|\bsR|}{\cos(2\kappa\aI)}, 
\end{equation} 
respectively. The rotations for $\bu$ and $\bv$ both start at $\bm_{-\infty}$ at $x\to-\infty$ and end at $\bm_{+\infty}$ at $x=\pm\infty$, thus the rotation directions are opposite for $\bu$ and $\bv$; see Fig.~\ref{fig:one_soliton}. Note that the rotation circle radius is related to the energy \eqref{eq:E1} as follows, $E_1=\pi R^2$. 

The velocity of the soliton is always in the range $-1 < v <1$ (this follows from \eqref{eq:m} and \eqref{eq:adot2}; note that $|v|=1$ is only possible if $\bn_0=\pm\bn_1\wedge\bn_2$, but this corresponds to the trivial solution $\bu(x,t)=\bv(x,t)=\bn_0$), and there are two limiting cases with $v=0$:  (i) $\bn_0\cdot(\bn_1\wedge\bn_2)=0$, (ii) $\aI=\delta\pm \eps$ with $\eps\downarrow 0$, in which case $v\to 0$, $\bm_{0}\to \mathbf{0}$ and $\bm_{+\infty}=-\bm_{\infty}$; in both cases (i) and (ii),  $\bu$ and $\bv$ rotate on a great circle of radius 1.

\paragraph{Soliton channels and chirality.} We propose to think of a soliton as consisting of two channels: the $\bu$-channel corresponding to the spin density $\bu$, and the $\bv$-channel corresponding to the spin density $\bv$. 
In a soliton, these two channels are tightly coupled, behave in a synchronised way (in that a change in one channel is always accompanied by a change in the other channel) and, generically, depending on parameters, either the $\bu$- or the $\bv$-channel dominates (in the sense that there is more spin rotation in one channel or the other). 
A simple way to see this is to compute the energy in the two channels, $E_{\bu}\coloneqq \int_{\R}\eps_{\bu}\, \dd{x}$ and $E_{\bv}\coloneqq \int_{\R}\eps_{\bv}\, \dd{x}$ with the energy densities $\eps_{\bu}$ and $\eps_{\bv}$ obtained from the ones in \eqref{eq:epsuepsv} by specializing to $N=1$:  the dominating channel has the larger energy. 
We found by an exact computation (see Appendix~\ref{app:eps1} for details):  
\begin{equation} 
\label{eq:EuEv} 
\frac{E_{\bu}-E_{\bv}}{E_{\bu}+E_{\bv}} = 2\left(1-\frac{\aI}{\delta} \right)
\quad (\delta/2<\aI<3\delta/2). 
\end{equation} 
Thus,  for $\delta/2<\aI<\delta$, the one-soliton is dominated by the $\bu$-channel and, for $\delta<\aI<3\delta/2$,  it is dominated by the $\bv$-channel; the limiting case $\aI=\delta$ corresponds to the special situation where both channels have equal weight and, in this case, the soliton is static: $v=0$.

The $\bu$- and $\bv$-channels of a soliton have, generically, a well-defined chirality, $\mathrm{ch}$,  which we define as the product of the propagation direction, $\mathrm{pd}\in\{+,-\}$, and rotation direction, $\mathrm{rd}\in\{+,-\}$: $\mathrm{ch}\coloneqq \mathrm{pd}\times \mathrm{rd}$. 
We define: $\mathrm{pd}= +$ for right-moving solitons ($v>0$), and $\mathrm{pd}=-$ for left-moving solitons ($v<0$); for static solitons with $v=0$, $\mathrm{pd}$ is not defined. 
Thus, $\mathrm{pd}$ is the same for both channels and, by \eqref{eq:adotgen}, it is equal to the sign of $(\bsR\wedge\bsI)\cdot\bm_{0}$. However, as discussed, the rotation direction is different for the two channels. 
We define and compute the rotation direction $\mathrm{rd}_{\bu}$ for the $\bu$-channel (the argument for the $\bv$-channel is similar). 

The vector $\bu$ starts from $\bu=\bm_{-\infty}$ at position $x\to-\infty$ and rotates towards $\bu=\bu_{\mathrm{c}}$ at $x=\aR$; we use the right hand-rule  to define: $\mathrm{rd}_{\bu}=+$ if the thumb of the right hand is parallel with $\bn$ (= vector from the origin to the center of the rotation circle) when its other fingers bend in the rotation direction from $\bu=\bm_{-\infty}$ to  $\bu=\bu_{\mathrm{c}}$, otherwise $\mathrm{rd}_{\bu}=-$; clearly, one can compute $\mathrm{rd}_{\bu}$ as the sign of $(\bm_{-\infty}\wedge\bu_{\mathrm{c}})\cdot\bn$, which is equal to the sign of $(\bm_{-\infty}\wedge\bu_{\mathrm{c}})\cdot\bv_{\mathrm{c}}=\bm_{-\infty}\cdot(\bu_{\mathrm{c}}\wedge\bv_{\mathrm{c}})$ by \eqref{eq:bn}, which is equal to the sign of 
\begin{equation} 
-\bsI\cdot(\bm_0\wedge\bsR)\left(\cot(\kappa\aI +\pi/4)-\cot(\kappa-\pi/4)\right) =-(\bsR\wedge\bsI)\cdot\bm_{0}\frac2{\cos(2\kappa\aI)}
\end{equation} 
by \eqref{eq:uCMvCM}. Thus, $\mathrm{ch}_{\bu}=-\sgn\left( 1/\cos(2\kappa\aI) \right)=+$, and by a similar argument for $\bv$, $\mathrm{ch}_{\bv}=-$. To summarize: 
\begin{equation} 
\label{eq:chuchv}
\begin{split} 
\mathrm{ch}_{\bu}=+,\\\mathrm{ch}_{\bv}=-, 
\end{split} 
\quad (\delta/2<\aI<3\delta/2,\quad \aI\neq\delta). 
\end{equation} 
Therefore, since for $\delta/2<\aI<\delta$ the soliton is dominated by the $\bu$-channel, the soliton has (mainly) chirality $+$ in this parameter regime, and, similarly, for $\delta<\aI<3\delta/2$, it has chirality $-$. 
It is interesting to note that this result is exactly the same as for the solitons of the non-chiral ILW equation (where chirality is equal to the propagation direction) \cite{berntson2020}. 

The arguments above also apply to the HWM equation in a limiting case: the HWM equations $\bu=\bu\wedge H\bu_x$ and $\bv=-\bv\wedge H\bv_x$ are chiral in that they only gives raise to solitons of chirality $+$ and $-$, respectively, as mentioned in the introduction. 

\subsection{Solution of constraints} 
\label{sec:constraints}

The constraints \eqref{eq:constraintsbs}--\eqref{eq:constraintsbm} are nonlinear equations that must be satisfied by the parameters appearing in our multi-soliton solutions \eqref{eq:solution}--\eqref{eq:ajsjt=0}. Analogous equations for the HWM equation \eqref{eq:HWM} were obtained in \cite{berntsonklabbers2020}, where a numerical scheme for their solution was proposed.
In this section, we show how  \eqref{eq:constraintsbs}--\eqref{eq:constraintsbm} can be solved exactly using linear algebra. 
This method can also be adapted to the constraints \cite[Eq.~(1.6)]{berntson2020} for the HWM equation.

To solve the constraints \eqref{eq:constraintsbs}--\eqref{eq:constraintsbm}, we simplify notation and write $a_{j}$  and $\bs_{j}$ for $a_{j,0}$ and $\bs_{j,0}$, respectively.  We take the complex conjugate of the second constraint in \eqref{eq:constraintsbs} and multiply by $\ii$ to write it as  
\begin{equation} 
\label{eq:secondconstraint} 
\bs^*_{j}\cdot \bigg( - \ii \sum_{k =1 }^{N} \bs_k\talpha(a^*_{j}-a_{k}) + \ii \sum_{k\neq j}^{N} \bs^*_k \alpha(a^*_{j}-a^*_{k})  \bigg)= \bs^*_{j}\cdot\bm_{0}. 
\end{equation} 
The general solution of $\bs_j^2=0$  is
\begin{equation}
\label{eq:constraint1generalsolution}
\bs_j=s_j\bn_{j,12},\quad \bn_{j,12}\coloneqq \bn_{j,1}+\ii \bn_{j,2}
\end{equation}
where $s_j\in\C$ is arbitrary and $(\bn_{j,1},\bn_{j,2})\in S^2\times S^2$ such that $\bn_{j,1}\cdot\bn_{j,2}=0$; see \eqref{eq:ssquare}--\eqref{eq:Uone}. Inserting this in \eqref{eq:secondconstraint} and dividing by $s^*_j$ we obtain 
\begin{equation} 
\label{eq:linearsystem}
- \sum_{k =1 }^{N} \ii\bn_{j,12}^*\cdot\bn_{k,12} \talpha(a^*_{j}-a_{k})s_k + \sum_{k\neq j}^{N} \ii\bn_{j,12}^*\cdot\bn_{k,12} \alpha(a^*_{j}-a^*_{k})s_k^* = \bn_{j,12}^*\cdot\bm_{0} 
\quad (j=1,\ldots,N)
\end{equation} 
which, together with the complex conjugate of these equations,  provides a linear system for the variables $s_j/m$ and $s_j^*/m$, $j=1,\ldots,N$; note that, at this point, only $\bn_0=\bm_{0}/m\in S^2$ but not $m=|\bm_{0}|$ is known. 

We now write \eqref{eq:linearsystem} in matrix form by defining
\begin{equation}
\label{eq:matrixdefinitions}
\begin{split}
A\coloneqq &\; (A_{jk})_{j,k=1}^N,\qquad A_{jk}=-\ii \bn_{j,12}^*\cdot \bn_{k,12}\talpha(a_j^*-a_k)/\kappa,  \\  
B\coloneqq &\;(B_{jk})_{j,k=1}^N,\qquad B_{jk}=\ii(1-\delta_{jk})\bn_{j,12}^*\cdot\bn_{k,12}^*{\alpha}(a_j^*-a_k^*)/\kappa, \\
C\coloneqq &\;(C_{j})_{j=1}^N,\qquad \quad  C_j=2\bn_0\cdot\bn_{j,12}^* , 
\end{split}
\end{equation}
which allows to write \eqref{eq:linearsystem} and its complex conjugate as the linear system
\begin{equation} 
\label{eq:system} 
\begin{split} 
AX+BX^*=&C,\\
A^*X^*+B^*X=&C^*, 
\end{split} 
\end{equation} 
for the unknown 
\begin{equation}
\label{eq:X}
X\coloneqq \; (X_{j})_{j=1}^N,\qquad \quad X_j = 2\kappa s_j/m ;
\end{equation}
note that our normalization is such that $A,B,C$  in \eqref{eq:matrixdefinitions} are independent of $\kappa$ provided the parameters $a_j$ are expressed in units of $\delta$; see \eqref{matrixdefinitions1}.  
We observe that $A^*_{jk}=A_{kj}$ and $B^*_{jk}=-B^*_{kj}$, which suggests to write \eqref{eq:system}  as in \eqref{combinedlinearsystem1}: the  $2N\times 2N$ matrix appearing in  \eqref{combinedlinearsystem1} is self-adjoint. 

Hence, for given $\bn_0\in S^2$, $a_j\in R^+_\delta$ and $(\bn_{j,1},\bn_{j,2})\in S^2\times S^2$ such that $\bn_{j,1}\cdot\bn_{j,2}=0$ for $j=1,\ldots,N$, the values of $X_j=2\kappa s_j/m$ can be determined by solving the linear algebra problem in \eqref{combinedlinearsystem1}. Subsequently, we can use \eqref{eq:constraintsbm} to compute $m$ as in \eqref{eq:msolution}. 
Similarly as in the case $N=1$, due to the $\mathrm{U}(1)$-gauge invariance in \eqref{eq:ssquare}--\eqref{eq:Uone}, it is actually more convenient to first pick $\bn_0$, $a_j\in R^+_j$ and $\be_{j,3}\in S^2$ for $j=1,\ldots,N$, and only then choose unit vectors $\be_{j,1}$ and $\be_{j,2}$ so that $\be_{j,3} =\be_{j,1}\wedge\be_{j,2}$: the solution does not depend on the latter choice. 

Clearly, the problem of determining $X$ has a unique solution unless the self-adjoint $2N\times 2N$ matrix in  \eqref{eq:constraintsbm} is singular; generically, this does not happen. 
However, it happens for  special parameter values. For example, for $N=1$,  this matrix is singular for $\aI=\delta$ and non-singular otherwise; however, our formulas above give a well-defined solutions of the constraints in the singular limits $\aI=\lim_{\eps\downarrow 0}\delta\pm\eps$: as discussed, the solutions obtained in these limits are special in that they describes solitons which are static and where both channels, $\bu$ and $\bv$, have equal weight. 
It would be interesting to study possible singular cases also for $N>1$, but this is left for future work. 

\begin{figure}[t]
\centering
\begin{tikzpicture}[scale=0.98]
\def\a{3.4};
\def\b{2.5};
\def\d{2.1};
\node at (0,\d) {\includegraphics[scale=0.76]{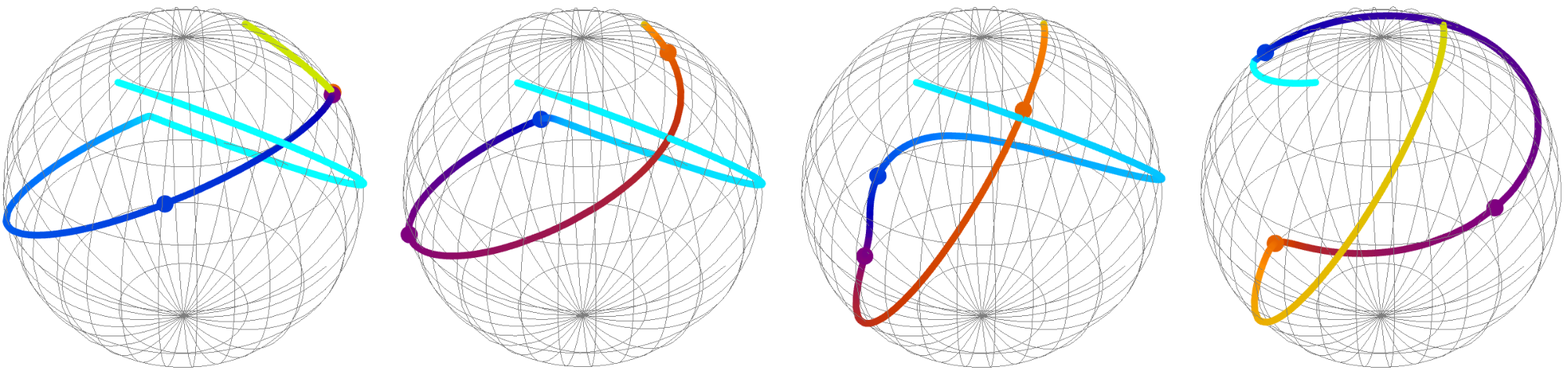}};
\node at (0,-\d) {\includegraphics[scale=0.76]{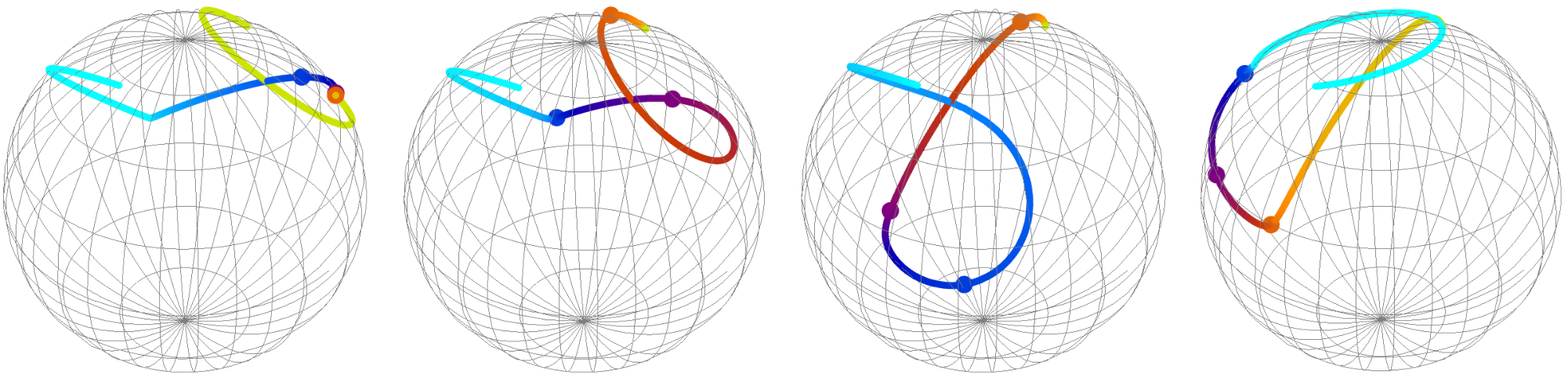}};
\node at (-0.05,-\d-\a) {\includegraphics[scale=1.3]{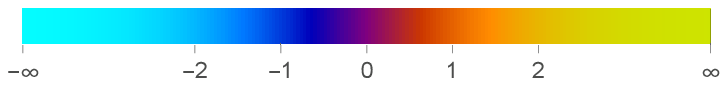}};
\node at (0,-\d-\a+0.7) {$x/\delta$};
\draw[gray] (-8,\d+\b) -- (8,\d+\b);
\draw[gray] (-8,-\d-\b+0.2) -- (8,-\d-\b+0.2);

\foreach \x in {0,4,8,12,16}
{
\draw[gray] (-8+\x,-\d-\b+0.2) -- (-8+\x,\d+\b);
};

\node at (-6,\d+\b-0.3) {\scriptsize $t/\delta=-5$};
\node at (-6+4,\d+\b-0.3) {\scriptsize $t/\delta=1/2$};
\node at (-6+8,\d+\b-0.3) {\scriptsize $t/\delta=5/2$};
\node at (-6+12,\d+\b-0.3) {\scriptsize $t/\delta=17/2$};

\node at (-7.7,-\d-\b+0.6) {
\tdplotsetmaincoords{50.638}{90-139}
\begin{tikzpicture}[tdplot_main_coords,font=\sffamily,scale=0.4]

\draw[-latex] (0,0,0) -- (1,0,0)  node[yshift=0pt, xshift=2.3pt]  {\tiny $x$};
\draw[-latex] (0,0,0) -- (0,1,0)  node[yshift=0pt, xshift=-1.8pt] {\tiny $y$};
\draw[-latex] (0,0,0) -- (0,0,1)  node[yshift=2.0pt, xshift=0pt]  {\tiny $z$};

\end{tikzpicture}
};

\end{tikzpicture}
\caption{Time evolution of the three-soliton solution with initial data \eqref{eq:initial_data_three_soliton} I: spatial dependence of $\bu(x,t)$ (upper spheres) and $\bv(x,t)$ (lower spheres) at four instances of time $t$, with colours indicating the position $x$ according to the legend on the bottom.
Note that the color gradient is non-linear, changing linearly only close to the origin $x=0$ but approaching a constant color away from the origin exponentially fast. For three distinguished points $x=-1$ (blue), $x=0$ (purple) and $x=1$ (red) on the $x$-axis, we have indicated the corresponding spins with a dot. The orientation of all plots is the same.
}
\label{fig:three_soliton}
\end{figure}

\subsection{Examples}
\label{sec:examples} 
We give examples of $N$-soliton initial data for $N=1,2,3$ constructed using the method in Section~\ref{sec:constraints}. To shorten our formulas, we find it convenient to give $\bX_j$ and $m$ in \eqref{eq:msolution} related to $\bs_j$ and $\bm_{0}$ in a simple way; see \eqref{eq:bsjbbb}. 
In all our examples, the reference direction is 
\begin{equation} 
\label{eq:bn0} 
\bn_0=(0,0,1). 
\end{equation} 
The results for $N=1$ and $N=2$ below are exact; the result for $N=3$ are numerical approximations to the exact result which are accurate up to the last  given digit. 

For $N=1$  and 
\begin{equation}
\label{eq:example12} 
a_1=\frac{3}{4}\ii\delta,\quad \bn_{1,3}=\frac{1}{\sqrt{2}}(0,-1,1),
\end{equation}
we obtain the following solution of \eqref{eq:constraintsbs}--\eqref{eq:constraintsbm}, 
\begin{equation}
\label{eq:initial_data_one_soliton2}
\bX_1 = \frac{1}{2}(0,1,1) -\ii\frac{1}{\sqrt{2}}(1,0,0); 
 \quad  m=\sqrt{\frac23} ;  
\end{equation}
this is exactly the example \eqref{eq:example1}.

For $N=2$, choosing 
\begin{equation}
\label{eq:example2}
\begin{split} 
a_1=\frac{3}{4}\ii\delta,\quad \bn_{1,3}=\frac{1}{\sqrt{2}}(0,-1,1), \\
 a_2=\frac{5}{4}\ii\delta, \quad  \bn_{2,3}=\frac{1}{\sqrt{2}}(1,0,-1), 
\end{split} 
\end{equation}
the solution of \eqref{eq:constraintsbs}--\eqref{eq:constraintsbm} is
\begin{equation}
\label{eq:initial_data_two_soliton}
\begin{split} 
\bX_1&= \frac17(8,10,10) + \ii\frac{\sqrt{2}}{7}(-10,4,4), \\
\bX_2&=-\frac17(10,8,10)-\ii \frac{\sqrt{2}}{7}(4,-10,4),\\
\end{split}
\quad m=\frac{1}{\sqrt{17}}.
\end{equation}

Finally, for $N=3$ and the choice 
\begin{equation}
\label{eq:example3} 
\begin{aligned}
a_1 &= \left(-3 + \frac{17}{20}\ii \right)\delta, \quad \bn_{1,3} = \frac{1}{\sqrt{2}}(0,-1,1) ,\\
a_2 &= \frac{4}{5} \ii \delta ,  \qquad\qquad\quad\;\; \bn_{2,3} = \frac{1}{\sqrt{2}}(1,0,-1) ,\\
a_3 &= \left(\frac{11}{10}+\frac{13}{10} \ii \right)\delta,  \quad \bn_{3,3} = \frac{1}{\sqrt{6}}(1,1,-2) ,
\end{aligned}
\end{equation}
the solution of   \eqref{eq:constraintsbs}--\eqref{eq:constraintsbm}   is 
\begin{equation}
\label{eq:initial_data_three_soliton}
\begin{split}
\bX_1 &= (-0.618461, 0.738925, 0.738925) -  \ii(1.04500, 0.437318, 0.437318),\\
\bX_2 &= (0.774004, 1.10235, 0.774004)+\ii(0.779477, - 1.09461, 0.779477),\\
\bX_3 &= -(0.195438, 1.03355, 0.614496)-\ii(1.09476, -0.410442,0.342159),
\end{split} 
\quad m=0.493378.
\end{equation}

\begin{figure}
\centering
\begin{tikzpicture}[scale=0.98]
\def\a{3};
\def\b{2.0};
\def\d{1.4};
\def\s{0.6};

\node at (-6,\d) {
\includegraphics[scale=0.45]{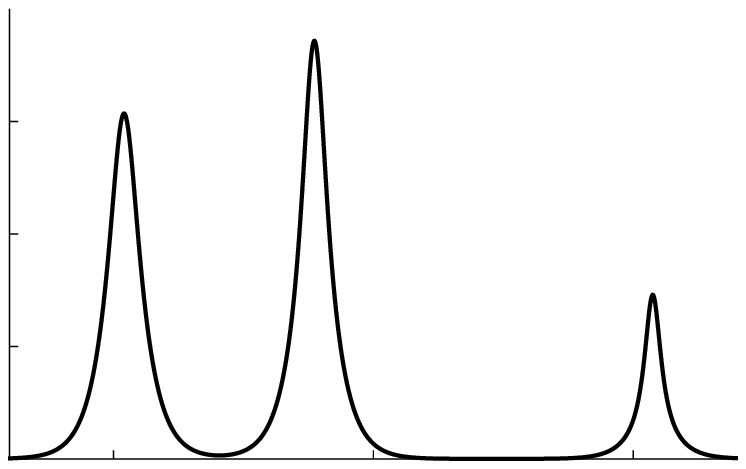}
};
\node at (-6,\d) {
\begin{tikzpicture}
\node at (1.9,-0.85) {\scalebox{0.6}{$x/\delta$}};
\node at (-1.4,1.5) {\small $\mathbf{\epsilon}$};
\node at (-1.5,-0.72) {\scalebox{\s}{$0$}};
\node at (-1.5,-0.2) {\scalebox{\s}{$2$}};
\node at (-1.5,0.3) {\scalebox{\s}{$4$}};
\node at (-1.5,0.82) {\scalebox{\s}{$6$}};

\node at (-0.92,-0.85) {\scalebox{\s}{$-5$}};
\node at (0.3,-0.85) {\scalebox{\s}{$0$}};
\node at (1.48,-0.85) {\scalebox{\s}{$5$}};
\end{tikzpicture}

};

\node at (-2,\d) {
\includegraphics[scale=0.45]{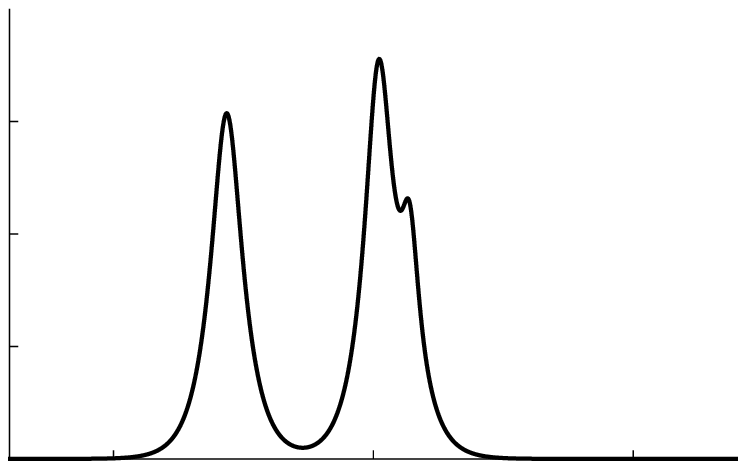}
};
\node at (-2,\d) {
\begin{tikzpicture}
\node at (1.9,-0.85) {\scalebox{0.6}{$x/\delta$}};
\node at (-1.4,1.5) {\small $\mathbf{\epsilon}$};
\node at (-1.5,-0.72) {\scalebox{\s}{$0$}};
\node at (-1.5,-0.2) {\scalebox{\s}{$2$}};
\node at (-1.5,0.3) {\scalebox{\s}{$4$}};
\node at (-1.5,0.82) {\scalebox{\s}{$6$}};

\node at (-0.92,-0.85) {\scalebox{\s}{$-5$}};
\node at (0.3,-0.85) {\scalebox{\s}{$0$}};
\node at (1.48,-0.85) {\scalebox{\s}{$5$}};
\end{tikzpicture}

};

\node at (2,\d) {
\includegraphics[scale=0.45]{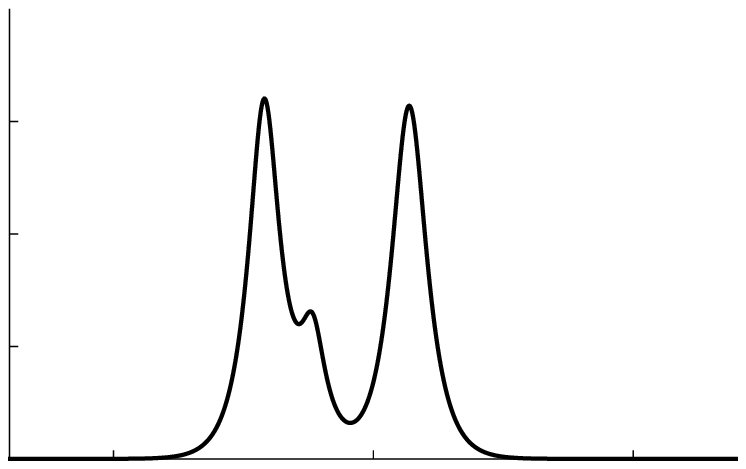}
};
\node at (2,\d) {
\begin{tikzpicture}
\node at (1.9,-0.85) {\scalebox{0.6}{$x/\delta$}};
\node at (-1.4,1.5) {\small $\mathbf{\epsilon}$};
\node at (-1.5,-0.72) {\scalebox{\s}{$0$}};
\node at (-1.5,-0.2) {\scalebox{\s}{$2$}};
\node at (-1.5,0.3) {\scalebox{\s}{$4$}};
\node at (-1.5,0.82) {\scalebox{\s}{$6$}};

\node at (-0.92,-0.85) {\scalebox{\s}{$-5$}};
\node at (0.3,-0.85) {\scalebox{\s}{$0$}};
\node at (1.48,-0.85) {\scalebox{\s}{$5$}};
\end{tikzpicture}

};

\node at (6,\d) {
\includegraphics[scale=0.45]{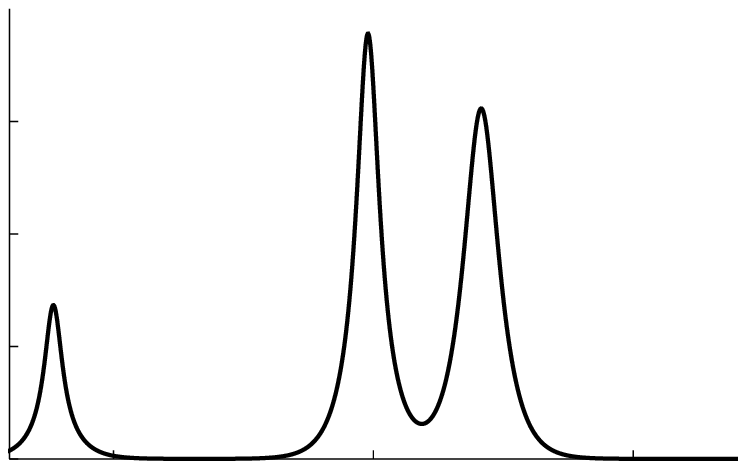}
};
\node at (6,\d) {
\begin{tikzpicture}
\node at (1.9,-0.85) {\scalebox{0.6}{$x/\delta$}};
\node at (-1.4,1.5) {\small $\mathbf{\epsilon}$};
\node at (-1.5,-0.72) {\scalebox{\s}{$0$}};
\node at (-1.5,-0.2) {\scalebox{\s}{$2$}};
\node at (-1.5,0.3) {\scalebox{\s}{$4$}};
\node at (-1.5,0.82) {\scalebox{\s}{$6$}};

\node at (-0.92,-0.85) {\scalebox{\s}{$-5$}};
\node at (0.3,-0.85) {\scalebox{\s}{$0$}};
\node at (1.48,-0.85) {\scalebox{\s}{$5$}};
\end{tikzpicture}

};

\node at (-6,-\d) {
\includegraphics[scale=0.42]{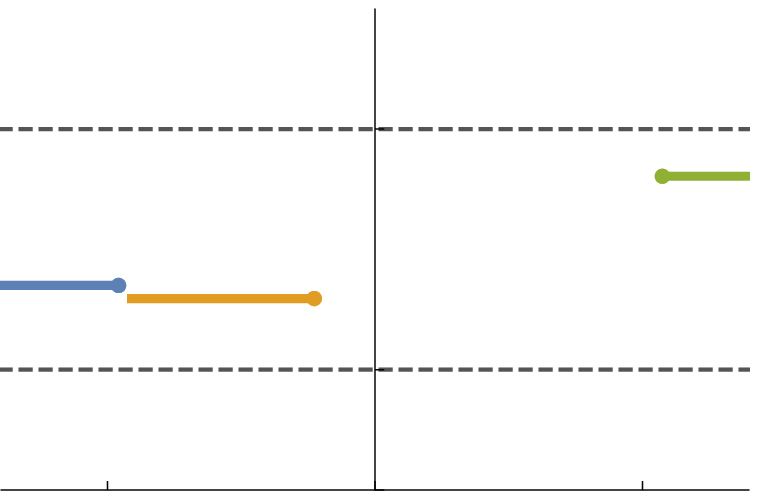}
};
\node at (-6,-\d) {
\begin{tikzpicture}
\node at (-1,-1) {};
\node at (1,1.8) {};
\node at (2.6,-0.8) {\scalebox{0.6}{$\aR/\delta$}};
\node at (1.2,1.4) {\scalebox{0.6}{$\aI/\delta$}};

\node at (-0.28,-0.8) {\scalebox{\s}{$-5$}};
\node at (0.95,-0.8) {\scalebox{\s}{$0$}};
\node at (2.1,-0.8) {\scalebox{\s}{$5$}};

\node at (0.75,-0.23) {\scalebox{\s}{$0.5$}};
\node at (0.75,0.8) {\scalebox{\s}{$1.5$}};
\draw (0.9,0.915) -- (1,0.915);
\draw (0.9,-0.115) -- (1,0.-0.115);

\end{tikzpicture}

};

\node at (-2,-\d) {
\includegraphics[scale=0.42]{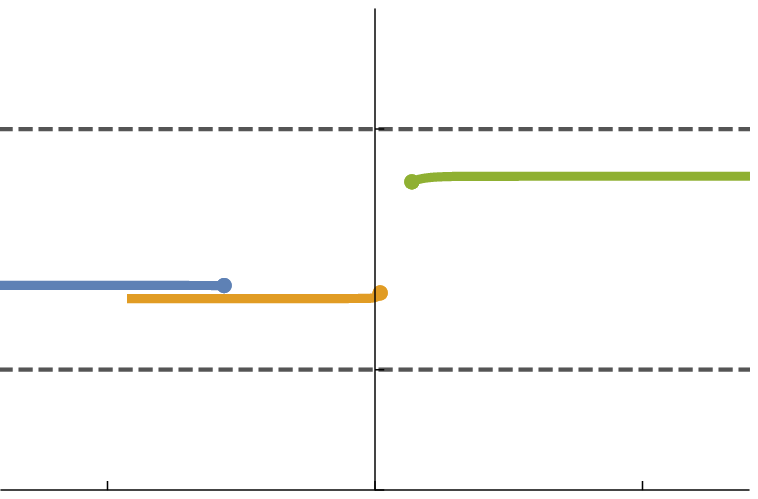}
};
\node at (-2,-\d) {
\begin{tikzpicture}
\node at (-1,-1) {};
\node at (1,1.8) {};
\node at (1,1.8) {};
\node at (2.6,-0.8) {\scalebox{0.6}{$\aR/\delta$}};
\node at (1.2,1.4) {\scalebox{0.6}{$\aI/\delta$}};

\node at (-0.28,-0.8) {\scalebox{\s}{$-5$}};
\node at (0.95,-0.8) {\scalebox{\s}{$0$}};
\node at (2.1,-0.8) {\scalebox{\s}{$5$}};

\node at (0.75,-0.23) {\scalebox{\s}{$0.5$}};
\node at (0.75,0.8) {\scalebox{\s}{$1.5$}};
\draw (0.9,0.915) -- (1,0.915);
\draw (0.9,-0.115) -- (1,0.-0.115);

\end{tikzpicture}

};

\node at (2,-\d) {
\includegraphics[scale=0.42]{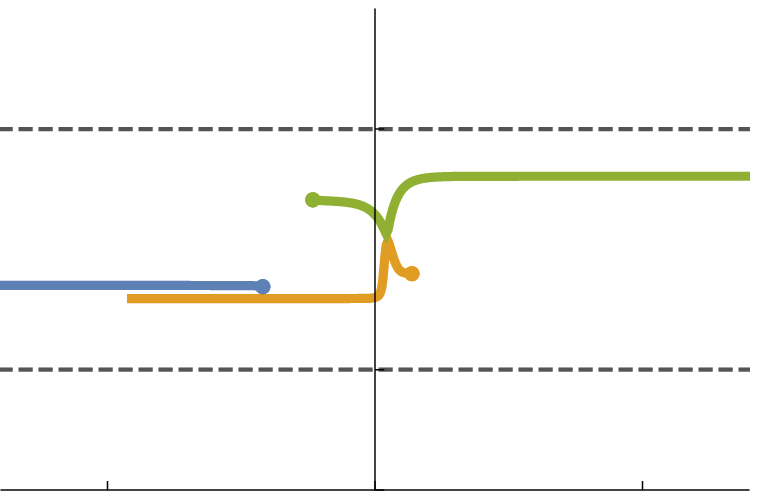}
};
\node at (2,-\d) {
\begin{tikzpicture}
\node at (-1,-1) {};
\node at (1,1.8) {};
\node at (1,1.8) {};
\node at (2.6,-0.8) {\scalebox{0.6}{$\aR/\delta$}};
\node at (1.2,1.4) {\scalebox{0.6}{$\aI/\delta$}};

\node at (-0.28,-0.8) {\scalebox{\s}{$-5$}};
\node at (0.95,-0.8) {\scalebox{\s}{$0$}};
\node at (2.1,-0.8) {\scalebox{\s}{$5$}};

\node at (0.75,-0.23) {\scalebox{\s}{$0.5$}};
\node at (0.75,0.8) {\scalebox{\s}{$1.5$}};
\draw (0.9,0.915) -- (1,0.915);
\draw (0.9,-0.115) -- (1,0.-0.115);

\end{tikzpicture}

};

\node at (6,-\d) {
\includegraphics[scale=0.42]{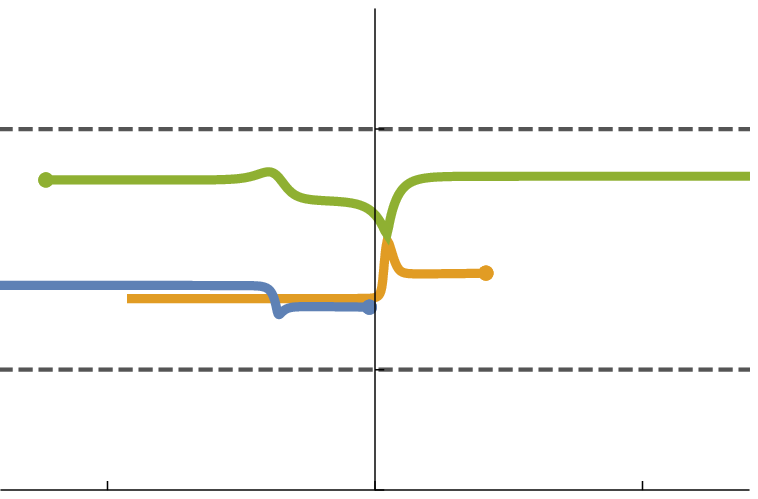}
};
\node at (6,-\d) {
\begin{tikzpicture}
\node at (-1,-1) {};
\node at (1,1.8) {};
\node at (1,1.8) {};
\node at (2.6,-0.8) {\scalebox{0.6}{$\aR/\delta$}};
\node at (1.2,1.4) {\scalebox{0.6}{$\aI/\delta$}};	

\node at (-0.28,-0.8) {\scalebox{\s}{$-5$}};
\node at (0.95,-0.8) {\scalebox{\s}{$0$}};
\node at (2.1,-0.8) {\scalebox{\s}{$5$}};

\node at (0.75,-0.23) {\scalebox{\s}{$0.5$}};
\node at (0.75,0.8) {\scalebox{\s}{$1.5$}};
\draw (0.9,0.915) -- (1,0.915);
\draw (0.9,-0.115) -- (1,0.-0.115);

\end{tikzpicture}

};

\draw[gray] (-8,\d+\b) -- (8,\d+\b);
\draw[gray] (-8,-\d-\b+0.2) -- (8,-\d-\b+0.2);

\foreach \x in {0,4,8,12,16}
{
\draw[gray] (-8+\x,-\d-\b+0.2) -- (-8+\x,\d+\b);
};

\node at (-6,\d+\b-0.3) {\scriptsize $t/\delta=-5$};
\node at (-6+4,\d+\b-0.3) {\scriptsize $t/\delta=1/2$};
\node at (-6+8,\d+\b-0.3) {\scriptsize $t/\delta=5/2$};
\node at (-6+12,\d+\b-0.3) {\scriptsize $t/\delta=17/2$};

\end{tikzpicture}
\caption{Time evolution of the three-soliton solution with initial data \eqref{eq:initial_data_three_soliton} II: energy density and location of poles at four instances of time $t$. 
At each time $t$, the energy density $\eps(x,t)$ (first row) and poles $a_j(t)=\aR_j(r)+\ii\aI_j(t)$ (blue, orange and green dots for $j=1,2,3$  at $t=-\infty$, respectively, in the second row) are shown. 
The plots visualize two soliton collisions: the first between $j=2$ and $j=3$ at time $t/\delta\approx 1.03$ and the second between $j=1$ and $j=3$ at time $t/\delta\approx 3.30$. 
As discussed in the last paragraph of Section~\ref{sec:Nsolitons}, the assignment of poles to solitons can be swapped in soliton collisions, and the colors blue, orange and green distinguish different solitons, i.e., the same color means the same soliton.
}
\label{fig:three_soliton_poles}
\end{figure}

As an example, we visualize the three-soliton solution with parameters \eqref{eq:example3} at four different instances of time in Figs.~\ref{fig:three_soliton} and \ref{fig:three_soliton_poles}: Fig.~\ref{fig:three_soliton} shows the spatial dependence of $\bu$ and $\bv$, and Fig.~\ref{fig:three_soliton_poles} the corresponding energy densities and pole positions. Below, we give physical interpretations and explanations of these figures.

\subsection{Physical properties of multi-solitons}
\label{sec:Nsolitons}
We give analytic arguments that allow to understand multi-soliton solutions in regions where the solitons are well-separated. 
We also describe observations, based on numerical computations, on which properties of solitons are preserved under interactions and which are not. 

\paragraph{Behavior in asymptotic regions.}  
We discuss $N$-soliton solutions, $N>1$, in asymptotic regions where the solitons are well-separated, i.e., 
\begin{equation} 
\label{eq:asregion} 
-\infty\ll \aR_{1}(t) \ll \aR_{2}(t) \ll \cdots \ll \aR_{N}(t) \ll \infty ; 
\end{equation} 
similarly as in the one-soliton case, the real part of the $j$-th pole, $\aR_j(t)$, can be identified with the center of soliton $j=1,\ldots,N$ at time $t$ (here an in the following, we use the notation in \eqref{eq:aRaI}).   
We give analytic arguments making precise in which sense a $N$-soliton solution can be approximated by a superposition of one-soliton solution in such a region. 
We also show that there are non-trivial long-range effects which affect the local vacua in the regions between two adjacent solitons; this is qualitatively different from corresponding results for the HWM equation \cite{berntsonklabbers2020}. 

Below, we interpret $A\ll B$    in    \eqref{eq:asregion}  as   $B-A>2d$    with    $d=\delta/2$ or $\delta$ (depending on the desired accuracy), and we write $A\approx B$ short for $A-B=O(\mathrm{e}^{-4d/\delta})$.
Thus, in practice, all approximations $\approx $ below are very accurate. 
Note that, for simplicity, we label solitons by the order of the real parts of the corresponding poles, without loss of generality. 

If \eqref{eq:asregion} holds, the interaction potentials in the CM equations of motion \eqref{eq:CM} are exponentially small, and therefore: 
\begin{equation} 
\label{eq:CMapprox} 
\bs_j(t)\approx \bs_j,\quad  a_j(t)\approx a_j+v_jt\quad  (j=1,\ldots,N)
\end{equation} 
for some time-independent $\bs_j$, $a_j$ and $v_j$ which characterize the $N$-soliton solution in the asymptotic region considered (and which, in general, can be computed by solving the  CM equations of motion \eqref{eq:CM} numerically). 
Moreover, using \eqref{eq:adot1real} and the non-trivial asymptotic behavior of the functions $\alpha(z)$ and $\tilde\alpha(z)$, 
\begin{equation} 
\label{eq:approx} 
\alpha(x+\ii y) \approx \begin{cases} +\kappa & (x\gg 0) \\ -\kappa & (x\ll 0) \end{cases} ,\quad \talpha(x+\ii y) \approx \begin{cases} +\kappa & (x\gg 0) \\ -\kappa & (x\ll 0) \end{cases} \quad (x,y\in\R), 
\end{equation} 
we obtain, in the asymptotic region \eqref{eq:asregion}, $\dot a_j(t)\approx v_j$ with 
\begin{equation} 
\label{eq:vj} 
v_j = -\ii \frac{\bs_j^*\wedge\bs_j}{\bs_j^*\cdot\bs_j}\cdot \bm_j = 2\frac{\bsR_j\wedge\bsI_j}{\bs_j^*\cdot \bs_j}\cdot\bm_j 
\end{equation}  
where 
\begin{equation} 
\label{eq:bmj} 
\bm_j = \bm_0 -\sum_{k=1}^{j-1}2\kappa\bsI_k + \sum_{k=j+1}^N 2\kappa \bsI_k\quad (j=1,\ldots,N), 
\end{equation} 
consistent with \eqref{eq:CMapprox}.

We write the $N$-soliton solution \eqref{eq:solution} as 
\begin{equation}
\label{eq:onesolitonN}
\left. \begin{array}{c} \bu(x,t)\\ \bv(x,t)\end{array} \right\} = \bm_0 -\sum_{j=1}^N 2 \, \im \left( \alpha(x-a_j(t)\pm \ii\delta/2 )\bs_j(t) \right) . 
\end{equation} 
For fixed $j=1,\ldots, N$, in the region 
\begin{equation} 
\label{eq:regionj} 
 \aR_j(t)-d <x <\aR_j(t)+d\quad (j=1,\ldots,N)  
\end{equation} 
of length $2d$ around the center of soliton $j$, our assumptions \eqref{eq:CMapprox} allow us to use  \eqref{eq:approx} to simplify \eqref{eq:onesolitonN}   as follows,   
\begin{equation}
\label{eq:onesolitonN1}
\left. \begin{array}{c} \bu(x,t)\\ \bv(x,t)\end{array} \right\} \approx  \bm_j - 2\, \im\left( \alpha(x-\aR_j-v_j t-\ii(\aI_j\mp\delta/2))\bs_j\right) 
\end{equation} 
where, again, $\bm_j$ \eqref{eq:bmj}  appears.  This makes manifest that {\em a $N$-soliton solution in a region \eqref{eq:regionj} can be well approximated by a one-soliton solution \eqref{eq:onesoliton1} with $\bm_0$, $\aR$, $\aI$, $\bsR$, $\bsI$ replaced by $\bm_j$, $\aR_j$, $\aI_j$, $\bsR_j$, $\bsI_j$, respectively}. 

\begin{figure}
\begin{center}
\begin{tikzpicture}[scale=0.98]
\def\a{2.8};
\def\b{1.35};
\def\c{2.5};
\draw (-8,\a) -- (8,\a);
\draw (-8,-\a) -- (8,-\a);
\draw (-8,\a) -- (-8,-\a);
\draw (8,\a) -- (8,-\a);
\draw[ultra thin] (-2.83,\a) -- (-2.83,-\a);
\draw[ultra thin] (2.83,\a) -- (2.83,-\a);

\node at (-5.33,0) {
\includegraphics[scale=0.6]{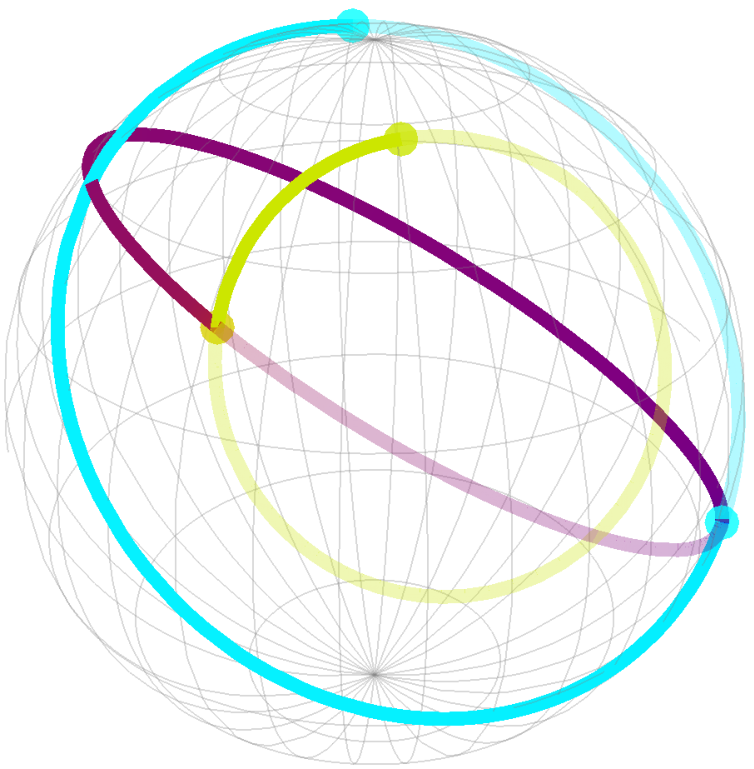}
};
\node at (0,0) {
\includegraphics[scale=0.6]{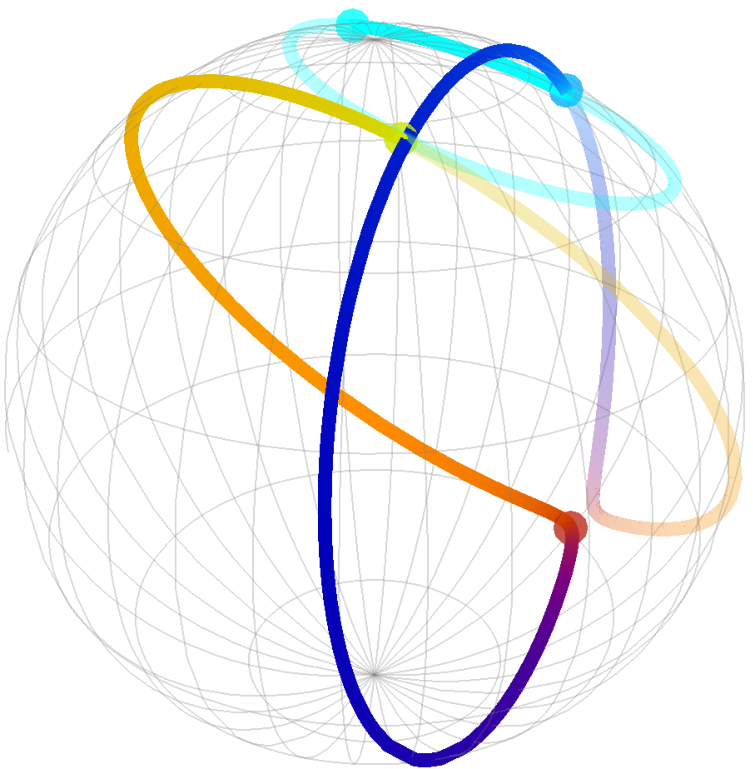}
};
\node at (5.33,0) {
\includegraphics[scale=0.6]{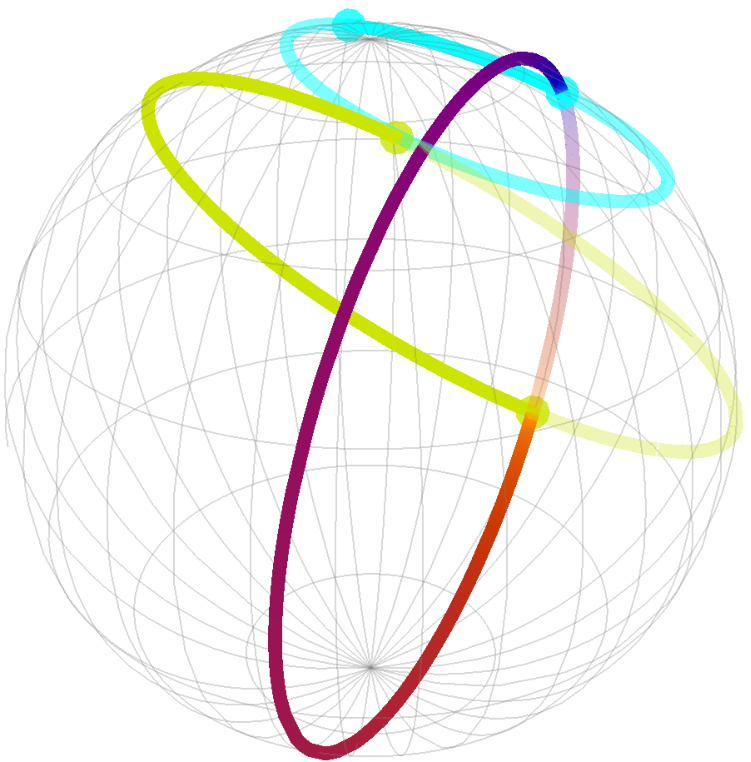}
};

\node at (-7.7,-\c+0.2) {
\tdplotsetmaincoords{57.29}{90-18}
\begin{tikzpicture}[tdplot_main_coords,font=\sffamily,scale=0.4]

\draw[-latex] (0,0,0) -- (1,0,0)  node[yshift=-1.8pt, xshift=1.8pt]  {\tiny $x$};
\draw[-latex] (0,0,0) -- (0,1,0)  node[yshift=-1.8pt, xshift=1.8pt] {\tiny $y$};
\draw[-latex] (0,0,0) -- (0,0,1)  node[yshift=2.0pt, xshift=0pt]  {\tiny $z$};

\end{tikzpicture}
};

\node at (0,-3.8){
\includegraphics[scale=1.2]{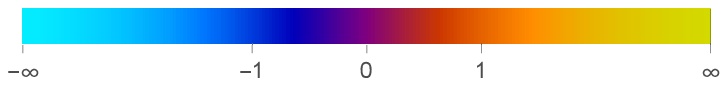}
};	
\node at (0,-3.8+0.65) {$x/\delta$};

\node at (-7.0,\c) {\scriptsize $t/\delta=-100$};
\node at (-7.45+5.33,\c) {\scriptsize $t/\delta=7$};
\node at (-7.0+10.67,\c) {\scriptsize $t/\delta=100$};

\end{tikzpicture}
\end{center}
\vspace{-20pt}
\caption{The spin configurations $\bu(x,t)$ (strong lines) and $\bv(x,t)$ (weak lines) for the three-soliton solutions with initial data \eqref{eq:example3}, at fixed times $t/\delta=\pm 100$ (far past and future) and $t/\delta=7$ when all solitons are well-separated. The individual rotation circles of the solitons are clearly visible. The colored dots indicate the local vacua $\bm_1^-=\bm_{-\infty}$, $\bm_j^+\approx \bm_{j+1}^-$ for $j=1,2$, and $\bm_3^{+}=\bm_{+\infty}$. The orientation of all three plots is the same, as indicated by
the coordinate system in the bottom left corner of the first plot.  The color of the image on $S^2$ indicates the value of $x$ according to the legend underneath. Note that, at $t/\delta=7$, solitons $2$  and $3$ are not as well-separated as the others and, therefore, the two circles at $\bm_2^+\approx \bm_3^-$ (indicated by the red dot) do not touch.}  
\label{fig:past_and_future}
\end{figure}

Thus, from our results on one-soliton solutions in Section~\ref{sec:onesoliton}, we immediately obtain the following description of the $N$-soliton solution in an asymptotic region: {\em Each soliton $j=1,\ldots,N$ is characterized by a rotation circle obtained by intersecting the sphere $S^2$ with the plane containing $\bm_j$ \eqref{eq:bmj} and spanned by the real three-vectors $\bsR_j$ and $\bsI_j$; circle $j$ contains the points
\begin{equation} 
\label{eq:bmjpm} 
\bm_j^\pm\coloneqq \bm_j\mp 2\kappa\bsI_j 
\end{equation} 
which correspond to the local vacua to the left ($\bm_j^-$) and right ($\bm_j^+$) of soliton $j$, respectively; since  
\begin{equation} 
\label{eq:bmjpm1} 
\bm^+_j=\bm^-_{j+1}\quad (j=1,\ldots, N-1), 
\end{equation} 
circles $j$ and $j+1$ intersect at the points in \eqref{eq:bmjpm1}, and these intersection points correspond to the local vacuum $\bu\approx \bv\approx \bm^+_j$ in the region 
\begin{equation} 
\aR_j(t)+d< x < \aR_{j+1}(t)-d \quad (j=1,\ldots,N-1)  
\end{equation} 
between soliton $j$ and soliton $j+1$; the local vacua $\bu\approx\bv\approx \bm_{\pm \infty}$ at $x\to \pm \infty$ are given by 
\begin{equation} 
\bm_{\pm \infty} = \bm_0\mp \sum_{j=1}^N 2\kappa\bsI_j
\end{equation} 
(note that $\bm_{-\infty}=\bm_1^-$ and $\bm_{+\infty}=\bm_N^+$); soliton $j$ starts at the local vacuum $\bu\approx \bv\approx \bm_j^-$ at $x \approx \aR_j(t)-d$ and, as one increases $x$, $\bu$ and $\bv$ rotate on circle $j$ in opposite directions towards the local vacuum  $\bu\approx \bv\approx \bm_j^+$ at $x \approx \aR_j(t)+d$; the centers and radii of these rotation circles are given by
\begin{equation} 
\label{eq:bnj} 
\bn_j(t)=\frac12\left( \bu(\aR_j(t),t)+ \bv(\aR_j(t),t)\right) \approx \bm_j -2\kappa\bsR_j\tan(2\kappa \aI_j) 
\end{equation} 
and
\begin{equation} 
\label{eq:Rj}
R_j(t) = \frac12\left| \bu(\aR_j(t),t)-\bv(\aR_j(t),t)\right|\approx   -\frac{2\kappa|\bsR_j|}{\cos(2\kappa\aI_j)}, 
\end{equation} 
respectively (see \eqref{eq:bn} and \eqref{eq:R}); soliton $j$ moves with velocity 
\begin{equation} 
\label{eq:dotaj}
\dot a^{\mathrm{R}}_j(t)\approx 2\frac{\bsR_j\wedge\bsI_j}{\bs_j^*\cdot \bs_j}\cdot\bm_j 
\end{equation} 
(see \eqref{eq:vj})  and has energy 
\begin{equation} 
\label{eq:Ej}
E_j(t)\coloneqq \int_{\frac12 [\aR_j(t)+\aR_{j-1}(t)]}^{\frac12[\aR_{j+1}(t)+\aR_{j}(t)]}\left(\eps_{\bu}(x,t) + \eps_{\bv}(x,t) \right)\, \dd{x}  \approx 
\frac{2\pi\kappa^2\bs_j\cdot\bs_j^* }{\cos^2(2\kappa\aI_j)}
\end{equation} 
(see \eqref{eq:E1}); soliton $j$ has chirality $+$ and $-$ for $\delta/2<\aI_j(t)<\delta$ and $\delta<\aI_j(t)<3\delta/2$, respectively (see \eqref{eq:EuEv}and \eqref{eq:chuchv}).} 

Note that the energies \eqref{eq:Ej} are defined such that $\sum_{j=1}^N E_j(t)=E$ equals the total energy independent of time.
Moreover, in an asymptotic region \eqref{eq:asregion}, the total energy density is well approximated by the following linear superposition of one-soliton energy densities, 
\begin{equation} 
\eps(x,t)\approx \sum_{j=1}^N E_{j,1} f(x-\aR_j-v_jt;\aI_j)
\end{equation} 
with the one-soliton energies \eqref{eq:Ej} and the distribution functions $f(x;\aI)$ given in \eqref{eq:eps1}. 

As an example, we show in Fig.~\ref{fig:past_and_future} the rotation circles of the three-soliton solution with parameters \eqref{eq:example3}  for three distinct asymptotic regions. Clearly, the orientations of these circles is different in different asymptotic regions.
\begin{figure}[t]
\begin{tikzpicture}
\def\a{6};
\def\b{0.7};
\def\s{0.75};
\draw (-8,\a) -- (8,\a);
\draw (-8,-\a) -- (8,-\a);
\draw (-8,\a) -- (-8,-\a);
\draw (8,\a) -- (8,-\a);
\draw (0,\a) -- (0,-\a);
\draw (-8,0) -- (8,0);

\node at (-4,\a/2) {
\includegraphics[scale=\b]{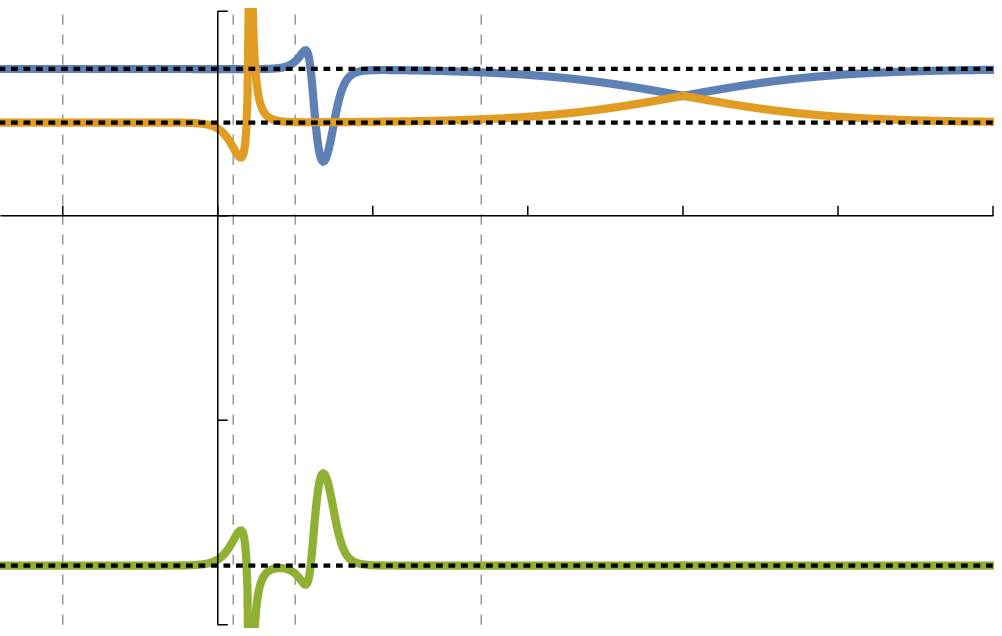}
};
\node at (-4,\a/2) {
\begin{tikzpicture}
\node at (-3.5,-2.2) {};
\node at (3.5,2.2) {};

\node at (-2.25,-2.15) {\scalebox{\s}{$-1$}};
\node at (-2.35,-0.75) {\scalebox{\s}{$-0.5$}};
\node at (-2.25,2.15) {\scalebox{\s}{$0.5$}};

\node at (-3.15,0.55) {\scalebox{\s}{$-5$}};
\node at (-2.1,0.55) {\scalebox{\s}{$0$}};
\node at (-0.9,0.55) {\scalebox{\s}{$5$}};
\node at (0.2,0.55) {\scalebox{\s}{$10$}};
\node at (1.3,0.55) {\scalebox{\s}{$15$}};
\node at (2.4,0.55) {\scalebox{\s}{$20$}};

\node at (3.3,0.52) {\scalebox{\s}{$t/\delta$}};
\end{tikzpicture}
};
\node at (4,\a/2) {
\includegraphics[scale=\b]{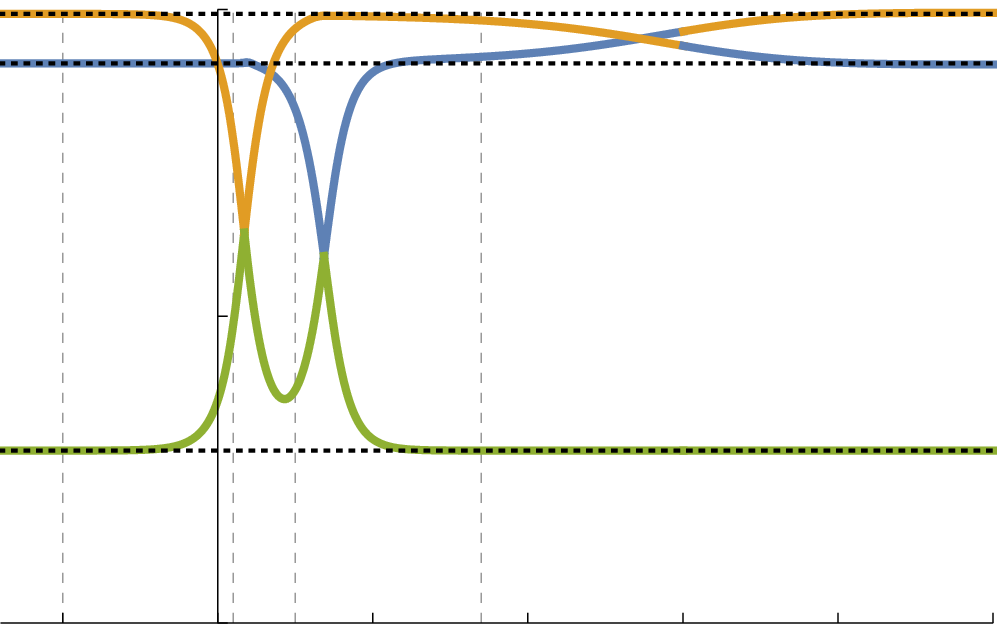}
};
\node at (4,\a/2) {
\begin{tikzpicture}
\node at (-3.5,-2.5) {};
\node at (3.5,2.5) {};

\node at (-2.1,-0) {\scalebox{\s}{$3$}};
\fill[white] (-2.15,2.1) rectangle ++(0.1,0.1); 
\node at (-2.1,2.15) {\scalebox{\s}{$6$}};

\node at (-3.15,-2.35) {\scalebox{\s}{$-5$}};
\node at (-2.,-2.35) {\scalebox{\s}{$0$}};
\node at (-0.9,-2.35) {\scalebox{\s}{$5$}};
\node at (0.2,-2.35) {\scalebox{\s}{$10$}};
\node at (1.3,-2.35) {\scalebox{\s}{$15$}};
\node at (2.4,-2.35) {\scalebox{\s}{$20$}};

\node at (3.3,-2.38) {\scalebox{\s}{$t/\delta$}};
\end{tikzpicture}
};

\node	 at (-4,-\a/2) {
\includegraphics[scale=\b]{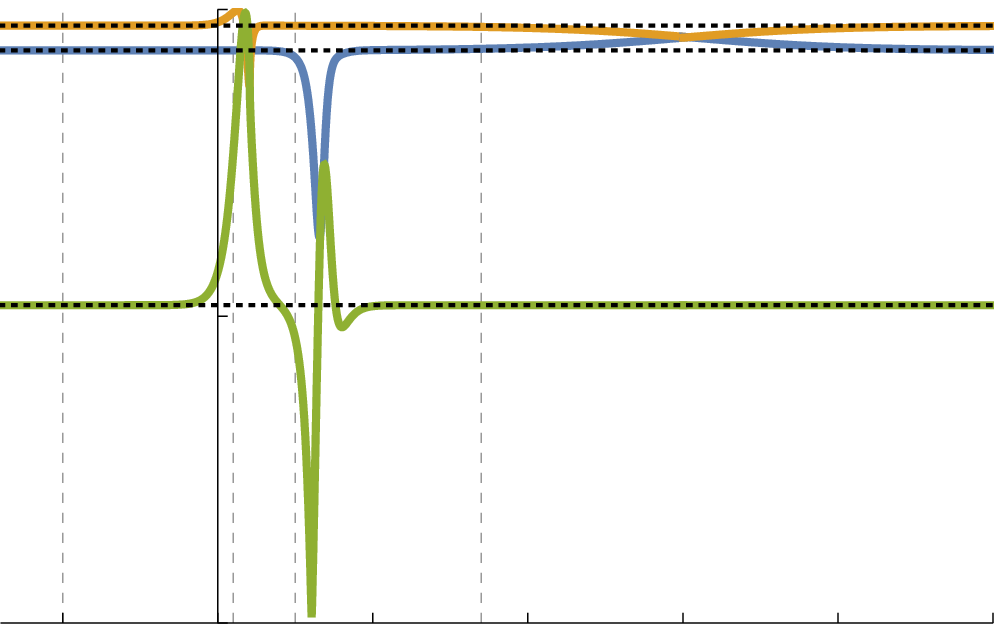}
};
\node at (-4,-\a/2) {
\begin{tikzpicture}
\node at (-3.5,-2.5) {};
\node at (3.5,2.5) {};

\node at (-2.2,-0.07) {\scalebox{\s}{$0.5$}};
\fill[white] (-2.15,2.1) rectangle ++(0.1,0.1); 
\node at (-2.1,2.2) {\scalebox{\s}{$1$}};

\node at (-3.15,-2.35) {\scalebox{\s}{$-5$}};
\node at (-2.,-2.35) {\scalebox{\s}{$0$}};
\node at (-0.9,-2.35) {\scalebox{\s}{$5$}};
\node at (0.2,-2.35) {\scalebox{\s}{$10$}};
\node at (1.3,-2.35) {\scalebox{\s}{$15$}};
\node at (2.4,-2.35) {\scalebox{\s}{$20$}};

\node at (3.3,-2.38) {\scalebox{\s}{$t/\delta$}};
\end{tikzpicture}
};

\node	 at (4,-\a/2) {
\includegraphics[scale=\b]{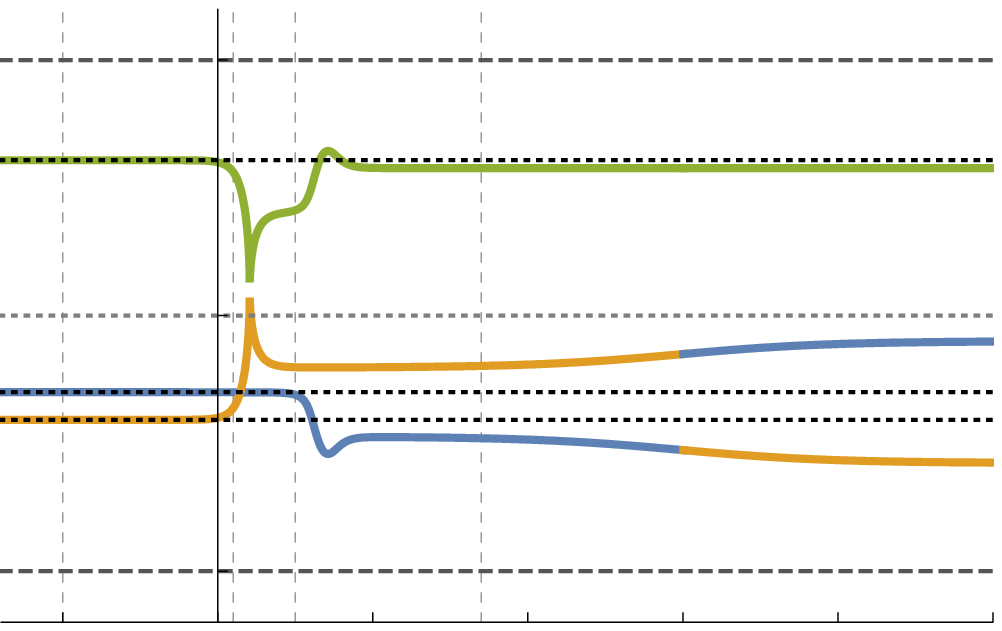}
};
\node at (4,-\a/2) {
\begin{tikzpicture}
\node at (-3.5,-2.5) {};
\node at (3.5,2.5) {};

\fill[white] (-2.39,-1.88) rectangle ++(0.34,0.1);
\node at (-2.2,-1.8) {\scalebox{\s}{$0.5$}};
\fill[white] (-2.15,-0.08) rectangle ++(0.1,0.1); 
\node at (-2.1,0) {\scalebox{\s}{$1$}};
\fill[white] (-2.35,1.8) rectangle ++(0.3,0.1); 
\node at (-2.2,1.83) {\scalebox{\s}{$1.5$}};

\node at (-3.15,-2.35) {\scalebox{\s}{$-5$}};
\node at (-2.,-2.35) {\scalebox{\s}{$0$}};
\node at (-0.9,-2.35) {\scalebox{\s}{$5$}};
\node at (0.2,-2.35) {\scalebox{\s}{$10$}};
\node at (1.3,-2.35) {\scalebox{\s}{$15$}};
\node at (2.4,-2.35) {\scalebox{\s}{$20$}};

\node at (3.3,-2.38) {\scalebox{\s}{$t/\delta$}};
\end{tikzpicture}
};

\node at (-4,\a-0.4) {\small (a): $\dot{a}^{\mathrm{R}}_j$};
\node at (4,\a-0.4) {\small (b): $E_j$};
\node at (-4,-0.4) {\small (c): $R_j$};
\node at (4,-0.4) {\small (d): $\aI_j/\delta$};
\end{tikzpicture}
\caption{Time evolution of four important single-soliton quantities for the three-soliton example with the initial data \eqref{eq:example3}: (a) velocities $\dot{a}^R_j(t)$ \eqref{eq:dotaj}, (b) energies $E_j(t)$ \eqref{eq:Ej}, (c) radii of rotation circles $R_j(t)$ \eqref{eq:Rj}, (d) imaginary parts of the poles $\aI_j(t)$ ($j=1,2,3$). The vertical dashed lines indicate the times $t/\delta=-5$, $1/2$, $5/2$ and $17/2$ when complementary information about this solution can be found in the Figs.~\ref{fig:three_soliton} and \ref{fig:three_soliton_poles}. The horizontal black dotted lines indicate the constant values in the asymptotic region $t\to-\infty$. 
In (d), we also indicate the boundaries and the middle of $R^+_\delta$ \eqref{eq:strip} (black dashed horizontal lines and gray dotted horizontal line, respectively). 
The colors blue, orange and green correspond to the three different solitons, as in Fig.~\ref{fig:three_soliton_poles}. 
}
\label{fig:properties}
\end{figure}
\paragraph{Soliton interactions.} 
At times when two solitons are close to each other, interactions are important, and the space-time evolution of $\bu$ and $\bv$ is complicated.  
In principle, by using an analytic solution of the hyperbolic spin CM system, one could study this by analytic means. 
However, this is a challenging enterprise beyond the scope of the present paper, and we therefore resorted to numerical methods to solve the spin-pole dynamics of this CM model. As a representative example, we choose the three-soliton solution with the parameters \eqref{eq:example3}; see Fig.~\ref{fig:three_soliton},\ref{fig:three_soliton_poles} and \ref{fig:past_and_future}. Clearly, in these figures, one can see that there are two soliton collisions at times $t/\delta\approx 1.03$ and $3.30$. 

As discussed in the previous paragraph, in a fixed asymptotic region \eqref{eq:asregion}, soliton $j=1,\ldots,N$ in a $N$-soliton solution has the following characteristic properties: 
(i) velocity $v_j$ \eqref{eq:vj}, (ii) local vacua before and after the soliton $\bm_j^\pm$ \eqref{eq:bmj}, (iii) spatial distribution of the energy density $f(x,\aI_j)$ \eqref{eq:eps1}, (iv) 
rotation circle centre $\bn_j$ \eqref{eq:bnj}, (v) rotation circle radius $R_j$ \eqref{eq:Rj}, (vi) energy $E_{j}$ \eqref{eq:Ej}, (vii) chirality, which is $+$ for $\delta/2<\aI_j<\delta$ and $-$ for $\delta<\aI_j<3\delta/2$, respectively (and undefined for $\aI_j=\delta)$. 

It is interesting to study what happens when two solitons collide. When this happens, the order of solitons changes, and it becomes more natural to assign labels $j=1,\ldots,N$ to solitons so as to allow for permutations (rather than using the labeling in \eqref{eq:asregion}): if the order of solitons is $j=1,\ldots,N$ before a time interval with soliton interactions, it will be $j=\sigma(1),\ldots,\sigma(N)$ after this time interval for some permutation $\sigma \in S_N$. It is important to know which of the soliton properties (i)--(vii) above are the same before and after collisions: only such properties can be regarded as characteristic attributes of a soliton which are the same in every asymptotic region. 

From Fig.~\ref{fig:past_and_future} it is clear that the rotation circle centres $\bn_j$ and the local vacua $\bm_j^\pm$ are different in different asymptotic regions: these are, certainly, not characteristic attributes of solitons. The same is true for the spatial distribution of the energy density $f(x,\aI_j)$: it is clear from Fig.~\ref{fig:three_soliton_poles} that the imaginary parts of the poles associated with a soliton can change; see also Fig.~\ref{fig:properties} (d).   However, the results in Fig~\ref{fig:properties} (a), (b) and (c) suggest that the velocity $v_j$, the energy $E_j$, the rotation radius $R_j$, and the chirality of a soliton are the same in different asymptotic regions: we believe these are characteristic attributes of solitons. 
We also believe that the chiralities of solitons are unchanged by soliton collisions.

It is interesting to note that the two different types of soliton collisions distinguished by the qualitative behaviour of the corresponding poles: (i) the solitons swap their poles in the collision, i.e., pole $a_j$ and $a_{j+1}$ correspond to soliton $j$ and $j+1$ before the collision but to soliton $j+1$ and $j$ after the collision, (ii)  the solitons keep their poles. For example, Fig.~\ref{fig:properties}(d) shows three soliton collisions where the first and second at times $t/\delta\approx 1.03$ and $t/\delta\approx 3.30$ clearly are of type (ii), while the third at time $t/\delta\approx 15.01$ is of type (i).
We speculate that collisions of solitons of the same chirality are type (i), whereas collisions of solitons of opposite chirality are type (ii) (note that this is consistent with the results in Fig.~\ref{fig:properties}(d): the solitons corresponding to orange, blue and green have chirality $+$, $+$ and $-$, respectively).

\section{Discussion and future directions}
\label{sec:discussion}
In this paper we have introduced a novel integrable equation which we named the non-chiral intermediate Heisenberg ferromagnet (ncIHF) equation. It generalizes the half-wave maps (HWM) equation \cite{zhou2015,lenzmann2018} in the same way as the non-chiral intermediate long-wave equation generalizes the Benjamin-Ono (BO) equation \cite{berntson2020}. 
We obtained the ncIHF equation as a continuum limit of an elliptic generalization of the Haldane-Shastry spin chain, and we demonstrated its integrability by constructing a Lax pair and a family of multi-soliton solutions. 
These results show that the ncIHF equation is related to the hyperbolic spin Calogero-Moser (CM) model in the same way as the BO equation is related to the rational CM model \cite{Polychronakos1995,Chen1979}. 
Our Lax pair and multi-soliton results generalize known results about the HWM equation in \cite{gerard2018} and \cite{berntsonklabbers2020}, respectively. 
In the former case, we prove a remarkable property of a matrix integral operator $\cT$ generalizing the Hilbert transform; in the latter case, we present a new method to solve constraints on the soliton parameters which improves the algorithm used in \cite{berntsonklabbers2020}. 

There are various possible future directions to be explored;
as discussed in the introduction, the present work and \cite{berntson2020} suggest that the HWM equation can be quantized to yield a chiral CFT which also provides a second quantization of the trigonometric spin CM system, thereby generalizing the results in \cite{carey1999}. Quantizing the ncIHF equation \eqref{eq:ncIHF} would provide a further generalization of that correspondence. Since it is typically easier to quantize a system of finite size, a good first step would be to generalize results from this paper to the periodic case, i.e., the case where hyperbolic functions are generalized to elliptic ones. We hope to come back to this in a future publication. 

Another way to generalize the current setting is by $q$-deformation. A $q$-generalization of the Haldane-Shastry (HS) spin chain \cite{uglov1995,lamers2018} can be obtained from the spin Ruijsenaars-Schneider (RS) many-body system \cite{kricheverzabrodin1995,arutyunov1998}, an integrability-preserving relativistic generalization of the spin CM system. In the nonrelativistic case, continuum HS dynamics are described by the HWM equation \cite{lenzmann2020}, whose solitons are governed by spin CM dynamics \cite{berntson2020}. In analogy, we expect that a continuum description of the $q$-HS spin chain would have solitons governed by spin RS dynamics. 

The directions above relate to the large-$\delta$ limit of the ncIHF equation \eqref{eq:ncIHF}, in which the hyperbolic functions degenerate to rational ones. By contrast, the (chiral) IHF equation \eqref{eq:IHF} is interesting in the small-$\delta$ limit as it reduces to the Heisenberg ferromagnetic (HF) equation in the limit $\delta\to 0$. We showed that the IHF equation can be obtained as a reduction from the ncIHF equation, which suggests that the multi-soliton solutions obtained in the present paper might reduce to solutions of the IHF equation and, by taking the limit $\delta\to 0$, to solutions of the HF equation. However, there is a mismatch of boundary conditions which makes this reduction difficult to use. We thus regard the construction of soliton solutions of the IHF equation as a particularly interesting open problem. 

Finally, it would be interesting to study the continuum dynamics of full spin CM models (rather than their infinite mass limits to spin chains), in generalization of the results for the BO equation in Ref.~\cite{abanov2009}. Proposals for such continuum equations have been constructed in \cite{kulkarni2009,xing2015}, but we believe that further interesting results in this direction remain to be found.

\section*{Acknowledgements}
BKB and EL thank Jonatan Lenells for collaboration on a closely related project. We are grateful to Patrick G\'{e}rard, Enno Lenzmann, Masatoshi Noumi, Junichi Shiraishi, Michael Stone, and Istv\'{a}n M. Sz\'{e}cs\'{e}nyi for inspiring and helpful discussions. 
BKB acknowledges support from the European Research Council, Grant Agreement No. 682537, and the Stiftelse Olle Engkvist Byggm\"{a}stare, Grant 211-0122. 
The work of RK was supported by the grant ``Exact Results in Gauge and String Theories'' from the Knut and Alice Wallenberg foundation. RK thanks \textsc{nordita}, where most of this work was conducted, for their hospitality. 

\appendix

\section{Hamilton formulation}
\label{app:EOM}
We derive two facts about the Hamilton formulation of the ncIHF equation: (i) the ncIHF equation is identical with the Hamilton equations $\bu_t=\{\cH,\bu\}$ and $\bv_t=\{\cH,\bv\}$  for the Hamiltonian and Poisson brackets in \eqref{eq:ncIHFHamiltonian} and \eqref{eq:ncIHFPoissonbrackets}, respectively (Section~\ref{subapp:HE}), (ii)  the Hamiltonians $\cH$ in   \eqref{eq:ncIHFHamiltonian}  and  \eqref{eq:ncIHFHamiltonian1} are the same for functions $\bu$ and $\bv$ obeying the asymptotic conditions \eqref{eq:BC} (Section~\ref{subapp:alternativecH}).

\subsection{Hamilton equations}
\label{subapp:HE}
To avoid technical issues explained in Appendix~\ref{subapp:realline} below, we give the argument in the periodic case; in this case, the Hamiltonian is as in \eqref{eq:ncIHFHamiltonian} but with $\int_{\R}$ replaced by $\int_{-L/2}^{L/2}$, and the Dirac delta in the Poisson brackets  \eqref{eq:ncIHFPoissonbrackets} are $L$-periodic. 

We note that, for $f,g,h$ any of the six functions $u^{a}$ and $v^{a}$ for $a=1,2,3$, one can simplify the computation of Poisson brackets using the identities
\begin{equation} 
\label{eq:Txidentity}
\begin{split} 
\int_{-L/2}^{L/2} f(x) \{ (T g_x)(x),h(x')\} \, \dd{x} = \int_{L/2}^{L/2} \{g(x),h(x')\}(T f_x)(x)\, \dd{x} , \\
\int_{L/2}^{L/2} f(x) \{ (\tT g_x)(x),h(x')\} \, \dd{x} = \int_{-L/2}^{L/2} \{g(x),h(x')\}(\tT f_x)(x)\, \dd{x} .
\end{split} 
\end{equation} 
To justify these identities, we use Fourier transform $\hat f(k)=\int_{-L/2}^{L/2}f(x)\ee^{-\ii kx} \, \dd x$ with $k\in(2\pi/L)\Z$, the Plancherel theorem, $\widehat{f_x}(k)=-\ii k\hat f(k)$, and the representations of $T$ and $\tT$ in Fourier space \eqref{eq:TT} to compute\footnote{We note in passing that $0\coth(0\delta)$ and $0/\sinh(0\delta)$ below should be interpreted as $0$ \cite[Proposition B.1]{berntsonlangmann2021}.} \begin{equation} 
\label{eq:Id} 
\begin{split} 
\int_{-L/2}^{L/2} f(x) (T g_x)(x)\, \dd{x} = &\frac1{L} \sum_{k\in(2\pi/L)\Z} \hat f(k) k\coth(k\delta)\hat g(-k)\, \dd{k}  = \int_{-L/2}^{L/2} (T f_x)(x)g(x)\, \dd{x} , \\
\int_{-L/2}^{L/2} f(x) (\tT g_x)(x)\, \dd{x} = &\frac1{L}  \sum_{k\in(2\pi/L)\Z} \hat f(k) \frac{k}{\sinh(k\delta)}\hat g(-k)\, \dd{k}  = \int_{-L/2}^{L/2} (\tT f_x)(x)g(x)\, \dd{x} .
\end{split} 
\end{equation} 
Inserting the definition of $T$ and $\tT$ in position space \eqref{eq:TTe}, one can use this to verify \eqref{eq:Txidentity}. 

We write the Hamiltonian \eqref{eq:ncIHFHamiltonian} as 
\begin{equation}
\label{eq:ncIHFHamiltonian2}
\mathcal{H}=-\frac1{2}\int_{-L/2}^{L/2} \left( u^c(x')(Tu^c_{x'})(x')+v^c(x')(Tv^c_{x'})(x') -u^c(x')(\tT v^c_{x'})(x') -v^c(x')(\tT u^c_{x'})(x') \right)\,\dd{x'}
\end{equation}
and use  the Poisson brackets \eqref{eq:ncIHFPoissonbrackets} and \eqref{eq:Txidentity} to compute (we suppress the time dependence of the functions),  
\begin{equation} 
\begin{split} 
u_t^a(x) = & \{\cH,u^{a}(x)\} \\
= & -\frac1{2}\int_{-L/2}^{L/2} \left( 2\{u^c(x'),u^{a}(x)\}(Tu^c_{x'})(x') -2 \{u^c(x'),u^{a}(x)\}(\tT v^c_{x'})(x')\right)\,\dd{x'}\\
= &  \int_{-L/2}^{L/2}\eps_{cab}\delta(x'-x)u^b(x) \left( (Tu^c_{x'})(x') - (\tT v^c_{x'})(x')\right)\,\dd{x'}\\
= & \eps_{abc} u^b(x)  \left( (Tu^c_{x})(x) - (\tT v^c_x)(x) \right) 
\end{split} 
\end{equation} 
equivalent to the first equation in \eqref{eq:ncIHF}; the derivation of the second equation in \eqref{eq:ncIHF}Â  is similar and thus omitted. 

The result in the real-line case is obtained in the limit $L\to\infty$. 

\subsection{Real-line case} 
\label{subapp:realline} 
It would be interesting to give the argument in Section~\ref{app:EOM} directly for real-line case. However, this is challenging due to the following technical complication: since $\bu$ and $\bv$ are $S^2$-valued, their limits cannot decay as $x\to\pm\infty$, and therefore their Fourier transforms exist only in a distributional sense. 
Thus, the $T$ and  $\tT$ transforms of $\bu$ and $\bv$ do not exist, and while the $T$ and $\tT$ transforms of $\bu_x$ and $\bv_x$ exist (since the differentiation removes the constant asymptotic terms), integrals like $\int_{\R}\bu\cdot T\bu_x\, \dd{x}$ are divergent; for these reasons, it is not even clear that the Hamiltonian  \eqref{eq:ncIHFHamiltonian} is well-defined (which is why we show this in Section~\ref{subapp:alternativecH}). 

To address this issue, one can use the decompositions 
\begin{equation}
\label{eq:bubvdecomposition}
\begin{split}
\bu(x,t)=&\;\bu_0(x,t)+\bm_0+s(x)\bn, \\
\bv(x,t)=&\;\bv_0(x,t)+\bm_0+s(x)\bn ,
\end{split}
\end{equation}
such that $\lim_{x\to\pm\infty}\bu_0(x,t)=\lim_{x\to\pm\infty}\bv_0(x,t)=0$, with $s(x)$ any function of $x\in\R$ such that $\lim_{x\to\pm \infty}s(x)=\pm 1$ (for example, $s(x)=\tanh(\kappa x)$), and $\bm_0$ and $\bn$ such that $\bm_{\pm\infty}=\bm_0\pm\bn$ (cf.\ \eqref{eq:BC}).\footnote{One could require that $s_x$, $\bu_0$ and $\bv_0$ are, for instance, Schwartz functions in $x$.} 
The technical issue described above becomes manifest in Fourier space: 
\begin{equation}
\label{eq:bubvdecomposition1}
\begin{split}
\hat \bu(k,t)=&\;\hat \bu_0(k,t)+2\pi\delta(k) \bm_0+\frac{\ii}{k}\hat s_0(k)\bn, \\
\hat \bv(k,t)=&\;\hat \bv_0(k,t)+2\pi\delta(k)\bm_0+\frac{\ii}{k}\hat s_0(k)\bn ,
\end{split}\quad (k\in\R)
\end{equation}
where $\hat s_0(k)$ is the Fourier transform of $s_x(x)$. 
Thus, to show that \eqref{eq:ncIHFHamiltonian} is well-defined, one can insert the decompositions \eqref{eq:bubvdecomposition} and verify that all terms which are ill-defined cancel; this can be done by straightforward but tedious computations. 

\subsection{Alternative form of the Hamiltonian}
\label{subapp:alternativecH} 
To obtain the Hamiltonian \eqref{eq:ncIHFHamiltonian}  from the Hamiltonian \eqref{eq:ncIHFHamiltonian1}, insert the identities
\begin{equation}
\label{eq:VTidentities}
\begin{split}
\pvint_{\R} V(x'-x)\left(\bu(x')-\bu(x)\right)^2\,\dd{x'}= & -2\pi\bu(x) \cdot (T\bu_x)(x)\\ &  -\kappa\left( \left(\bm_{+\infty}-\bu(x)\right)^2 + \left(\bm_{-\infty}-\bu(x)\right)^2\right) , \\
\int_{\R} \tV(x'-x)\left(\bu(x')-\bv(x)\right)^2\,\dd{x'} = &-2\pi\bu(x)\cdot (\tT\bv_x)(x) \\ &  -\kappa\left( \left(\bm_{+\infty}-\bv(x)\right)^2 + \left(\bm_{-\infty}-\bv(x)\right)^2\right), 
\end{split}
\end{equation}
and similar ones with $\bu$ and $\bv$ interchanged, and observe that the boundary terms add up to zero. 

We justify the first identity in \eqref{eq:VTidentities}: we insert $V(x'-x)=-\partial_{x'}\alpha(x'-x)$ and use integration by parts to compute, recalling  \eqref{eq:alpha} and \eqref{eq:TTalpha},  
\begin{equation} 
\begin{split} 
\mathrm{LHS} =& -\lim_{\eps\downarrow 0}\left(\int_{-\infty}^{x-\eps}+\int_{x+\eps}^\infty\right)\left(\partial_{x'}\alpha(x'-x) \right) \left(\bu(x')-\bu(x)\right)^2\,\dd{x'} = \\
 &-\lim_{\eps\downarrow 0}\left( \left. \alpha(x'-x)\left(\bu(x')-\bu(x)\right)^2\right|_{x'=-\infty}^{x-\eps} +\left. \alpha(x'-x)\left(\bu(x')-\bu(x)\right)^2\right|_{x'=x+\eps}^{\infty} \right) \\
&+ \pvint_{\R} \alpha(x'-x)2\left(\bu(x')-\bu(x)\right)\cdot\bu_{x'}(x')\, \dd{x'}\\
= & 
-\kappa\left( \left(\bm_{+\infty}-\bu(x)\right)^2 + \left(\bm_{-\infty}-\bu(x)\right)^2\right)- 2\bu(x)\cdot \pvint_{\R} \alpha(x'-x) \bu_{x'}(x')\, \dd{x'} = \mathrm{RHS}
\end{split} 
\end{equation} 
since $\alpha(\pm\eps)\left(\bu(x\pm\eps)-\bu(x)\right)^2\to 0$ as $\eps\downarrow 0$ for differentiable functions $\bu(x')$ (we used that $\lim_{x\to\pm\infty}\alpha(x)=\pm\kappa$ and $\bu(x')\cdot \bu_{x'}(x')=0$; the latter is implied by  $\bu(x')^2=1$). The computation giving the second identity in  \eqref{eq:VTidentities} is similar but simpler and thus omitted. 

\section{Details on the Heisenberg ferromagnet limit}
\label{app:HFlimit}
We derive two results about local limits: (i) we show that the the IHF equation $\bu_t=\bu\wedge T\bu_x$  reduces to the HF equation $\bu_t=\bu\wedge\bu_{xx}$ in the  limit $\delta\to 0^+$ (Appendix~\ref{subapp:IHF}), (ii) we argue that the ncIHF equation \eqref{eq:ncIHF} does not have a well-defined such limit  (Appendix~\ref{subapp:ncIHF}). 

To obtain these results, we use the following $\delta$-expansions of the integral operators in \eqref{eq:TTh} \cite[Appendix B]{berntsonlangmann2020}, 
 \begin{equation}
 \label{eq:Texpansion}
\begin{dcases}(Tf)(x)=-\frac1{2\delta}\int_{-\infty}^x f(z)\,\dd{z}+\frac1{2\delta}\int_{x}^{\infty} f(z)\,\dd{z}+\frac{\delta}3 f^\prime(x)+\mathrm{O}(\delta^3), \\
(\widetilde{T}f)(x)=-\frac1{2\delta}\int_{-\infty}^x f(z)\,\dd{z}+\frac1{2\delta}\int_{x}^{\infty} f(z)\,\dd{z}-\frac{\delta}6 f^\prime(x)+\mathrm{O}(\delta^3) ,
\end{dcases}
\qquad \text{as } \delta\to 0^+. 
\end{equation}
 They, together with the boundary conditions \eqref{eq:BC}, imply 
\begin{equation}
\label{eq:Tuexpansion}
\begin{dcases}
(T\bu_x)(x)=-\frac{1}{\delta}\bu(x)+\frac{1}{2\delta}(\bm_{-\infty}+\bm_{\infty})+\frac{\delta}{3}\bu_{xx}(x)+\mathrm{O}(\delta^3), \\
(\tT\bu_x)(x)=-\frac{1}{\delta}\bu(x)+\frac{1}{2\delta}(\bm_{-\infty}+\bm_{\infty})-\frac{\delta}{6}\bu_{xx}(x)+\mathrm{O}(\delta^3),
\end{dcases}\qquad \text{as } \delta\to 0^+, 
\end{equation}
and similarly for $\bv$. 

\subsection{Intermediate Heisenberg ferromagnet equation}   
\label{subapp:IHF} 

We insert the $\delta$-expansion of $T\bu_x$ in \eqref{eq:Tuexpansion} into the IHF equation  $\bu_t=\bu\wedge T\bu_x$ and divide by $\delta/3$ to obtain 
\begin{equation} 
\frac{3}{\delta}\bu_t = \frac{3}{2\delta^2}\bu\wedge(\bm_{-\infty}+\bm_{\infty}) + \bu\wedge\bu_{xx}+ \mathrm{O}(\delta^2). 
\end{equation} 
This shows that, if we scale time: $t\delta/3\to t$, and impose the condition $\bm_{-\infty}=-\bm_{\infty}$ equivalent to $\lim_{x\to\pm\infty}\bu(x,t)=\pm \bm_{\infty}$, we can take the limit $\delta\to 0^+$ and obtain the HF equation $\bu_t=\bu\wedge \bu_{xx}$. 

\subsection{Non-chiral intermediate Heisenberg ferromagnet equation}
\label{subapp:ncIHF} 
We insert the $\delta$-expansions in \eqref{eq:Tuexpansion}  into the ncIHF equation \eqref{eq:ncIHF} and divide by $\delta/6$ to obtain 
\begin{equation}
\begin{split}
\frac{6}{\delta}\bu_t=&\;  \frac{6}{\delta^2} \bu\wedge\bv+\bu\wedge(2\bu_{xx}+\bv_{xx}) + \mathrm{O}(\delta^2),  \\
\frac{6}{\delta}\bv_t=&\; \frac{6}{\delta^2}\bu\wedge\bv-\bv\wedge(2\bv_{xx}+\bu_{xx}) + \mathrm{O}(\delta^2).
\end{split}
\end{equation}
This shows that, even if we scale time: $t\delta/6\to t$, the terms $\frac{6}{\delta^2} \bu\wedge\bv$ spoil the possibility to take the limit $\delta\to 0^+$. 

One can try to also scale the spatial variables,  $x\to sx$ with $s>0$, to introduce a second scaling parameters $s$; however, at closer inspection, one finds that this does not help: there is no non-trivial limit of $\delta$ and $s$ leading to a coupled system of HF-type equations. 

\section{Functional identities}
\label{app:propalpha}
We give well-known functional identities satisfied by the special functions $\alpha(z)$ in \eqref{eq:alpha} and $V(z)$ in \eqref{eq:V} (see, e.g., \cite{calogero1975}; we write them as needed in our construction of multi-soliton solutions of the ncIHF equation).

For $z \in \C$, $a, b \in \mathbb{C}$ such that $a\neq b$, $\delta>0$, and $\kappa=\pi/2\delta$, the following identities hold true, 
\begin{equation}
\label{eq:identities}  
\begin{aligned}
\alpha(z-a)\alpha(z-b)=&\; \alpha(a-b)\big(\alpha(z-a)-\alpha(z-b)\big) +\kappa^2, \\
\alpha(z-\ii\delta)=&\;\alpha(z+\ii\delta), \\
V(z)=&\;-\alpha^\prime(z)=\alpha(z)^2-\kappa^2,\\
\alpha(z-a)\alpha^\prime(z-b)=&\; - \alpha(a-b)\alpha^\prime(z-b) +\alpha^\prime(a-b)\big(\alpha(z-a)-\alpha(z-b)\big),\\
\alpha(-z)=&\;-\alpha(z),\quad V(-z)=V(z)
\end{aligned}
\end{equation} 
(they all can be verified by elementary methods). 

\section{Conservation of spin and energy}
\label{app:verification}
We derive two facts about conservation laws of the ncIHF equation \eqref{eq:ncIHF} stated in the main text: (i) we show by direct computations that the total spin $\bS$, defined in \eqref{eq:totalspin}, is conserved (Section~\ref{subapp:verificationbS}), (ii) we derive the relation \eqref{eq:Itwo} between the conservation law $\cI_2$ in \eqref{eq:In}, the Hamiltonian $\cH$ in \eqref{eq:ncIHFHamiltonian}, and $\bS$ in \eqref{eq:totalspin}, which implies energy conservation (Section~\ref{subapp:verificationcH}).

\subsection{Spin conservation} 
\label{subapp:verificationbS}
To avoid technicalities, we give the argument in the periodic case where $\bS=\int_{-L/2}^{L/2}(\bu-\bv)\,\dd{x}$.
A direct proof in the real-line case is outlined further below. 

\paragraph{Periodic case.}  We compute, inserting \eqref{eq:ncIHF}, 
\begin{equation}
\begin{split} 
\bS_{t}\coloneqq \frac{\dd}{\dd{t}}\bS
=&\;\int_{-L/2}^{L/2}(\bu_t-\bv_t)\,\dd{x} \\
=&\; \int_{-L/2}^{L/2} \left(\bu\wedge T\bu_x-\bu\wedge \tT\bv_x+\bv\wedge T\bv_x-\bv\wedge\tT\bu_x\right)\,\dd{x} . 
\end{split}
\end{equation}
We use that the operators $T\partial_x$ and $\tT\partial_x$ are self-adjoint (see \eqref{eq:Id}) to write this as 
\begin{equation} 
\begin{split} 
\bS_{t}
=&\;\int_{-L/2}^{L/2}\left((T\bu_x)\wedge\bu-(\tT\bu_x)\wedge\bv+(T\bv_x)\wedge\bv-(\tT\bv_x)\wedge \bu\right)\,\dd{x} \\
=&\; -\int_{-L/2}^{L/2} \left(\bu\wedge T\bu_x-\bu\wedge \tT\bv_x+\bv\wedge T\bv_x-\bv\wedge\tT\bu_x\right)\,\dd{x} . 
\end{split}
\end{equation}
We have two expressions for $\bS_{t}$ that differ by a sign. We conclude $\bS_{t}=\mathbf{0}$. 

The result for the real-line case can be obtained by taking the limit $L\to\infty$. 

\paragraph{Real-line case.} It would be interesting to prove $\bS_t=\mathbf{0}$ directly in the real-line case. 
This can be done by inserting the decompositions \eqref{eq:bubvdecomposition} into 
\begin{equation} 
\begin{split} 
\bS_{t}=&\; \int_\R \left(\bu\wedge T\bu_x-\bu\wedge \tT\bv_x+\bv\wedge T\bv_x-\bv\wedge\tT\bu_x\right)\,\dd{x} 
\end{split} 
\end{equation} 
and showing by straightforward but tedious computations that all terms which are ill-defined cancel, noting that $T-\tT$ in Fourier space, i.e., 
\begin{equation} 
\ii\coth(k\delta) - \frac{\ii}{\sinh(k\delta)}=\ii\tanh(k\delta/2) 
\end{equation} 
(see \eqref{eq:TTe}), vanishes linearly with $k$ as $k\to 0$. 
After that, one can use a similar argument as in the periodic case to prove $\bS_t=\mathbf{0}$. 

\subsection{Energy conservation} 
\label{subapp:verificationcH}

Following \cite{gerard2018}, we show that $\cI_2$ is a linear combination of Hamiltonian \eqref{eq:ncIHFHamiltonian} and the square of the the total spin \eqref{eq:totalspin}.

We compute, using \eqref{eq:KL}, \eqref{eq:In}, $\mathrm{tr}_{\C^2}(\ua\ub) = 2\ba\cdot\bb$ for three-vectors $\ba$ and $\bb$,  and that $\alpha(z)$ and $\talpha(z)$ are odd functions, 
\begin{equation}
\begin{split} 
\cI_2= \frac{2}{\pi^2}\int_{\R^2} \Bigl[ &\alpha(x'-x)^2\left(\left(\bu(x')-\bu(x)\right)^2+\left(\bv(x')-\bv(x)\right)^2\right)\\
- &\talpha(x'-x)^2\left(\left(\bu(x')-\bv(x)\right)^2+\left(\bv(x')-\bu(x)\right)^2\right) \Bigr]\,\dd{x'}\,\dd{x}.
\end{split} 
\end{equation}
By inserting the identities $\alpha(z)^2=V(z)+\kappa^2$ and  $\talpha(z)^2=\tV(z)+\kappa^2$  we obtain $\cI_2$ as a sum of two terms: the first is
\begin{equation}
\label{eq:alphaidentities}
\begin{split} 
\frac{2}{\pi^2}\int_{\R^2} \Bigl[ &V(x'-x)\left(\left(\bu(x')-\bu(x)\right)^2+\left(\bv(x')-\bv(x)\right)^2\right)\\
- &\tV(x'-x)\left(\left(\bu(x')-\bv(x)\right)^2+\left(\bv(x')-\bu(x)\right)^2\right) \Bigr]\,\dd{x'}\,\dd{x} =\frac{8}{\pi}\cH , 
\end{split} 
\end{equation}
recalling formula  \eqref{eq:ncIHFHamiltonian1} for the ncIHF Hamiltonian $\cH$ (note that the principal value prescription of the second integral in \eqref{eq:ncIHFHamiltonian1} can be ignored for differentiable functions $\bu$ and $\bv$); the second is 
\begin{equation}
\begin{split} 
\frac{2}{\pi^2}\kappa^2 \int_{\R^2} \Bigl[ &\left(\left(\bu(x')-\bu(x)\right)^2+\left(\bv(x')-\bv(x)\right)^2\right)\\
- &\left(\left(\bu(x')-\bv(x)\right)^2+\left(\bv(x')-\bu(x)\right)^2\right) \Bigr]\,\dd{x'}\,\dd{x} \\
= - \frac{4}{\pi^2}\kappa^2 \int_{\R^2} &\left( \bu(x')\cdot\bu(x)+\bv(x')\cdot\bv(x)-  \bu(x')\cdot\bv(x)+\bv(x')\cdot\bu(x)\right)\,\dd{x'}\,\dd{x}  \\
= - \frac{4}{\pi^2}\kappa^2 \int_{\R^2} &\left(\bu(x')-\bv(x')\right)\cdot \left(\bu(x)-\bv(x)\right) \,\dd{x'}\,\dd{x} = - \frac{4}{\pi^2}\kappa^2 \bS^2, 
\end{split} 
\end{equation}
recalling that $\bS=\int_{\R}\left( \bu-\bv\right)\,\dd{x}$ is the total spin (see also \eqref{eq:totalspin}). By adding these results we obtain \eqref{eq:Itwo}. 

\section{Proof of generalized Cotlar identity}
\label{app:Cotlar}
We prove the generalized Cotlar identity \eqref{eq:Cotlargen}.

Inserting the definition of the component-wise product \eqref{eq:Hadamard},  we find that \eqref{eq:Cotlargen} is equivalent to 
\begin{equation} 
\begin{split} 
\left(\begin{array}{c} T(f_1g_1)-\tT(f_2g_2) \\ \tT(f_1g_1)-T(f_2g_2) \end{array}\right) 
= \left(\begin{array}{c} {[}T(f_1)-\tT(f_2){]}g_1 \\ {[}\tT(f_1)-T(f_2){]}g_2 \end{array}\right)
+  \left(\begin{array}{c} f_1{[}T(g_1)-\tT(g_2){]} \\ f_2{[}\tT(g_1)-T(g_2){]} \end{array}\right)  \\
+ \left(\begin{array}{c}  T({[}T(f_1)-\tT(f_2){]} {[}T(g_1)-\tT(g_2){]})-\tT( {[}\tT(f_1)-T(f_2){]} {[}\tT(g_1)-T(g_2){]})\\  \tT({[}T(f_1)-\tT(f_2){]} {[}T(g_1)-\tT(g_2){]})-T( {[}\tT(f_1)-T(f_2){]} {[}\tT(g_1)-T(g_2){]})\end{array}\right) 
\end{split} 
\end{equation} 
for all $\C$-valued functions $f_j,g_j$ of $x\in\R$ with well-defined Fourier transforms (in the classical, not distributional, sense). 
This is equivalent to 8 equations: for each component of the two-vector and each pair of functions $f_j,g_{j'}$, $j,j'=1,2$, there is one equation. 
Writing down these equations and simplifying the notation by writing $f,g$ short for $f_j,g_{j'}$,  we obtain the following 
4 distinct equations: 
\begin{equation} 
\label{eq:Cotlargen1}
\begin{split} 
&T(fg) -T(f)g-fT(g) -T(T(f)T(g))+\tT(\tT(f)\tT(g))=0,\\
&\tT(fg) +T(\tT(f)\tT(g))-\tT(T(f)T(g))=0,\\
&\tT(f)g +T(\tT(f)T(g))-\tT(T(f)\tT(g))=0,\\
&f\tT(g) +T(T(f)\tT(g)) - \tT((\tT(f)T(g))=0
\end{split}
\end{equation} 
(each of these equations appears twice). Thus, to prove \eqref{eq:Cotlargen}, we have to verify that the equations in \eqref{eq:Cotlargen1} hold true, for arbitrary functions $f,g$ with well-defined Fourier transforms. 

We recall that the action of $T$ and $\tT$ on Fourier transformed functions $\hat{f}(k)=\int_{\R} f(x)\ee^{-\ii kx}\,\dd{x}$  is as follows, 
$\widehat{(Tf)}(k)=\hat{T}(k)\hat f(k)$ and  $\widehat{(\tT f)}(k)=\hat{\tT}(k)\hat f(k)$ with 
\begin{equation} 
\label{eq:TktTk}
\hat{T}(k)\coloneqq \ii\coth(k\delta),\quad \hat{\tT}(k)\coloneqq \frac{\ii}{\sinh(k\delta)} .
\end{equation} 
Using also that the point-wise product in position space corresponds to convolution in Fourier space, we can write the first identity in \eqref{eq:Cotlargen1} as  
\begin{equation} 
\begin{split} 
\int_{\R}\hat{f}(k-k')\hat{g}(k')\left[ \hat{T}(k) - \hat{T}(k-k')-\hat{T}(k') 
-\hat{T}(k)\hat{T}(k-k')\hat{T}(k')+\hat{\tT}(k)\hat{\tT}(k-k')\hat{\tT}(k') \right]\,\dd{k'}  = 0, 
\end{split} 
\end{equation} 
which holds true for arbitrary functions $f,g$ if and only if the expression in the square brackets vanishes. 
By this and a similar argument for the other three identities in \eqref{eq:Cotlargen1} we conclude that \eqref{eq:Cotlargen1} is true if and only if 
\begin{equation} 
\begin{split} 
&\hat{T}(k) -\hat{T}(k-k')-\hat{T}(k')-\hat{T}(k)\hat{T}(k-k')\hat{T}(k')+\hat{\tT}(k)\hat{\tT}(k-k')\hat{\tT}(k')=0,\\
&\hat{\tT}(k) +\hat{T}(k)\hat{\tT}(k-k')\hat{\tT}(k')-\hat{\tT}(k)\hat{T}(k-k')\hat{T}(k')=0,\\
&\hat{\tT}(k-k') +\hat{T}(k)\hat{\tT}(k-k')\hat{T}(k')-\hat{\tT}(k)\hat{T}(k-k')\hat{\tT}(k')=0,\\
&\hat{\tT}(k') +\hat{T}(k)\hat{T}(k-k')\hat{\tT}(k')-\hat{\tT}(k)\hat{\tT}(k-k')\hat{T}(k')=0, 
\end{split}
\end{equation} 
for all non-zero $k,k'\in\R$ such that $k\neq k'$. 
We insert \eqref{eq:TktTk}, use the short-hand $p=k\delta$, $p'=k'\delta$,  and multiply each equation with the factor $-\ii \sinh(p)\sinh(p-p')\sinh(p')$ to obtain 
\begin{equation} 
\label{eq:coshsinh}
\begin{split} 
&\cosh(p)\sinh(p-p')\sinh(p') -\sinh(p)\cosh(p-p')\sinh(p')-\sinh(p)\sinh(p-p')\cosh(p')\\ & +\cosh(p)\cosh(p-p')\cosh(p')-1=0,\\
&\sinh(p-p')\sinh(p') -\cosh(p)+\cosh(p-p')\cosh(p')=0,\\
&\sinh(p)\sinh(p') - \cosh(p)\cosh(p')+\cosh(p-p')=0,\\
&\sinh(p)\sinh(p-p') -\cosh(p)\cosh(p-p')+\cosh(p')=0.
\end{split}
\end{equation} 
The first of these equations can be written as $\cosh(p-p')[\cosh(p)\cosh(p')-\sinh(p)\sinh(p')]-\sinh(p-p')[\sinh(p)\cosh(p')-\cosh(p)\sinh(p')]=1$, 
which holds true due to the following well-known identities,  
\begin{equation} 
\label{eq:coshsinh1}
\begin{split} 
\cosh(z)\cosh(z')-\sinh(z)\sinh(z')=& \cosh(z-z'),\\
\sinh(z)\cosh(z')-\cosh(z)\sinh(z')= & \sinh(z-z'), \\
\cosh(z)^2-\sinh(z)^2=&1, 
\end{split} 
\end{equation} 
for $z,z'\in\C$. The remaining three equations in \eqref{eq:coshsinh} are equivalent to the first identity in \eqref{eq:coshsinh1}. 
This completes the proof of \eqref{eq:Cotlargen}.

\section{One-soliton energies: computational details}
\label{app:eps1}
We give details on the derivation of the results in \eqref{eq:eps1}--\eqref{eq:E1} and  \eqref{eq:EuEv}.

We set $N=1$ and insert $\tilde V(a-a^*)=\tV(2\ii\aI)=-\kappa^2/\cos^2(2\kappa\aI)$ into \eqref{eq:EN} to obtain $E_1$ in \eqref{eq:E1}. 
This suggests to write \eqref{eq:epsuepsv} for $N=1$ as 
\begin{equation} 
\left. \begin{array}{c} 
\eps_{\bu}(x)\\
\eps_{\bv}(x) \end{array}\right\} = E_1 f_\pm(x-\aR;\aI),\quad f_\pm(x;\aI) = \pm \frac1\pi \im\left(\alpha(x-\ii(\aI \mp \delta/2) )  \right) .
\end{equation} 
We insert $\alpha(z)=\kappa\coth(\kappa z)$ to obtain  
\begin{equation} 
\begin{split} 
f_\pm(x;\aI) = & \pm \frac{\kappa}{\pi}\frac{\sin(2\kappa\aI\mp\pi/2)}{\cosh(2\kappa x)-\cos(2\kappa\aI\mp\pi/2)}\\
= &  - \frac{\kappa}{\pi}\frac{\cos(2\kappa\aI)}{\cosh(2\kappa x)\mp \sin(2\kappa\aI)}, 
\end{split} 
\end{equation}  
which both are $\geq 0$ since $\cos(2\kappa\aI)<0$ for $\delta/2<\aI<3\delta/2$. This gives the result in \eqref{eq:eps1} since 
\begin{equation} 
f_+(x;\aI)+f_-(x;\aI) =   - \frac{2\kappa}{\pi}\left( \frac{\cos(2\kappa\aI)}{\cosh(2\kappa x)- \sin(2\kappa\aI)} + \frac{\cos(2\kappa\aI)}{\cosh(2\kappa x)+ \sin(2\kappa\aI)}\right) 
\end{equation} 
is equal to $f(x;\aI)$ in  \eqref{eq:eps1}. 

One can compute the fraction of the total energy in the $\bu$- and $\bv$-channels using the following  exact integrals: 
\begin{equation} 
\label{eq:uvweight}
\int_{\R} f_\pm(x;\aI)\,\dd{x} = \frac12 \pm \left(1-\frac{\aI}{\delta} \right)\quad (\delta/2<\aI<3\delta/2); 
\end{equation} 
this implies \eqref{eq:EuEv}. For the convenience of the reader, we mention that \eqref{eq:uvweight} can be verified by the change of variables $z=\ee^{2\kappa x}$, transforming the integral on the left-hand side to 
\begin{equation} 
-\frac{\cos(\alpha)}{ \pi}\int_0^\infty \frac{\dd{z}}{(z\pm\ii\ee^{\ii\alpha})(z \mp \ii\ee^{-\ii\alpha})}\quad (\alpha=2\kappa\aI), 
\end{equation} 
which can be computed using a partial fraction decomposition of the integrand.

\bibliographystyle{unsrt}	

\bibliography{BKL3}

\end{document}